\documentclass[useAMS,usenatbib]{mn2e}

\usepackage{aas_macros}
\usepackage{graphicx}
\usepackage{ulem}
\usepackage{lscape}
\usepackage{mathptmx}

\def\Lya{\mbox{Ly$\alpha$}}
\def\lymana{\mbox Lyman-$\alpha$}
\def\Lyb{\mbox{Ly$\beta$}}
\newcommand{\lya}{Ly$\alpha$}
\newcommand{\lyb}{Ly$\beta$}

\newcommand{\K}{{\,{\rm K}}}
\newcommand{\cm}{{\rm cm}}
\newcommand{\kms}{\,\,{\rm km}\,{\rm s}^{-1}}

\def\HI{\mbox{H\,{\sc i}}}
\def\CII{\mbox{C\,{\sc ii}}}
\def\CIII{\mbox{C\,{\sc iii}}}
\def\CIV{\mbox{C\,{\sc iv}}}
\def\NV{\mbox{N\,{\sc v}}}
\def\NIII{\mbox{N\,{\sc iii}}}
\def\OI{\mbox{O\,{\sc i}}}
\def\OVI{\mbox{O\,{\sc vi}}}
\def\MgII{\mbox{Mg\,{\sc ii}}}
\def\AlII{\mbox{Al\,{\sc ii}}}
\def\AlIII{\mbox{Al\,{\sc iii}}}
\def\SiII{\mbox{Si\,{\sc ii}}}
\def\SiIII{\mbox{Si\,{\sc iii}}}
\def\SiIV{\mbox{Si\,{\sc iv}}}
\def\FeII{\mbox{Fe\,{\sc ii}}}
\def\NeVIII{\mbox{Ne\,{\sc viii}}}

\def\HIss{{\rm {H\,\sc{I}}}}

\title[Probing the Circumgalactic Medium at High-Redshift]{Probing the Circumgalactic Medium at High-Redshift Using Composite BOSS Spectra of Strong  Lyman-$\alpha$ Forest Absorbers}
\author[Pieri M. M. et al.]
{Matthew M. Pieri$^{1,2}$\thanks{E-mail:matthew.pieri@port.ac.uk},
 Michael J. Mortonson$^{3,8}$,
 Stephan Frank$^{4}$,
 Neil Crighton$^{5,6}$,
 \newauthor
David H. Weinberg$^{4}$,
 Khee-Gan Lee$^{6}$,
 Pasquier Noterdaeme$^{7}$,
 Stephen Bailey$^{8}$, \newauthor
Nicolas Busca$^{9}$,
 Jian Ge$^{10}$,
 David Kirkby$^{11}$,
 Britt Lundgren$^{12}$,
 Smita Mathur$^{4}$,
  \newauthor
 Isabelle Paris$^{13}$,
 Nathalie Palanque-Delabrouille$^{7}$,
 Patrick Petitjean$^{7}$,
  James Rich$^{14}$,   \newauthor
Nicholas P. Ross$^{8}$,
 Donald P. Schneider$^{15}$
 and Donald G. York$^{16}$\\
 $^{1}$Institute of Cosmology \& Gravitation, University of Portsmouth, Dennis Sciama Building, Portsmouth PO1 3FX, UK\\
 $^{2}$CASA, Department of Astrophysical and Planetary Sciences, University of Colorado, 389-UCB, Boulder, CO, USA 80309\\
$^{3}$Space Sciences Lab and Department of Astronomy, University of California, Berkeley, CA 94720, USA\\
$^{4}$Department of Astronomy and CCAPP, Ohio State University, Columbus, OH 43210, USA\\
$^{5}$Centre for Astrophysics and Supercomputing, Swinburne University of Technology, PO Box 218, Victoria 3122, Australia\\
$^{6}$Max-Planck-Institut f\"{u}r Astronomie, K\"{o}nigstuhl 17, 69117 Heidelberg, Germany\\
$^{7}$Institut d'Astrophysique de Paris, CNRS-UPMC, UMR 7095, 98bis bd Arago, 75014, Paris, France noterdaeme@iap.fr\\
$^{8}$Lawrence Berkeley National Lab, 1 Cyclotron Rd, Berkeley CA, 94720, USA\\
$^{9}$APC, Universit\'{e} Paris Diderot-Paris 7, CNRS/IN2P3, CEA, Observatoire de Paris, 10, rueA. Domon \& L. Duquet,  Paris, France\\
$^{10}$Astronomy Department, University of Florida, 211 Bryant Space Science Center, Gainesville, FL 32611-2055\\
$^{11}$Department of Physics and Astronomy, University of California, Irvine, 4129 Frederick Reines Hall, Irvine, CA 92697-4575, USA\\
$^{12}$Department of Astronomy, University of Wisconsin, 475 North Charter Street, Madison, WI 53706, USA\\
$^{13}$Departamento de Astronom\'ia, Universidad de Chile, Casilla 36-D, Santiago, Chile\\
$^{14}$CEA, Centre de Saclay, IRFU,  F-91191 Gif-sur-Yvette, France\\
$^{15}$Department of Astronomy and Astrophysics, and Institute for Gravitation and the Cosmos, The Pennsylvania State University, University Park, PA 16802\\
$^{16}$Department of Astronomy and Astrophysics and the Enrico Fermi Institute, The University of Chicago, 5640 South Ellis Avenue, Chicago, Illinois, 60615, USA\\
}
\begin{document}

\date{Accepted xxxx}

\pagerange{\pageref{firstpage}--\pageref{lastpage}} \pubyear{x}

\maketitle

\label{firstpage}

\begin{abstract}

We present composite spectra constructed from a sample of 242,150  \lymana\ (\lya) forest absorbers at redshifts $2.4<z<3.1$ identified in quasar spectra from the Baryon Oscillation Spectroscopic Survey (BOSS) as part of Data Release 9 of the Sloan Digital Sky Survey III. We select forest absorbers by their flux  in bins $138\kms$ wide (approximately the size of the BOSS resolution element). We split these absorbers into five samples spanning the range of flux $-0.05 \le F<0.45$. 
Tests on a smaller set of high-resolution spectra show that our
three strongest absorption samples would probe circumgalactic regions
(projected separation $< 300$ proper kpc and $|\Delta v| < 300\kms$)
in about 60\% of cases for very high signal-to-noise ratio.
Within this subset, weakening \lya\ absorption is associated with decreasing purity of circumgalactic selection once BOSS noise is included. Our weaker two \lya\ absorption samples are dominated by the intergalactic medium. 

We present composite spectra of these samples and a catalogue of measured absorption features from \HI\ and 13 metal ionization species, all of which we make available to the community. 
We compare measurements of seven Lyman series transitions in our composite spectra to single line models and obtain further constraints from their associated excess Lyman limit opacity. This analysis provides results consistent with column densities over the range $14.4 \la \log (N_{\rm HI}) \la 16.45$. We compare our measurements of metal absorption to a variety of simple single-line, single-phase models for a preliminary interpretation. Our results imply clumping on scales down to $\sim 30$ pc and near-solar metallicities in the circumgalactic samples, while high-ionization metal absorption consistent with typical IGM densities and  metallicities is visible in all samples.

\end{abstract}


\section{Introduction}

Since the birth of the first stars, the formation and evolution of galaxies has been intertwined with the evolution of intergalactic and circumgalactic media. These reservoirs of gas feed star formation and galaxy assembly, but also reflect the history of star formation by virtue of the presence and properties of heavy elements.
Knowledge of the degree of enrichment by metals in the intergalactic medium (IGM) and circumgalactic medium (CGM) allows exploration of the energetics of mechanical outflows from galaxies (e.g. \citealt{1999ApJ...513..142M, 2002ApJ...574..590S, 2007ApJ...658...36P, 2010ApJ...725.2087P}).
These outflows are a key component in our understanding of galaxy formation as they have the potential to heat and eject gas from dark matter halos, suppressing star formation, or increase the cooling rate by depositing metals, promoting star formation (e.g. \citealt{2001ApJ...560..599A, 2006MNRAS.373.1265O, 2010MNRAS.402.1536S, 2013MNRAS.430.3213B}). The pattern of elemental abundances present may also reveal the stellar populations that give rise to them. The abundance of individual ionization species of these elements allows investigation of the extragalactic UV background (due to the impact of photoionization), and the density and temperature of the gas (due to collisional ionization and recombination). A large number of metal species are required to measure all of these effects and break degeneracies between them \citep{2008ApJ...689..851A, 2010ApJ...724L..69P}. In this paper, we apply a novel analysis technique; we stack quasar spectra on the rest-frame locations of \lya\ absorbers (\citealt{2010ApJ...724L..69P}, P10 hereafter) in the enormous spectral database of the Baryon Oscillation Spectroscopic Survey (BOSS, \citealt{2013AJ....145...10D}). This affords the derivation of constraints on the metal enrichment and ionization state of intergalactic and circumgalactic gas.
 
The distribution of gas on large scales and its metal properties may be probed by measuring intervening absorbers along the line of sight to bright background quasars. The aptly named \lymana\ forest is a crowded distribution of absorption lines associated with trace amounts of neutral hydrogen (or \HI) in this largely photoionized medium \citep{1965ApJ...142.1633G, 1971ApJ...164L..73L}. 
It provides us detailed information on the density of structure over a significant range of redshifts for each quasar spectrum.
  Most of the \lymana\ (\lya) forest is associated with moderate overdensities ($\rho/\bar{\rho}\sim1$--10) and traces filamentary structure on large scales, but some strong forest absorbers along with Lyman limit systems and damped \lymana\ systems (DLAs) are thought to be associated with galaxies and the circumgalactic medium (e.g. \citealt{2011MNRAS.412L.118F, 2011MNRAS.418.1796F}).
  
These spectra also allow measurement of a variety of metal species in principle, but in practise only a limited number are detectable in the \lya\ forest. Three-times-ionized carbon or \CIV\ (e.g. \citealt{1987ApJ...315L...5M, 1995AJ....109.1522C, 2003ApJ...596..768S, 2006ApJ...638...45P}) and five-times-ionized oxygen or \OVI\ (e.g. \citealt{2000ApJ...541L...1S, 2004MNRAS.347..985P, 2004ApJ...606...92S, 2010ApJ...716.1084P}) are particularly well-observed. Many other weak lines are challenging to detect, but progress has been made towards this goal (P10) by stacking \lya\ forest lines in Sloan Digital Sky Survey II Data Release 5 (SDSS-II DR5; \citealt{2000AJ....120.1579Y,2007ApJS..172..634A}) quasar spectra.
  
Associations between galaxies and intervening absorbers are easier to establish at low-redshifts where the galaxies are readily identified (e.g. \citealt{ 2011Sci...334..948T,2012ApJ...760...49L,2013ApJ...763..148S,2013arXiv1304.6716N}). Significant challenges exist in surveying galaxies at high-redshift ($z>2$), particularly since much of this population is expected to be composed of faint, dwarf galaxies (e.g. \citealt{2007ApJ...662L...7P}, A. Rahmati et al in prep). Efforts are underway to build large samples of galaxies using the Lyman break technique (known as `Lyman break galaxies' or LBGs).  Samples such as the Keck Baryon Structure Survey (KBSS, \citealt{2010ApJ...717..289S, 2012ApJ...750...67R, 2012ApJ...751...94R}) and the VLT LBG survey \citep{2011MNRAS.414...28C} are being used to study the interaction of galaxies and the IGM/CGM, but these surveys have yet to show many weaker lines in optically thin gas. Progress has also been made by using observations of galaxies to explore the properties of their own circumgalactic regions \citep{2005ApJ...621..227M, 2010ApJ...717..289S}, but ambiguities exist between small-scale galactic effects and larger scale circumgalactic absorption.
 
We present an alternative approach using absorbers as a proxy for galactic and circumgalactic regions in order to construct a much larger sample and so have greater sensitivity to weaker metal transitions. This approach is not without precedent; \citet{2006ApJ...638...45P} used strong \CIV\ lines as a proxy for galaxy locations motivated by observations of LBGs  and nearby quasars \citep{2003ApJ...584...45A}. This approach was primarily used to investigate the existence of metals far from known galaxies.  \citet{2013ApJ...770..138L} use partial Lyman limit systems and Lyman limit systems as a proxy for galaxies at $z\la1$. In addition P10 stacked strong, blended \lya\ forest lines in the SDSS-II DR5 sample and found indications of circumgalactic conditions in the resultant composite spectra. 

We return to the analysis of P10 with methodological modifications, a larger data set from Data Release 9 \citep{2012ApJS..203...21A} of SDSS-III \citep{2011AJ....142...72E}, more extensive results, and investigations of our selection function using simulated spectra and high-resolution spectra with LBG proximity from the VLT LBG survey.
 We present a catalogue of composite spectra for varying \lya\ absorber strength, each with measurements of seven Lyman series lines, Lyman limit opacity, and 26 metal transitions from 7 elements and 13 ionization species. We also publish our composite spectra online to enable alternative measurements of the absorption profiles. We perform a preliminary interpretation of our measurements including constraints on gas metallicity and absorber sizes. In future work, we will carry out direct comparisons to the predictions of cosmological hydrodynamic simulations.

This paper is structured as follows. In Section~\ref{data}, we describe the data used in this analysis. In Section~\ref{lya selection}, we describe our selection  of \lya\ absorbers and investigate the kind of systems and environments it picks out, appealing to both simulations and  the VLT LBG analysis. Section~\ref{stacking procedure} sets out the method we used to produce composite spectra of these absorbers. In Section~\ref{results}, we describe our results including characterisation of \HI\ absorption, measurements of metals, and comparison to simple models. Section~\ref{discussion} provides a discussion of our results and it is followed by a summary. 

We assume a solar pattern of elemental abundances taken from \citet{1989GeCoA..53..197A} for our interpretation. There are a number of alternative choices available, but the differences are small on the scale of the effects discussed here.

\section[]{Data}
\label{data}

The Baryon Oscillation Spectroscopic Survey is one of four spectroscopic surveys that make up SDSS-III \citep{2011AJ....142...72E}, all conducted from the 2.5-meter Sloan telescope \citep{2006AJ....131.2332G} at Apache Point Observatory.
The BOSS instrument \citep{2013AJ....146...32S} consists of two double spectrographs each with a blue channel and a red channel.
We use the SDSS-III Data Release 9 \citep{2012ApJS..203...21A} high-redshift quasar sample, selected using the techniques
and data outlined in \citet{2012ApJS..199....3R} (see also \citealt{2011ApJ...729..141B}).
The sample of high-redshift quasars was compiled by visual inspection of quasar candidates as outlined in \cite{2012A&A...548A..66P}. 

The resolution of these spectra varies over  $R\approx 1650$ to $R\approx 2500$ and is broadly lower (higher) at the low (high) wavelength end of both the blue and red channel of each BOSS spectrograph. In the portion of the spectra we use for selecting \Lya\ absorbers, it varies over $1650\la R\la  2150$. We assume a constant value of R=2000 when fitting line properties, and this assumption will lead to small systematic errors in our line widths;
however, we do not draw detailed conclusions from line widths in this paper. The BOSS spectrograph provides coverage over the wavelength range 3450--10400\,\AA, although the signal-to-noise ratio (S/N) declines at the blue end, and much of the redward end of the spectra is discarded due to strong sky lines.

The SDSS spectroscopic pipeline  \citep{2012AJ....144..144B} performs the basic data reduction and calibration of quasar spectra.
For our analysis we make use of several, publicly available, value added products from \cite{2013AJ....145...69L} including principal component analysis (PCA) continua, a correction of spectral residuals (associated with a small error introduced by imperfect masking of Balmer lines in spectroscopic standard stars used for flux calibration) and zero-velocity calcium H+K absorption lines (likely associated with the Milky Way halo), a supplementary mask of sky lines in addition to those provided by the SDSS pipeline, and corrections for known biases in the pipeline estimated noise. There are 54,468 quasars in the \cite{2013AJ....145...69L} sample with redshifts providing intergalactic \lya\ forest coverage ($z_{\rm q}> 2.15$).

In order to arrive at a sample well-suited to the purpose of selecting absorbers in a meaningful way, we implement a quality cut described in Section~\ref{lya selection}, and as a result, only 30,555 of these spectra provide a portion of the forest in the desired redshift range of sufficient quality, with a total redshift path of $\Delta z =  3409$. This section also explains our decision to rebin our sample by a factor of 2 compared to the SDSS pipeline, giving $138\kms$ wide wavelength bins.
DLAs from the \cite{2012A&A...547L...1N} catalogue were excluded from the analysis. We reject DLAs in order to improve the homogeneity of our selection of strong, blended Lya forest lines and simplify interpretation (although we note that their omission had negligible impact due to their scarcity in the sample). Their cores were masked using the value added flags in  \cite{2013AJ....145...69L}. In addition, the wings of these lines were masked where the provided fits indicated a greater than 5\% flux decrement. These cuts produce a sample of 242,150 \lya\ forest absorbers in $138\kms$ wide bins in a renormalised flux range from $-0.05 \le F<0.45$.

We use two different methods for renormalising the quasar continua to probe intervening absorbers, one for our selection of \lya\ absorbers to stack and another for creating spectra to stack. PCA continua \citep{2012AJ....143...51L} are used for the selection of absorbers to stack, but these continua are limited to 1030-1600\,\AA\ in the quasar rest-frame. This approach is adequate for \lya\ selection since we only stack \lya\ absorbers that are free from contamination by higher order Lyman lines.

We perform an absorption-rejecting cubic spline fit to estimate the continuum level over the full spectral range. The PCA continua provide more robust absolute measurements of \lya\ absorbers, but our spline fits are of sufficient quality for the stacking of spectra, where it is only necessary that they be smooth with respect to the width of absorbers and that they remove the bulk of the quasar properties. Any remaining quasar properties will contribute only weak stochastic error in our composite spectra due to our pseudo-continuum fitting of stacked spectra (see Section~\ref{stacking procedure}).

We fit a cubic spline to nodes measured as the median flux in 20\,\AA\ wide chunks. We discard any blueward band with a 100-pixel ($6900\kms$) boxcar smoothed S/N $<1$ as our absorption rejection procedure performs poorly where the noise is greater than the signal. We iteratively refine the median flux of these nodes by rejecting negative deviations from the spline greater than the spectral error estimate in the forest and twice the spectral error outside the forest. We do not fit within $\pm 15$\,\AA\ of \lya\  and \CIV\ transitions in the quasar rest-frame since these emission line centres are too sharp to produce a good fit. We then discard points within $\pm 20$\,\AA\ of these features, an extra 5\,\AA\ padding, because spline fits can produce unstable results when presented with sharp edges.

\section{Lyman $\alpha$ absorber selection}
\label{lya selection}

Given the sensitivity of stacking analyses to the selection of objects, we have developed our selection with the aid of simulated spectra which are a close reproduction of those present in the BOSS \lya\ survey. These simulations are described in detail in our future paper (M. Mortonson et al. in prep), but we will briefly summarise them below (Section~\ref{prodsim}).  Our selection function is imposed on us predominantly by the resolution of the data. Our goal is to explore the statistical properties of systems selected by this procedure.

We select \lya\ absorbers in the redshift range $2.4<z<3.1$. This choice is motivated by our desire to obtain high quality spectral coverage of the Lyman limit at the blue limit and  \MgII\ at the red limit in stacked spectra.
The spectral resolution in our sample equates to a full width half maximum (FWHM) of 2--2.6 pixels ($138$--$179\kms$) in the standard SDSS spectral wavelength solution.
Because of the critical impact of S/N on our absorber selection, we rebin the spectra by a factor of 2 before selection to reduce noise, with only minimal loss of resolving power. This provides us a wavelength bin size of $138\kms$, and has the additional benefit of precluding double-counting of absorption from the same resolution element.

We use only the flux in these rescaled bins (with a quality cut described below) to select systems for stacking.  We present five samples of \lya\ absorbers for stacking, with $(0.1n-0.05) \le F < (0.1n+0.05)$, where $n=0,1,\ldots,4$. Our simulations described in Section~\ref{simselec} indicate that this value of $\Delta F$ is a good reflection of our ability to produce distinct samples when an appropriate quality cut is used. In order of increasing flux (and hence decreasing absorption) our samples include 11441, 25128, 42034, 65045 and 98502 absorbers.

\begin{figure}
\begin{center}
\includegraphics[angle=0, width=0.47\textwidth]{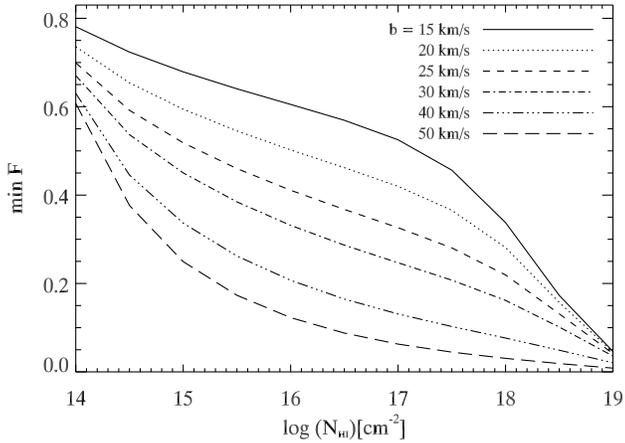}
\end{center}
\caption{The minimum \lya\ forest flux reached in single line models at BOSS resolution in a $138\kms$ bin.}
\label{blendingmod}
\end{figure}

At BOSS resolution with $138\kms$ bins, a \lya\ forest line of typical width with no damping wings does not reach saturation, regardless of its column density. 
The distribution of Doppler parameters in the \Lya\ forest is well characterised by a Gaussian with median $b\approx 30\kms$ and $\sigma=10 \kms$ that is cropped below $b=$15--20$\kms$ (e.g \citealt{1995AJ....110.1526H,2012ApJ...750...67R}). As Figure \ref{blendingmod} shows, only single lines with $\log N_{\rm HI} \ga 18$ or unusually high Doppler parameters ($>40 \kms$) reach a minimum flux below $F=0.15$. Therefore, fluxes below this limit generally arise from blends of lines, and even at higher fluxes blends of weak lines may dominate by number over isolated strong lines.
The 2-pixel rebinning increases the degree to which blending affects the selection. We discuss the significance of this selection function further in relation to simulations in Section~\ref{simselec} and circumgalactic regions in Section~\ref{selectiongalaxies}.

As we show in Section~\ref{results}, the inferred \HI\ column in our samples are characteristic of saturated lines that don't generate significant opacity at the Lyman limit in individual high-resolution spectra. This effect makes such systems challenging to identify without appealing to \lyb\ observations \citep{2002MNRAS.335..555K}, which are themselves difficult to confidently distinguish from \lya\ absorbers. However, our sample successfully probes this regime.

Our sample of absorbers is drawn from wavelengths $1041 {\rm \,\AA} < \lambda <1185 {\rm \,\AA}$  in the quasar rest-frame. We make this choice to eliminate the selection of \OVI\ absorbers and absorbers proximate to the quasar (within 7563$\kms$). We assume that all remaining absorbers arise due to the \lymana\ transition and so are available for selection. As we discuss in \S \ref{measuremet} this assumption is not always valid, but typically the cases where it does fail are distinct from the effects we intend to measure, and the breakdowns themselves provide useful information about metal interlopers.

\subsection{Production of simulated spectra}
\label{prodsim}
	
The simulated Ly$\alpha$ spectra we use to study issues related to 
absorber selection are generated by computing the transmitted flux fraction 
$F$ along random sightlines through an N-body+smoothed particle 
hydrodynamic (SPH) simulation in a $100\,h^{-1}\,{\rm Mpc}$ box 
with $2\times 576^3$ particles (with equal numbers of gas and dark matter 
particles). The simulation was run 
with a version of the code {\sc GADGET}-2 \citep{2005MNRAS.364.1105S} with several 
modifications described by \citet{2008MNRAS.387..577O}, including modeling 
of galactic outflows with a momentum-driven wind model. The cosmological 
parameters assumed for the simulation are consistent with results from
the Five-Year Wilkinson Microwave Anisotropy Probe (WMAP5, \citealt{2009ApJS..180..225H}).
For the analysis presented here, we use only the $z=2.5$ simulation output. 

For each sightline through the simulation volume, we run the 
{\sc SPECEXBIN} code \citep{2006MNRAS.373.1265O}, which computes the 
gas density, temperature, metallicity, and velocity contributed by 
all SPH particles intersecting the line of sight. It utilises those 
quantities to compute the redshift-space optical depth along the sightline
for Ly$\alpha$ and several metal lines, assuming ionization equilibrium
to determine the ionization fraction for a given density and temperature.
In this paper, we use simulations only to calibrate Ly$\alpha$ absorber
selection, so we only use the Ly$\alpha$ optical depth $\tau_{{\rm Ly}\alpha}$
from {\sc SPECEXBIN}; simulated spectra that include additional Lyman 
series lines and metal lines will be studied in our forthcoming comparison paper 
(M. Mortonson et al., in prep.), which will also describe the simulation itself in greater detail. 

To produce simulated spectra that match the absorber redshift distribution,
resolution, and noise properties of the BOSS DR9 sample, we use 
measurements from the \citet{2013AJ....145...69L} sample of quasar spectra and 
value added products. For each high-redshift quasar in this sample, we 
divide the observed Ly$\alpha$ forest wavelength range (specifically, 
$1041{\rm \,\AA} < \lambda < 1185 {\rm \,\AA}$ in the quasar rest-frame) into 
segments with length equal to the simulated sightlines described above, 
i.e.\ $100\,h^{-1}\,{\rm Mpc}$. Typically, 3--5 such segments are required to 
cover the Ly$\alpha$ forest for each quasar. For each segment, we randomly 
draw one of the simulated $100\,h^{-1}\,{\rm Mpc}$ sightlines and shift it 
by a random offset in the line of sight direction using the periodic 
boundary conditions of the simulation volume. We multiply 
all $\tau_{{\rm Ly}\alpha}$ values along the sightline by a factor
$\tau_{\rm eff}(z)/\tau_{\rm eff,sim}$, where $\tau_{\rm eff,sim}=0.233$ is 
the effective Ly$\alpha$ optical depth measured in the $z=2.5$ simulation
and $\tau_{\rm eff}(z)$ is the observed power-law relation from 
\citet{2008ApJ...681..831F} evaluated at the appropriate redshift
for Ly$\alpha$ absorption at the observed wavelength. 
All of the rescaled sightlines are combined 
into a single Ly$\alpha$ forest spectrum of transmitted flux
$F_{\rm sim} = \exp(-\tau_{{\rm Ly}\alpha})$ for each quasar.

We smooth the resulting spectrum $F_{\rm sim}(\lambda)$ 
to BOSS resolution by convolving it with a Gaussian with width equal
to the actual wavelength-dependent dispersion for the quasar spectrum from 
the DR9 sample. We then resample the smoothed flux on the SDSS 
wavelength solution; we designate the smoothed and pixelized (but still 
noise-free) flux fraction $F_0$.

Finally, we add simulated noise to the spectrum: $F=F_0+N$. For each pixel, we 
draw $N$ from a Gaussian with wavelength- and flux-dependent variance
\begin{equation}
\label{eq:noise_variance}
\sigma_F^2(\lambda) = k(\lambda)\,\frac{F_0(\lambda)C_{\rm obs}(\lambda)+
S_{\rm obs}(\lambda)} {C_{\rm obs}(\lambda)^2}\,,
\end{equation}
where $C_{\rm obs}(\lambda)$ and $S_{\rm obs}(\lambda)$ are the estimated 
continuum flux and sky background flux, respectively, from the 
actual quasar spectrum in the \citet{2013AJ....145...69L} catalog, 
and at each observed wavelength $\lambda$, $k(\lambda)$ is the coefficient 
that best fits the relation in equation~(\ref{eq:noise_variance}) as 
measured in the BOSS DR9 sample of quasar spectra (replacing $F_0$ with 
the observed flux fraction $F_{\rm obs}$ and using the estimated noise variance 
for $\sigma_F^2$).

\begin{figure}
\begin{center}
\includegraphics[angle=0, width=0.46\textwidth]{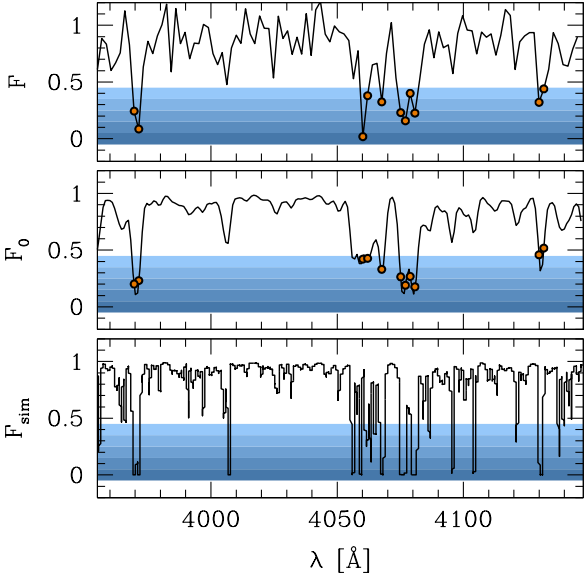}
\end{center}
\caption{
A portion of simulated \lya\ forest spectrum representing a typical BOSS spectrum (S/N $= 4.1$), with BOSS resolution, noise, and our 2-pixel rebinning is presented in the {\it top panel}. The shaded horizontal blue regions indicate the flux intervals used for absorber selection and the absorbers selected within these ranges are shown as red dots. The same simulated spectrum without noise or rebinning is displayed in the {\it middle panel} and the noiseless locations of selected absorbers are indicated by red dots. The {\it bottom panel} again shows the same spectrum, this time at the full resolution of the simulation. The shaded blue regions in the bottom two panels are displayed to aid comparison with the top panel.
}
\label{simulationexample}
\end{figure}

\subsection{Simulation motivated selection}
\label{simselec}

Figure \ref{simulationexample} provides an illustration of the typical spectral characteristics that form part of our selection. The bottom panel presents a portion of spectrum with simulation resolution (reproducing data with fully resolved lines) and no noise. The middle panel shows the same portion of simulated spectrum with BOSS resolution, our 2-pixel rebinning, and no noise. Comparison of these two panels confirms that low flux at high-resolution is not sufficient
to produce low flux at BOSS resolution; a degree of blending is also required. The top panel displays the same spectrum as the middle panel with the addition of typical noise in our sample. Note that we treat each wavelength bin that falls into our target flux range as an absorber. As is clear in this figure, adjacent strongly absorbed wavelength bins arise from extended structure in velocity space, and not thermal wings of lines.

Our key metric for \lya\ selection quality is the successful recovery of noiseless \lya\ flux in BOSS spectra. We maintain this quality standard by discarding selected absorbers in portions of the forest with a 100-pixel ($6900 \kms$) boxcar smoothed S/N per pixel below a required value. We perform this quality test by selecting simulated absorbers with noise for our five selection samples and examining the distribution of noiseless fluxes that went into this sample. 

\begin{figure}
\begin{center}
\includegraphics[angle=0, width=0.49\textwidth]{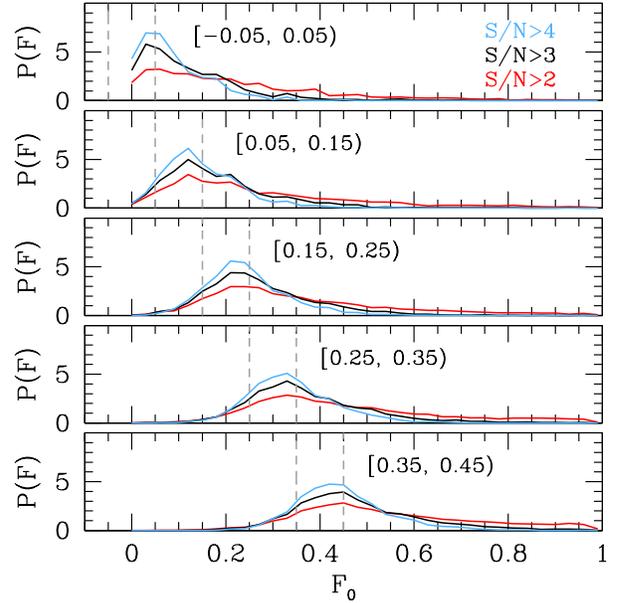}
\end{center}
\caption{The simulated \Lya\ recovery of true flux given noisy data with a varying band-wise S/N quality cut. The panels from top to bottom show the selection of absorbers within the range shown in the label with bracketed range in simulated spectra with realistic noise added. This value is compared to the `true' noise free flux in those spectra for three different S/N thresholds.}
\label{selectiontest}
\end{figure}

Figure \ref{selectiontest} shows the results of this test for all our selection samples, whose boundaries are shown by dashed lines. In each case, the band-wise S/N $> 3$ provides acceptable noiseless flux recovery (the most inclusive cut that allows the peak in the $P(F)$ of true flux ($F_0$) to lie within the selected flux range in all cases), and we adopt this as our preferred quality cut. This figure also demonstrates that there will always be a large contribution from higher flux pixels when selecting strong absorbers, as there are many more weak absorbers that may enter the selection due to noise (see Section \ref{selectiongalaxies} for further discussion). Our lowest absorption sample (top panel) is a special case intended to recover pixels with saturated absorption. Since the true flux is never negative, the distribution of noiseless fluxes in this selection is necessarily highly asymmetric, and the selection is strongly contaminated by weaker absorbers. Note that while our simulation does include systems with $N_{\rm HI} > 10^{16} \cm^{-2}$ that dominate this sample (as we shall show in Section~\ref{char hi}), more work is needed to compare the detailed incidence of such systems to observations.

\subsection{Comparison between \HI\ absorption and galaxies}
\label{selectiongalaxies}

The results from KBSS \citep{2012ApJ...750...67R, 2012ApJ...751...94R} provide useful indications of the relationship between our strong \lya\ absorbers and Lyman break galaxies (LBG) at redshifts $z=$2--3. The KBSS surveyed for LBGs in the fields of 15 bright quasars with Keck/HIRES spectra enabling absorption measurements characteristics of circumgalactic gas on scales of 50--3000 proper kpc (pkpc). \citet{2012ApJ...750...67R} argue that the CGM is best characterised by portions of quasars absorption spectra within 300~pkpc transverse separation of LBGs and $\pm300\kms$ along the line of sight. They find that nearly half of \HI\ absorbers with $\log N_{\rm HI} > 15.5$ arise in circumgalactic regions (their figure 30) and that absorbers with $\log N_{\rm HI} > 14.5$ cluster strongly with LBGs.

\citet{2012ApJ...751...94R} take a pixel optical depth approach in their investigation. In lines of sight separated from LBGs by $<130$~pkpc, the median $\tau_{\rm Ly\alpha} \approx 10$ on scales of $\pm150 \kms$ along the line of sight was measured. This scale is the size of the BOSS resolution element, and 70\% of the pixels in these regions have $2 < \tau_{\rm Ly\alpha} < 50 $. \citet{2012ApJ...751...94R} argue that the regions probed are the virialised halos of
 LBGs and that the relatively high velocities along the line of sight correspond to peculiar motions of this gas.

We will present evidence from Lyman series and Lyman limit absorption that our three strongest absorber samples ($F<0.25$) are associated with typical \HI\ column densities of $10^{15.55} \cm^{-2}$ and above. The KBSS results suggest that these populations will probe circumgalactic regions. To address this expectation more directly in the context of our dataset, we have examined data from the VLT LBG survey \citep{2011MNRAS.414....2B,2011MNRAS.414...28C,2013MNRAS.430..425B} and a forthcoming VLT/FORS2 survey (N. Crighton et al., in preparation), which measure spectroscopic redshifts of BX-selected \citep{2004ApJ...607..226A} and LBG-selected galaxies in the fields of 13 bright quasars with high-resolution echelle spectra. We also include the 10 closest pairs in the \citet{2012ApJ...750...67R} sample using high-resolution QSO spectra from the Keck archive. These bright quasars were selected for their high incidence of local LBGs.

\begin{figure}
\begin{center}
\includegraphics[angle=0, width=0.5\textwidth]{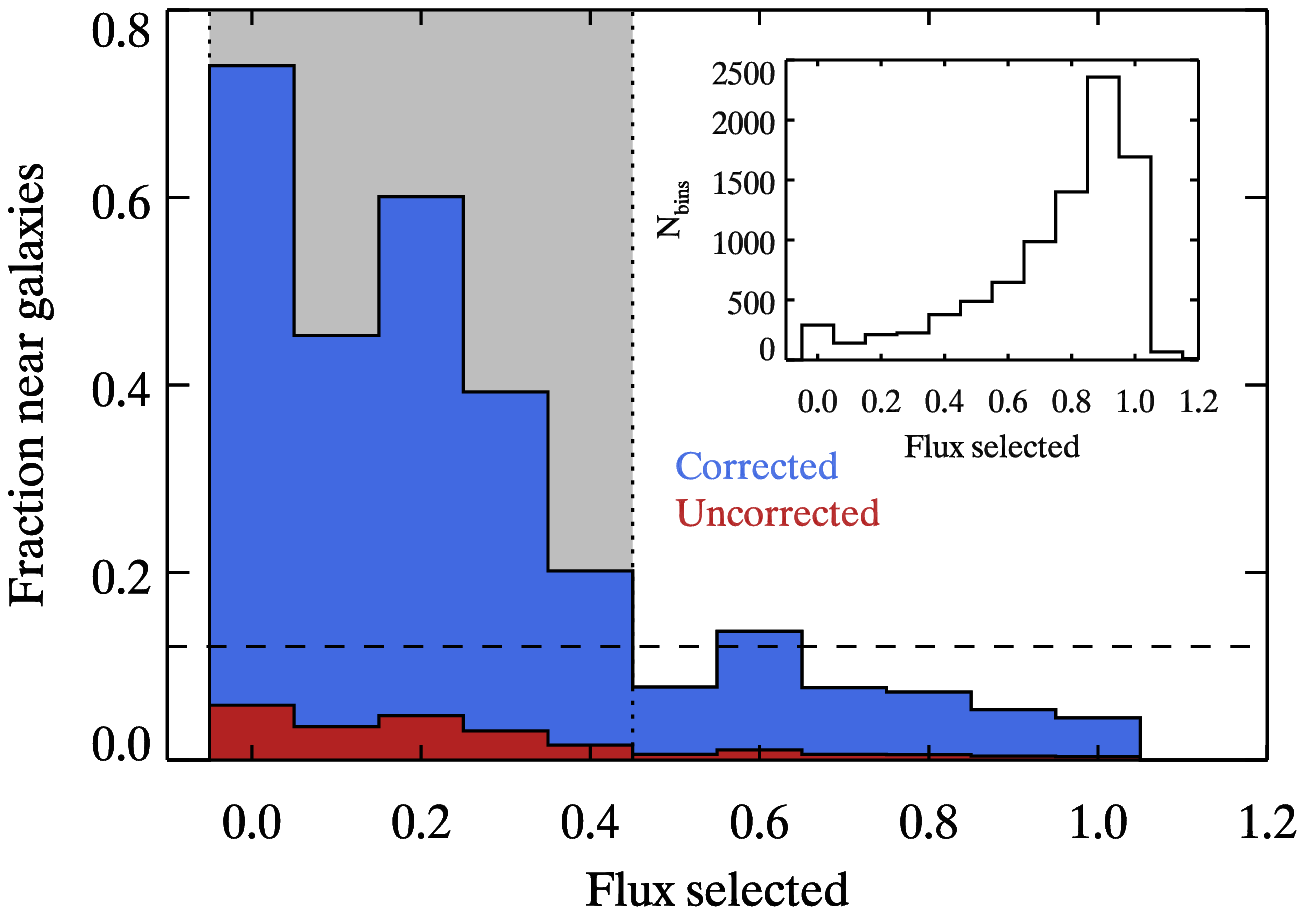}
\end{center}
\caption{The fraction of $138\kms$ wide wavelength bins proximate to LBGs as a function of flux, measured using data from the VLT LBG survey \citep{2011MNRAS.414...28C} and a subset of the KBSS sample \citep{2012ApJ...750...67R}. We convolve the VLT spectra to BOSS resolution; compared to the actual BOSS spectra they are effectively noiseless.  The red and blue histograms show the fraction of $138\kms$ bins in intervals of flux that lie `near an LBG', which is defined to be within 300~pkpc and $\pm 300\kms$ along the line of sight. The red histogram shows the measured fraction and the blue histogram shows the fraction after a correction for the completeness of the LBG sample has been applied.  The grey band indicates the range of fluxes for our five intervals of selected \lya\ flux. The inset shows the total number of wavelength bins in each flux interval. The dashed line indicates the completeness corrected fraction of galaxies for a random distribution of galaxies.
}
\label{nearlbgs}
\end{figure}

We convolve these high-resolution, high S/N quasar spectra with a Gaussian broadening profile to match BOSS resolution, and rebin them to a wavelength scale with bin width $128\,\kms$. We include bins in the range 1041--1185\,\AA\ in the quasar rest-frame and within $2.1<z<3.$. We then investigate the fraction of wavelength bins that meet our \lya\ flux selection criteria and are near an LBG, which we define as being within 300~pkpc transverse separation and $\pm 300 \kms$ along the line of sight. The results are shown in Figure \ref{nearlbgs}.

We apply completeness corrections based on an assessment of the completeness of the LBG sample as a whole. We derive a total volume density of  $2.5<z<3.5$ Lyman-break selected galaxies from \citet{2008ApJS..175...48R} by assuming all LBGs down to $0.2 M*$ show \lya\ absorption similar to that observed around our galaxy sample with nearby sightlines to bright QSOs. This is then compared with the volume inside a cylinder of radius 300~pkpc and the length of a forest line of sight, to obtain the total expected number of LBGs near the line of sight. Our results are then rescaled to bring the total number of LBGs in line with this expectation. Operationally, this involves multiplying the fraction near galaxies for each \Lya\ absorption sample by a factor 12.7.
Our observed sample of galaxies are in the redshift range $2.1<z<3.$ compared to our \lya\ forest range of $2.4<z<3.1$, but the UV galaxy luminosity function shows little evolution from redshift 2--3, and the mean flux evolution over this range is weak compared to differences in \lya\ flux near and far from LBGs. The main uncertainty in this completeness correction is the faintest magnitude limit and uncertainties in the faint-end slope of the $z\sim2.5$ luminosity function. These lead to a potential systematic error of a factor of two in our inferred complete LBG number densities. Our volume estimates do not take clustering into account, which will tend to overestimate the correction factor. However we expect this effect to be small compared to the uncertainty in the faint magnitude cutoff.

Compared to actual BOSS spectra, the convolved VLT and Keck spectra are effectively noiseless and so analogous to $F_0$ rather than $F$ in Figures~\ref{simulationexample} and \ref{selectiontest}. We find that, where the noiseless flux is $F_0< 0.25 $, the selected absorber is near an LBG on average 60\% of the time. The probability is roughly flat over this flux range and indicates that over the strongest three flux intervals of our BOSS analysis, we would be probing the CGM roughly half the time, in the absence of noise.
 Wavelength bins with fluxes in the range $0.75<F_0< 1.05$ show only a weak trend in proximity to LBGs. Wavelength bins with a flux of $F_0< 0.25 $ are ten times more likely to be near to LBGs than bins with $F_0 > 0.75$.  Bins with  $F_0< 0.25 $  are five times more likely to reside near LBGs than for a random distribution of LBGs, as indicated by the horizontal dashed line in Figure~\ref{nearlbgs}.

In combination with Figure~\ref{selectiontest}, we see that these flux intervals show decreasing purity of CGM selection for increasing flux due to a greater contribution from noisy non-CGM absorbers. Using these simulations we can quantify the degree to which we select  absorbers in this flux range: the probability that the true flux is $F_0< 0.25 $  for our five samples is (with decreasing \lya\ absorption strength) 90\%, 79\%, 49\%, 13\% and 2\%. If CGM regions are selected 60\% of the time when $F_0< 0.25$, but never for higher fluxes, the purity of CGM selection would be 54\%, 47\%, 29\%, 8\% and1\%. This calculation is clearly an underestimate as the incidence of CGM regions does not drop to zero, nor does it decline as a step function.

It should be noted that the CGM regions, as characterised in the above way,  are frequently not directly associated with the LBGs observed. Simulations by \citet{2014MNRAS.438..529R} obtain good agreement with Figure 30 of \citet{2012ApJ...750...67R}, but find that this is only achieved by including less massive objects with lower SFRs which are below the detection limit of observations. The observed LBGs have virial radii of $\sim 100$~pkpc, but they cluster with these dwarf galaxies creating an extended CGM region out to $\sim 300$~pkpc. Therefore, our inferences about circumgalactic regions should not be regarded as a measurement of regions associated with detected LBGs themselves but rather the LBGs and their low luminosity neighbours.

\subsection{Line list modelling}
\label{linelist}

We have compared our selection function to hydrodynamic simulations and observed sight-lines near LBGs. Here we complete the picture by exploring the range of column densities of systems selected. We achieve this by reconstructing model spectra of five quasars from the line list of \citet{2002MNRAS.335..555K}, which was compiled in order to investigate evolution in the line number density. This was achieved by generating a Voigt profile for each line using the quoted column density, b-parameter and redshift using the ESO MIDAS data reduction package. We produced high resolution mock spectrum of each quasar, smoothed them to reproduce BOSS resolution and rebinned them to our 138 km/s wavelength bins. This allowed us to reproduce our absorber selection function and explore the HI column densities of lines that contributed to these absorbers, again in essentially noiseless spectra.

We argued in the previous two sections that our strongest three absorption samples are characterised as circumgalactic by virtue of their selection of noiseless $F_0<0.25$ wavelength bins on 138 $\kms$ scales, and they do so with varying degrees of efficiency due to BOSS noise. We identified 78 of these wavelength bins over a redshift path of $\Delta z =2.39$ and found that this sample typically consists of blends of three lines which are distinct in high resolution spectra. The highest column density line of the blend was in the range $10^{14} \cm^{-2} < N_{\rm HI}< 10^{16} \cm^{-2}$ in 94\% of cases. This indicates that our sample is dominated by blends of strong forest lines.

The number of wavelength bins with $F<0.25$ over our total redshift path $\Delta z =  3409$ is 82412 (for our three strongest \Lya\ absorption samples plus those bins with $F<-0.05$). This corresponds to an incidence rate of $l(z) =24.2$ with SDSS noise, compared to $l(z) =31.3$ of effectively noise free data from our line list model spectra. The difference can be accounted for by a 77\% completeness of $F<0.25$ systems in our sample - a value which is  consistent with the signal-to-noise of our sample.

\citet{2002MNRAS.335..555K} state that they are not able to confirm the presence of $N_{\rm HI}> 10^{17} \cm^{-2}$  lines. They reconstruct the full \Lya\ forest, but in these cases they will underestimate the column. If we conservatively assume that all systems with $N_{\rm HI}> 10^{17} \cm^{-2}$  are sufficiently blended with lower column lines such that they are all selected, we can assess their maximal potential impact.  We also assume that they are selected independently and are not blended with each other. \citet{2013ApJ...775...78F} find that the incidence per unit redshift of lines with a Lyman limit opacity above $\tau_{LL} =2$ (i.e. $N_{\rm HI}> 10^{17.5} \cm^{-2}$)  is $l(z) =1.21$. This would mean that there are 2.9 systems in our line list model spectra i.e. 3.7\% of the sample.
Note that this analysis includes the contribution of super Lyman limit systems (or sub-DLAs) and DLAs. Since we have excluded the Noterdaeme et al. (2012) DLA sample (with $N_{\rm HI}> 10^{20.3} \cm^{-2}$) from our analysis, the above constitutes a 10\% over-estimate of the incidence rate of high Lyman limit opacity systems in our sample (taking the relative incidence rate of \citealt{2013ApJ...765..137O}).

We may also consider the contribution of partial Lyman limit systems with $\tau_{\rm LL} > 0.5$ (i.e. $N >10^{16.9} \cm^{-2}$) using the \citet{2013ApJ...765..137O} incidence rate of such systems; $l(z)=2.0$ in this more inclusive range. Again, we can conservatively assume 100\% selection in our line list model blended only with non-LL systems. This would provide 4.8 systems and so 6.1\% of the sample. 

It is clear even in a conservative assessment, that systems with significant Lyman limit opacity ($\tau_{\rm LL} > 0.5$ ) are a small minority of systems selected. However, we recognise that a small minority of systems with strong metal signal could have a significant impact on measurements of metal lines in our composite spectra. We discuss the the impact of this effect in Section \ref{llincomp}.

\section{Stacking procedure}
\label{stacking procedure}

We follow the procedure described in P10 to stack absorbers in the \Lya\ forest that are selected as described in the previous section. All absorbers are assumed to arise due to the \lymana\ transition \footnote{ See Section \ref{measuremet} for a discussion of where  this assumption breaks down.}, and so are assigned a suitable redshift, $z_{abs}$, and shifted to the absorber rest-frame. In the process the whole quasar absorption spectrum is treated as arising due to gas at the same redshift and so shifted by a factor of $1+z_{\rm abs}$ to the absorber rest-frame. Absorption that is correlated with the selected \lya\ absorbers adds coherently to the stack, while uncorrelated absorption adds incoherently to produce noise.

\begin{figure}
\begin{center}
\includegraphics[angle=0, width=.49\textwidth]{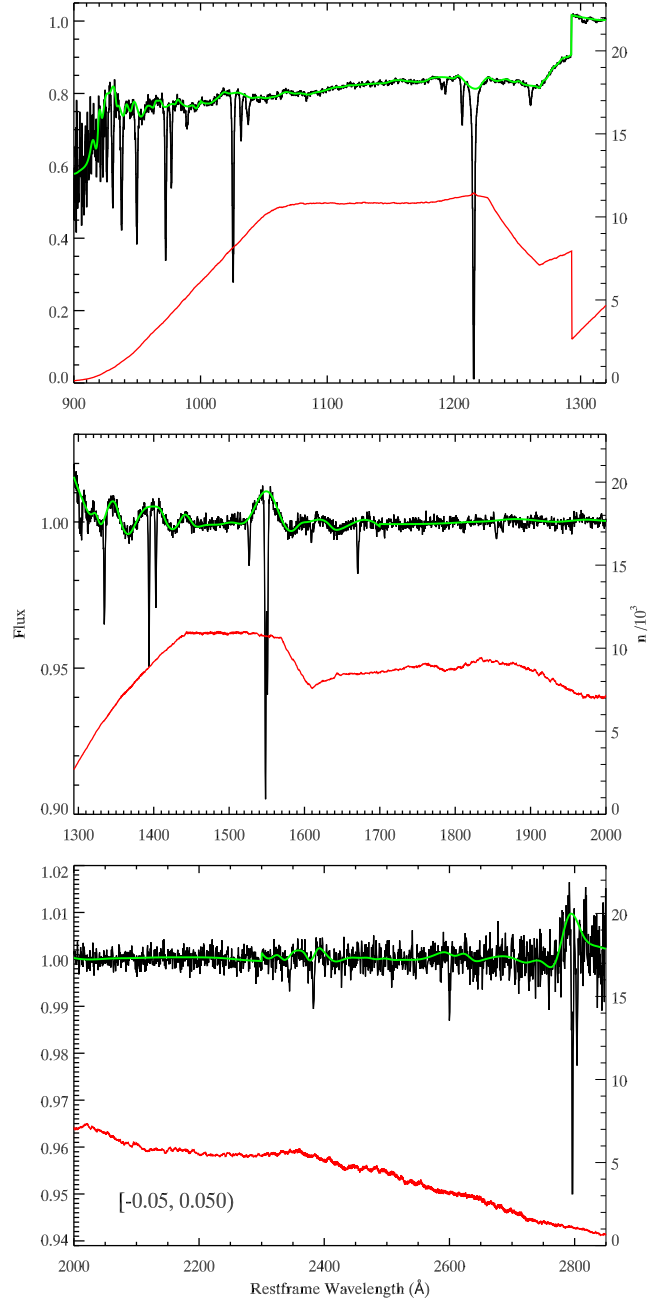}
\end{center}
\caption{The stacked spectrum of \Lya\ absorbers selected with flux between $-0.05 \le F<0.05$. Plotted is the median of the stack of spectra. This stacked spectrum has not been pseudo-continuum fitted and so includes broad absorption by uncorrelated absorbers. The pseudo-continuum is overlaid as a green curve. The red curve, with the y-scale on the right hand side, shows the number of pixels (in units of $10^3$) that went into the stacked spectrum at each point. Note the scale of the y-axes in each panel.}
\label{stack}
\end{figure}

This procedure is repeated for all selected absorbers, sometimes using the same quasar spectrum more than once, and a stack of spectra is obtained. This wavelength grid of the stacked spectrum is set by the standard SDSS wavelength solution, 
\begin{equation}
10^{-4}= \log_{10} \lambda_{i+1}-\log_{10} \lambda_{i} 
\label{sol}
\end{equation}
where $\lambda$ is the wavelength in \AA.
If we select an absorber at $\lambda_1$ and assume it is a \lya\ forest absorber, we have a redshift $z_1$ given by
\begin{equation}
\lambda_1=\lambda_\alpha (z_1 +1).
\end{equation}
Assuming that the adjacent pixel in the composite at wavelength $\lambda^\prime>\lambda_\alpha$ also measures the properties of the same gas (and so is at the same redshift) then
\begin{equation}
\lambda_2=\lambda^\prime (z_1 +1).
\end{equation}
Substituting these into equation (\ref{sol}),
\begin{eqnarray}
10^{-4} &=& \log_{10}[\lambda^\prime (z_1 +1)] -\log_{10} [\lambda_\alpha (z_1 +1)] \\
&=& \log_{10}\lambda^\prime - \log_{10} \lambda_\alpha  
\end{eqnarray}
Hence the pixel-spacing of the wavelength solution in the stacked spectrum is identical to the data where the data has a uniform $\Delta \log\lambda$ solution. The wavelength solution of the stacked spectrum is fully described by this pixel-spacing and the rest-frame wavelength of the system stacked (\Lya\ in this paper). Consequently the operation of stacking is a simple process of shifting by the number of pixels separating the marker location in the quasar spectrum and \lya\ in the rest-frame in the stack, with no interpolation. We rebin our spectra for \lya\ selection by a factor of 2 to reduce noise. However,  we do not rebin the spectra before stacking. As a consequence, 1215.67\,\AA\ in the wavelength solution for the composite spectrum is centred between two pixels. In the results that follow, the full profiles of lines in the composite spectra are measured without rebinning, and the central fluxes of lines are measured by averaging the two nearest pixels in the composite (see Section~\ref{char hi}).

\begin{figure*}
\begin{center}
\includegraphics[angle=0, width=.99\textwidth]{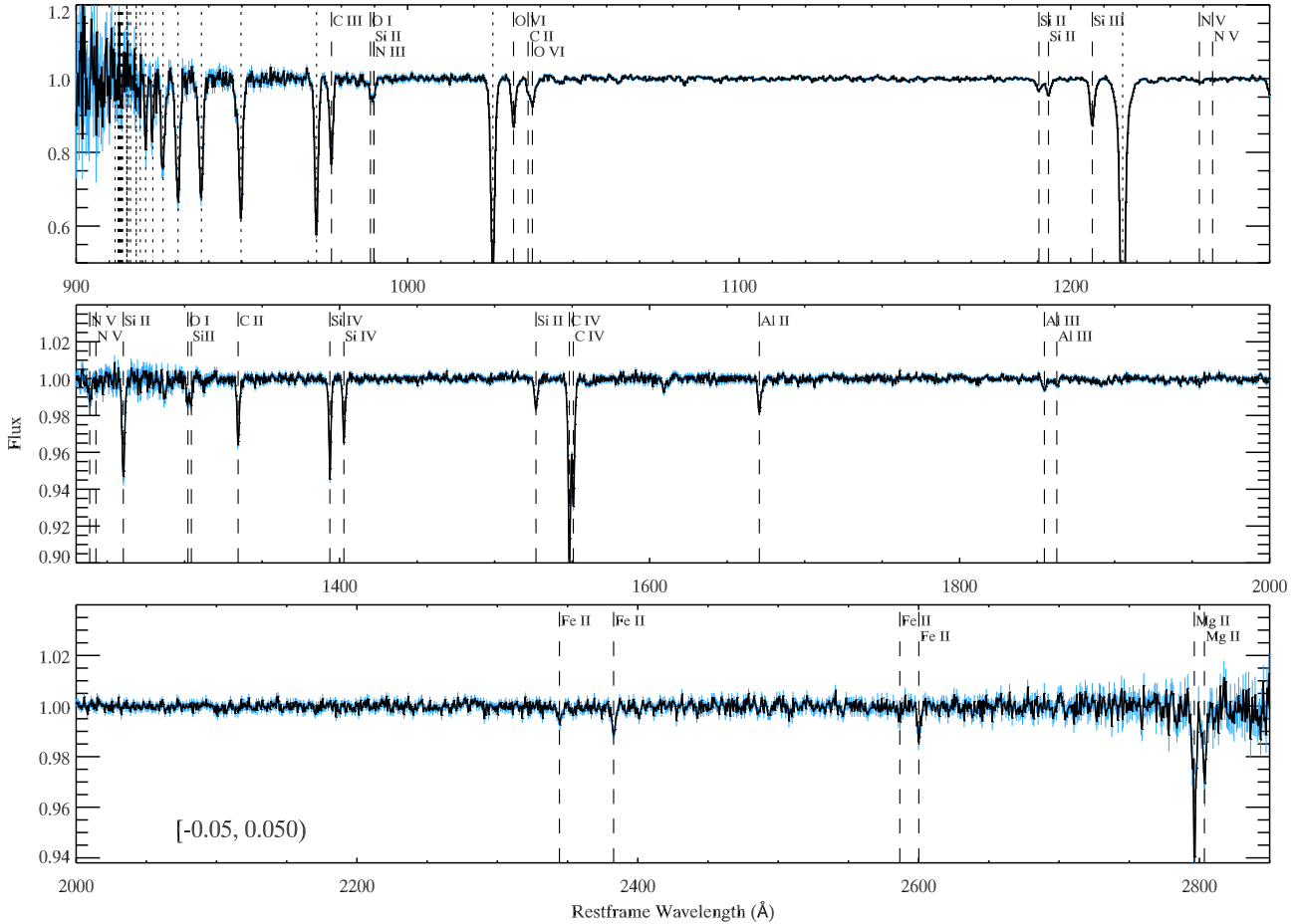}
\end{center}
\caption{Composite spectrum of \lya\ absorbers selected with flux between $-0.05 \le F<0.05$ produced using the median statistic. Error bars are shown in blue. Vertical dashed lines indicate metal lines identified and dotted vertical lines denote the locations of the Lyman series. All lines measured for all \lya\ samples are presented in Figures~\ref{am_fits} and \ref{med_fits}. Note the scale of the y-axis in each panel: this is our lowest S/N composite spectrum and yet we measure absorption features with depth as small as 0.0005.}
\label{composite}
\end{figure*}

The next step is to characterise this stack of spectra with a chosen statistic. We use two statistics: the median and the 3\% clipped arithmetic mean (i.e. discarding the highest 3\% and the lowest 3\% fluxes). In both cases we determine the error estimate in a pixel by bootstrapping the stack of spectra. 
 
Redward of 1215.67\,\AA\ in the composite spectrum, the stack of the spectra transitions from entirely within the forest to entirely outside the forest. Noise due to uncorrelated contaminating absorption is the same order of magnitude as the corrected pipeline error estimate in the forest-contaminated portion of the stack of spectra and is subdominant outside of the forest. There is significant pixel-to-pixel covariance in composite spectra due to this contaminating absorption, since contaminating lines extend over resolution element scales.
This transition region is dictated by the redshift of the selected \lya\ absorbers with respect to the emission redshift of the quasars in whose spectra they reside. We may choose to require that only forest-free regions contribute to some portion of this transition region by eliminating the contribution of some absorbers from that portion of the stack. We do so where the reduction of noise from eliminating contaminating absorption exceeds the increase of noise from having fewer spectra. In practice this balance occurs at wavelengths $>1293$\,\AA. 
In effect, for portions of the composite spectrum with wavelength $\lambda_{\rm comp} > 1293$\,\AA, we require that $(1+z_{\rm abs})> (1+z_{\rm q})\lambda_\alpha/\lambda_{\rm comp} $.
Figure~\ref{stack} presents the resultant stacked spectrum, and a discontinuity due to this condition is clearly visible.

The red line in Figure~\ref{stack} shows the number of spectra that contribute to the stacked spectrum at each point. The decline in counts at the red and blue ends are a consequence of  limited spectral coverage in individual spectra. 
The lowest redshift absorbers stacked do not have coverage at the blue end and the highest redshift absorbers do not have coverage in the red end. As a consequence the mean redshift is higher by $\Delta z = 0.2$
at the blue end of the stacked spectrum compared to the red end in our results.

These raw stacked spectra show smooth flux trends related to the mean flux decrement of largely uncorrelated absorbers; these are clearly undesirable features when producing composite spectra of selected absorbers. There are also smooth trends where the stacked spectrum exceeds unity by around 1\%. These deviations in our stacked spectra as shown in Figure~\ref{stack} (and not our composite spectra of forest absorbers such as that shown in Figure~\ref{composite}) are a consequence of smooth systematic errors in spline fitting of individual spectra.

In producing our final composite spectra, we correct both these effects by fitting this `pseudo-continuum' and removing it in keeping with the manner in which these uncorrelated absorbers contaminate the desired signal. 
In P10, we argued that contaminating absorption typically occurs within the same resolution element, but would not  overlap with the measured transition if fully resolved.
Hence, their effect is more accurately described as additive in flux decrement rather than additive in optical depth.
As a result we increase the flux globally by the flux decrement of these regions in an additive manner, i.e. the desired composite spectrum is 
\begin{equation}
F=F_s + (1-F_c),
\end{equation}
where $F_s$ is the raw stacked spectrum and $F_c$ is the pseudo-continuum. We fit this pseudo-continuum by manually selecting nodes and taking a spline of these nodes. An example pseudo-continuum is shown as a green line in Figure~\ref{stack}.  We mask known absorption lines from this process to prevent misleading the eye. This pseudo-continuum fitting procedure introduces errors  of order the size of the error estimate in the composite spectrum.  In order to take this effect into account, we sum these two sources of error in quadrature in the error estimate carried forward for the composite spectra, increasing the noise-per-pixel by $\sqrt {2}$.

Broad absorption features produced by clustering are difficult to disentangle from the pseudo-continuum. In particular, the \Lya\ feature in the strongest composite spectra shows signal out to $\sim 3000 \kms$ (in agreement with \citealt{2006ApJS..163...80M} and \citealt{2013arXiv1306.5896P}). We fit through these large-scale features in producing the pseudo-continuum, but this is a somewhat subjective exercise. This effect leads to an uncertainty in the measurement of the strongest \Lya\ features in our analysis, but we demonstrate in the following section that our inference of \HI\ column is not sensitive to the measurement of \Lya\ or \Lyb\ in composite spectra.

Figure \ref{composite} shows an example composite spectrum in full. This spectrum is a composite produced using the median stack for \Lya\ absorbers selected with flux between $-0.05 \le F<0.05$. Our two statistics have complementary uses. In the limit where contaminating absorption purely occurs as a sum of flux decrements (where lines are not intrinsically coincident, but lie within the resolution element), the subtraction of the pseudo-continuum for the arithmetic mean is a mathematically ideal operation. The median is a more outlier resistant statistic and so can provide a useful gauge of absorbing population size in tandem with the arithmetic mean when the range of the absorber strengths is broader than the noise. A joint measurement of absorber strength and population size using the full distribution of pixel fluxes in stacked spectra will be presented in a future work.

\section{Results}
\label{results}

The BOSS resolution element is broader than the width of observed \lya\ forest and metal lines in high-resolution spectra, so even in a single pixel of the stacked spectrum the absorption arises from gas that is clustered.
Furthermore, the absorption features in our composite spectra are substantially broader than the width dictated by the resolution element in all cases. All Lyman series features have a FWHM $>300\kms$, in cases where it can be reliably measured, and all our metal lines show a contribution from clustering at least as large as BOSS instrumental broadening across the full profiles (see Tables \ref{med_fits} and \ref{am_fits}). 

As a consequence we may integrate over the full line profiles and so include all measured clustering, or limit our integration to the minimum degree of clustering set by the $138\kms$ scale of our \lya\ selection bin (which is approximately the FWHM of the resolution element). 
We adopt the full line profile approach as our fiducial method, denoted by subscript $f$, and also consider measurements of the central wavelength bin of our features and denote them with subscript $c$.

\begin{figure}
\begin{center}
\includegraphics[angle=0, width=0.42\textwidth]{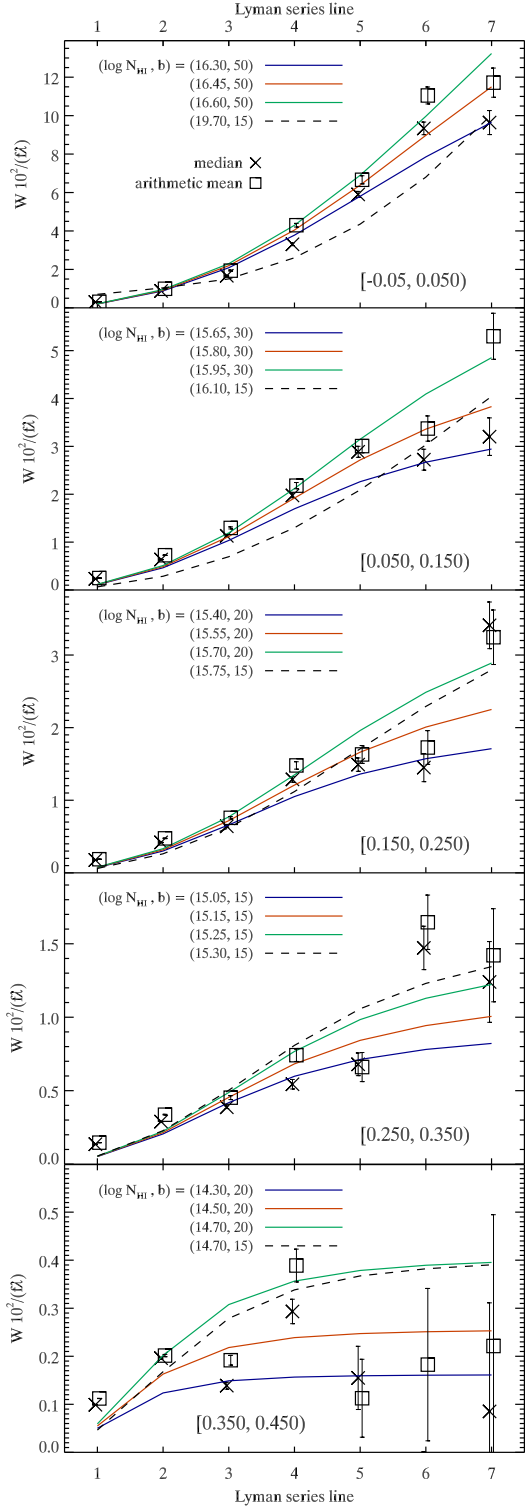}
\end{center}
\caption{
The equivalent width over the full profile of Lyman series lines $W_f$ with a scaling of  $(f\lambda)^{-1}$ that yields a 
constant quantity for optically thin lines.  Seven Lyman series lines are shown for our three strongest samples using both the median and arithmetic mean statistics.  Single line models for full line profiles are presented in order to characterise the allowed range of column densities and Doppler parameters assuming that a single line dominates the total column density.}
\label{lymanseries_fullline}
\end{figure}

\subsection{\HI\ Absorption}
\label{char hi}

We characterise the \HI\  absorption by measuring Lyman series lines present in the stacked spectra and complement this with a measurement of the Lyman limit opacity provided by a modified stacking approach.

\subsubsection{Lyman series lines}
\label{lymanserieslines}

We measure the absorber rest equivalent width ($W$) as shown in Figure \ref{lymanseries_fullline}. This measurement is normalised by the oscillator strength ($f$) and the wavelength ($\lambda$) in order to produce a quantity that asymptotes to a constant value when the lines in the series become unsaturated. This quantity is measured for each of seven Lyman series lines for both composite spectra produced using the median and the arithmetic mean statistic. Precision measurements of the equivalent width are challenging for the \lya\ line (and \lyb\ line to a lesser degree) due to systematic errors in distinguishing the pseudo-continuum from these broad features produced by clustering as discussed in the previous section.

We produce model line profiles using the analysis package VPFIT\footnote{http://www.ast.cam.ac.uk/$\sim$rfc/vpfit.html} and measure their normalised equivalent width in the same manner. Our aim is to use the Lyman series to constrain the characteristic \HI\ column density of the systems of our \lya\ absorber composites.This task would be straightforward if the higher-order lines showed a clear plateau in $W_f/f\lambda$ indicating that these lines were unsaturated, in which case one could simply find the column density of models that show the same plateau. Unfortunately, the situation is not always so clear, because the high-order lines remain marginally saturated in the strongest absorption samples. We must therefore infer \HI\ column densities more qualitatively.

Models are selected to provide reasonable agreement with the highest order well-measured Lyman lines. We do not require a match to the \lya\ or \lyb\ lines because weak blended lines raise their equivalent widths  while having little effect at higher orders. For each absorber flux interval we show solid curves indicating the range of column densities allowed for a Doppler parameter choice that roughly matches the observed trend in $W_f/f\lambda$ for higher order lines without overproducing \lya\ or \lyb\ absorption. We adopt the column density of the central curve as the inferred column density characteristic of that absorber sample, and the comparison indicates a typical uncertainty of $\Delta \log N_{\rm HI} = \pm 0.15$. For comparison we also display our upper limit on the column densities by relaxing our requirement that broad agreement with lines $n=$3--7 be obtained. Here we only require consistency with $n=6$ and $n=7$ measurements (while treating lower order lines as upper limits) and allow the Doppler parameter to take its lowest value seen in high-resolution observations, $b=15 \kms$. Little increase in upper limit on the column density is derived from this test in all but the strongest absorption sample despite the fact that these models do not reproduce the observed trends well.

In some cases Lyman series lines measured in the arithmetic mean composite indicate a significantly higher HI column density than the median composite. This indicates that the absorbing population is not uniform and that there are a minority of selected systems that show stronger absorption. These systems would have a greater impact on a less outlier resistant statistic such as the arithmetic mean. Rather than choose a preferred statistic we quote allowed values for $N_{\rm HI}$ that span the range indicated by both statistics. 

We report our inferred column densities in Table~\ref{hicolumns} and plot the $N_{\rm HI}$ values in Figure~\ref{finalhicolumns}. If we use the central bin fluxes instead of the full velocity range, inferred column densities are lower by $\sim0.5$dex.

\begin{table}
 \caption{Inferred \HI\ column densities from Lyman series measurements.}
 \label{hicolumns}
 \begin{tabular}{@{}lcc}
  \hline
  Flux selected & $\log N_{f, \rm HI} (\cm^{-2})$  & $\log N_{c, \rm HI}(\cm^{-2})$ \\
  \hline
$[-0.05, 0.05)$ &$16.45 \pm0.15  $ &$15.95\pm0.15$ \\
$[0.05, 0.15) $&$15.80 \pm0.15   $ & $15.55\pm0.15$\\
 $[0.15, 0.25)$ &$15.55 \pm0.15      $ & $15.1\pm0.1$\\
 $[0.25, 0.35)$ &$15.15 \pm0.1  $ & $14.6\pm0.15$\\
 $[0.35, 0.45)$ &$14.5 \pm0.2       $ & $14.1\pm0.2$\\
\hline
 \end{tabular}
 \end{table}
  
\subsubsection{Lyman limit opacity}
\label{llopacity}

\begin{figure}h
\begin{center}
\includegraphics[angle=0, width=0.49\textwidth]{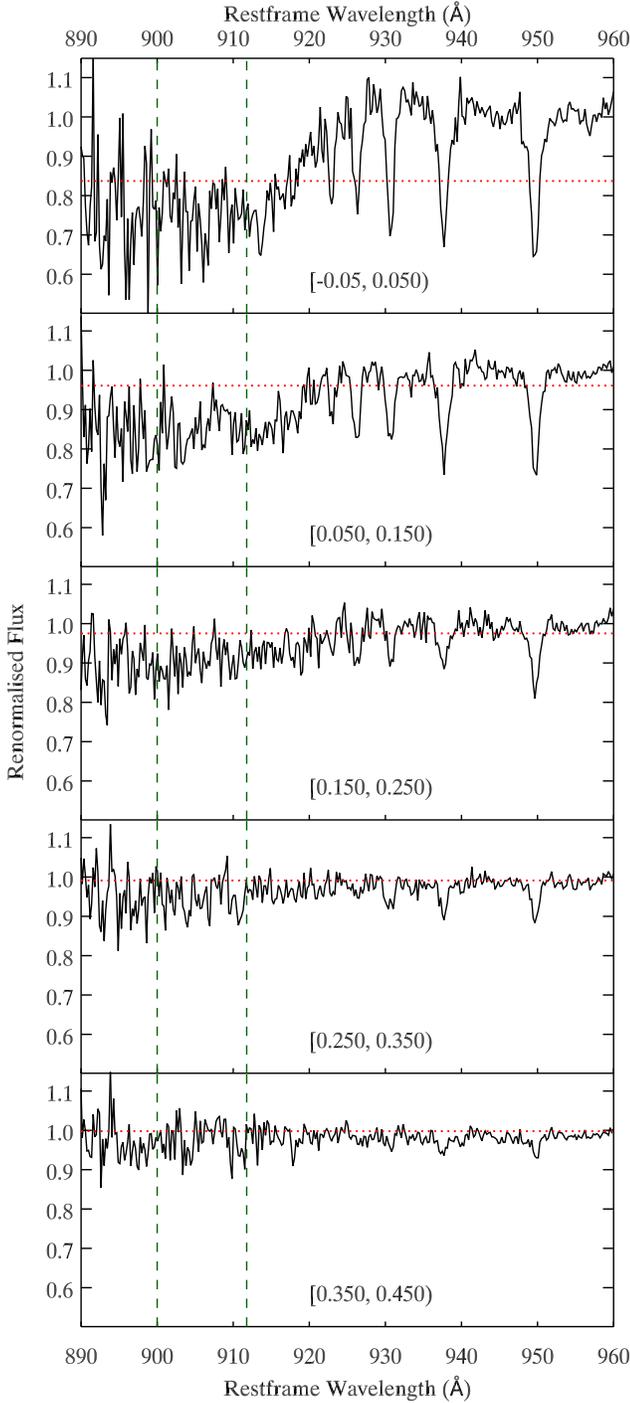}
\end{center}
\caption{Renormalised composite spectra showing the excess absorption at the Lyman limit that arises upon selecting \lya\ absorption as described in Section~\ref{lya selection}. The portion of the composites used to measure the Lyman limit opacity are designated by the vertical green dashed lines. The expected Lyman limit flux decrement in this region inferred from our Lyman series analysis of full line profiles (see Section~\ref{lymanserieslines}) is indicated by the horizontal red dotted line.}
\label{llmeasure}
\end{figure}

Higher order Lyman lines are a valuable measure of high column density \HI, but have limitations when lines become saturated. Hence, we combine  constraints from the previous section with a measurement of the Lyman limit opacity because the opacity is simply proportional to $N_{\rm HI}$, albeit integrated along the line of sight.
Here we stack our selected absorbers and measure the mean opacity blueward of the Lyman limit (912\,\AA) to characterise the highest column density population in our sample. In order to measure the Lyman limit we must omit the continuum fitting described in \S\ref{data} as this would subtract the desired signal. Equally, it is not sufficient to simply stack unfitted quasar spectra, as a number of unwanted effects will modify the measured opacity at the Lyman limit: weighting due to quasar S/N, peculiarities in the wavelength dependent throughput of the BOSS quasar spectra, departures from self-similarity in the quasar SED power-law, and the cumulative Lyman limit opacity of the average \Lya\ forest (Prochaska et al. 2011).

Our main stacking approach includes no weighting and thus it is necessary to replicate this feature and preclude any unwanted quasar-flux-based weighting. We achieve this by normalising each quasar spectrum by a scalar value such that the mean flux in a band with $1041 {\rm \,\AA} < \lambda <1185 {\rm \,\AA}$ in the quasar rest-frame is set to the mean flux in the forest given by \citet{2008ApJ...681..831F}.

We then stack the \Lya\ sample described in Section~\ref{lya selection} using these renormalised spectra and the method described Section~\ref{stacking procedure}. In order to take into account all the remaining unwanted effects listed above, we produce a baseline stack measuring them independently of the desired signal. This is done by stacking pixels with the mean forest flux across our target forest sample ($\bar{F}=0.7645$) in our chosen redshift range. We select pixels over a $\Delta F = 0.04$ range around this mean such that there are always more pixels in the baseline than in our main \Lya\ samples. This approach provides an appropriate baseline assuming that our main samples and these mean flux baseline pixels are both randomly distributed over $2.4<z<3.1$. We therefore renormalise our Lyman limit composite spectrum by this baseline composite spectrum. 
Figure~\ref{llmeasure} presents the resultant renormalised composite spectra for each of our \lya\ samples.

\begin{figure}
\begin{center}
\includegraphics[angle=0, width=0.47\textwidth]{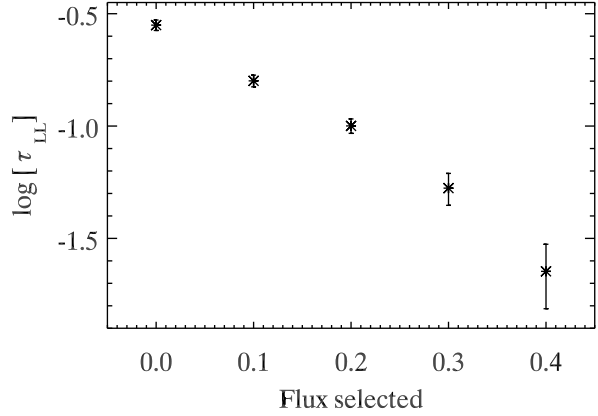}
\end{center}
\caption{The  Lyman limit optical depth measured in composite spectra presented in Figure~\ref{llmeasure} between 900\,\AA\ and the Lyman limit. The mean Lyman limit opacity of the \lya\ forest has been corrected for; however, the measurement is a sum of the Lyman limit opacity of selected \lya\ absorbers and the opacity of large-scale structures correlated with them.}
\label{lymanlimitopacity}
\end{figure}

We measure the mean Lyman limit opacity between 900\,\AA\ and the Lyman limit for each of our selected absorbers using this method; these results are shown in Figure \ref{lymanlimitopacity}. The error in these measurements is established by bootstrapping the sample of renomalised fluxes in these windows. The bootstrap element is 3 pixels wide to encompass the FWHM of the resolution element. Figure~\ref{llmeasure} denotes the region to be utilised for the measurement of Lyman limit opacity bounded by the green vertical dashed lines. An additional, residual stochastic error may be present as a consequence of small differences in the realisation of the unwanted effects listed above between the sample and baseline stacks. 

\begin{figure}
\begin{center}
\includegraphics[angle=0, width=0.5\textwidth]{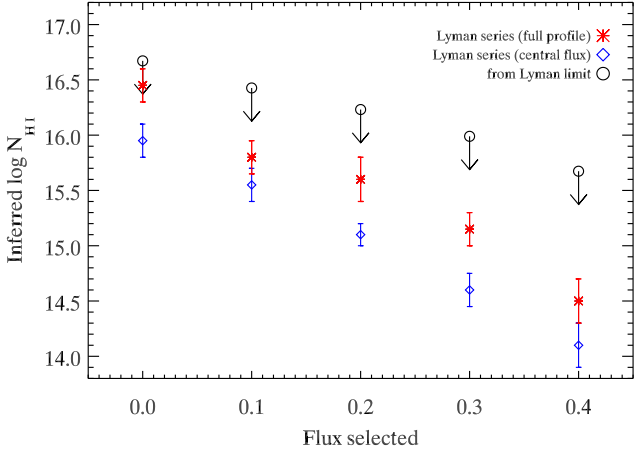}
\end{center}
\caption{The \HI\ column density of our selected \lya\ absorbers inferred from Lyman series lines (described in Section~\ref{lymanserieslines} and listed in Table~\ref{hicolumns}), with upper limits derived from the  Lyman limit opacity described in Section~\ref{llopacity}. Lyman series are measured within the central $138\kms$ wavelength bin (blue diamonds) and over the full line (red stars). 
For the Lyman limit opacity we plot $+1\sigma$ values as upper limits,
since they are a sum of the selected absorbers and their associated large-scale structure. }
\label{finalhicolumns}
\end{figure}

We can infer the \HI\ column density from these measurements using the approximation $N_{\rm HI} \approx  \tau_{LL} 10^{17.2} {\rm cm ^{-2}} $  \citep{2010ApJ...718..392P}. These values are greater than the column densities preferred by Lyman series measurements, as one can judge visually from Figure~\ref{llmeasure}; however, for every absorber we stack we also select a large-scale excess in forest absorption over $\sim 3000\kms$ scales \citep{2006ApJS..163...80M,2013arXiv1306.5896P}. This associated excess absorption will raise the Lyman limit opacity and so the inferred column density by comparison with our baseline stack. An overestimate in the column density of our \lya\ absorber sample is to be expected here, although it is not clear whether this effect is sufficient to explain the difference. In Figure~\ref{finalhicolumns}, the 1$\sigma$ upper envelope of column densities derived from the Lyman limit opacity are treated as upper limits on \HI\ column density of the systems stacked.

While the impact of clustered absorption is difficult to judge without extensive investigation, these Lyman limit opacities clearly provide an upper limit on the \HI\ column directly associated with our stacked absorbers. Both these upper limits and our Lyman series analysis exclude the possibility that Lyman Limit systems and DLAs dominate any samples, consistent with the independent argument based on line list modelling in Section~\ref{linelist}. While Figure~\ref{lymanseries_fullline} suggests that the Lyman series measurements for the strongest absorption sample marginally allow a column density consistent with that of Lyman limit systems, our Lyman limit opacities rule out these systems representing the dominant population.

\subsection{Metal Absorption}
\label{measuremet}

Our analysis of Lyman series absorption was developed to account for the important contribution of saturation in the composite spectra. Since there is no indication of saturation in any of our metal species with multiple transitions, we take a simpler approach here. We treat each metal transition from the same species as an independent measurement of column density.

We measure full line profiles by Voigt profile fitting with only the redshift fixed (to $z=0$ since the spectra are already in the rest-frame). As for our fits of \HI,  we also perform line centre fits, i.e. we take the two nearest pixels to the rest-frame wavelength of the metal line of interest, average them, and obtain an error estimate by summing the errors of the two pixels in quadrature.  As described in Section~\ref{char hi}, this is done in order to bound the range of viable integration paths lengths for each line in the composite spectrum. The central bin provides no absorption profile to fit, so we must assume a b-parameter. Since the central bin \HI\ analysis provides plausible single line b-parameters, we take these values
($25\kms$ for $-0.05\le F < 0.05$ and $15\kms$ for all other absorption samples),
which corresponds to the minimum integration scale that we can plausibly investigate in the context of metal lines that show clear clustering and/or complexes. 
 For these central bin measurements, we construct a suite of models and take our column densities and their upper and lower error bars from the central bin fluxes and error estimates.
We perform our full profile fits and construct our central bin models using both
VPFIT and the ESO MIDAS data reduction package and find excellent agreement between them. We show resultant column densities, and measured or assumed b-parameters in the Tables in Appendix~\ref{metalcat}.

As discussed in P10, there is a significant contribution to the composite spectra from the misidentification of \SiIII\ as \lya\ in the selection of absorbers for stacking. This misidentification produces an effect we have dubbed `\SiIII\ shadows' whereby many metal lines show a signal offset in wavelength by a factor of $(1216/1207)$. It is possible to infer the size of the selected \SiIII\ absorbing population as a fraction of our selected systems by measuring the strength of the \CIII\ shadow line associated with such systems in the arithmetic mean composite. For example, in our strongest absorption sample the selected \SiIII\ absorbers must be typically at or near saturation in order to meet our selection requirements. In such cases, the associated \CIII\ absorption must be of similar strength given their similar ionization characteristics and their oscillator strengths (assuming a broadly solar abundance pattern). In the composite this \CIII\ shadow line shows a flux decrement of 2\%. Each case of \CIII\ saturation must be diluted by approximately 50 more cases where no absorption is seen, and therefore only 2\% of selected systems are \SiIII\ absorbers. This is marginally significant in the context of the quoted error in our measured column densities, but is negligible in the context of our current interpretation.
We can use this approach to rule out significant selection of \CIV\ and \MgII\ lines in our sample as no associated \SiIV\ or \FeII\ respectively are seen in our composite spectra.

All metal lines labelled in Figure~\ref{composite} are measured for both the arithmetic mean and median composite spectra with some notable exceptions.   It is typically sufficient to disregard lines associated with \SiIII\ shadows as they lie in unused portions of the composite spectrum; however, the \SiIII\ shadow of the strongest member of the \SiIV\ doublet contaminates the weaker member of this doublet. Therefore we do not present results from the weaker doublet line. Additionally we do not provide measurements of the \NIII\ line because it is a complex blend with \SiII\ and an \OI\ doublet. We forgo presenting any potentially misleading results as these challenging measurements are the only available transition of \NIII. The \SiII\ and \OI\ transitions associated with this blend are also excluded in our analysis.

Each line is measured independently in order to make consistency checks possible for both full line and centre fits. However, we must modify our approach in cases of line blends. Our approach for full line fits is illustrated in Figures~\ref{am_fits} and \ref{med_fits}, where the red curve shows the fit region. For the blending of the \CIV\ doublet we fit each line separately and omit the central overlap region. Where two lines are blended such that it is not possible to measure them independently we perform a joint fit. We make this modification for the blend of \CII\ and \OVI\ at $\approx1037$\,\AA\ and the blend of \OI\ and \SiII\ at $\approx1303$\,\AA.

We make no special arrangements for measuring the central flux when lines are blended. We do, however, require that the adjacent line has a negligible contribution to the flux in the central bin. This situation is true in all cases but one: we discard the measurement of the central wavelength bin for \CII\ due to its blend with \OVI. 

We present our metal line measurements in Figure~\ref{metalcolumns} and the Tables in Appendix~\ref{metalcat}, conservatively excluding all measurements with less than $5\sigma$ significance. The central flux approach shows fewer measurements as a consequence of this restriction and the fact that this approach discards some signal. Overall the narrow velocity range integration of the central bin approach provides results roughly 0.5 dex lower than the full profile measurements. See Section~\ref{discussion} for a discussion of the significance of this difference.

Where multiple transitions of the same species are available they provide a useful alternative indication of the error in measuring lines. Their broadly good agreement indicates that we are not observing saturated metal absorbers that are diluted by noisy selection. Even a small population of very strong ($F\la0.5$) but not fully saturated metal lines is ruled out. For example, the ratio of doublet line strengths of 2:1 in lithium-like ions (e.g. \OVI) is a ratio in opacities and not flux; hence a very strong doublet that is diluted by non-detections to appear as a weak doublet in the composite spectrum would have a measurably different doublet ratio from that of a ubiquitous, weak doublet.

We treat measurements of the arithmetic mean and the median as two measurements of the same systems. The comparison between these two statistics is, to a limited extent, an indication of population size amongst the stack, particularly outside of the \lya\ forest where the distribution of contaminating absorption is much larger than the stack. The question of absorber population sizes (i.e. what fraction of the selected \lya\ systems are contributing to a given metal line) will be addressed in a future publication.

\begin{figure*}
\begin{center}
\includegraphics[angle=0, width=1.\textwidth]{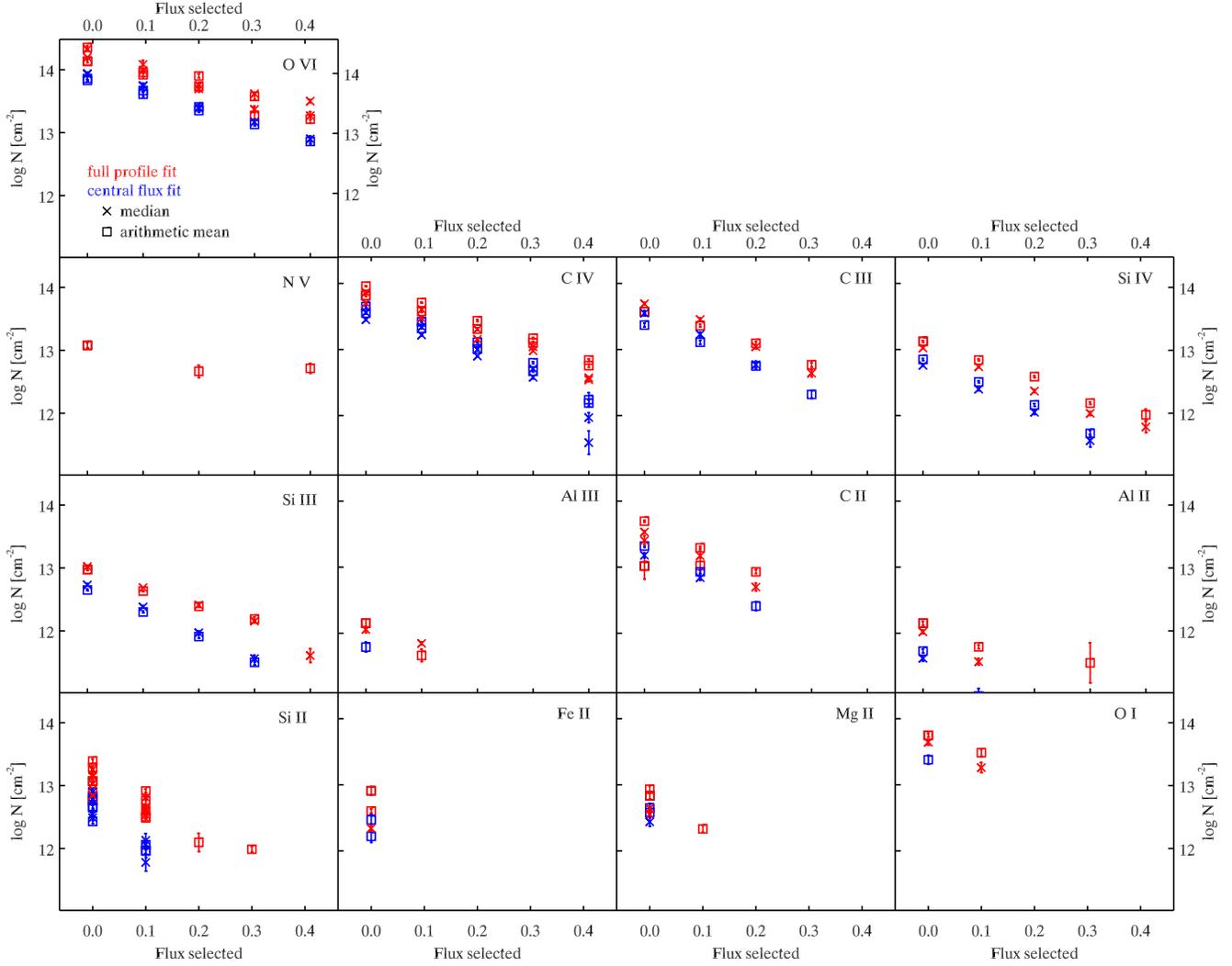}
\end{center}
\caption{Metal species column densities with respect to stacked \lya. Each panel displays a different metal species ordered from highest to lowest ionization potential, going first from left to right and then from top to bottom. Line measurements from both the arithmetic mean and median composite spectra are presented for both the full line profile and the central wavelength bin. Multiple transitions of a given species are used where available.}
\label{metalcolumns}
\end{figure*}

\subsection{Comparison to simple models}
The column densities of metal species in the intergalactic and circumgalactic media depend upon metallicity, abundance patterns and ionization. Changes in metallicity of the  observed gas will scale all our measured columns globally, while changes in abundance pattern will scale our measurements element-by-element. Changes in ionization will modify metal line column densities in a set of broad, but constrained, ways, and it is upon ionization that we focus here. We treat metallicity as a free parameter and assume solar abundance patterns in this publication.

The ionization of all these elements is dictated by gas density, temperature, ionising radiation, and geometry. Geometry is only of significance when gas is self-shielded. We neglect this effect since, as we demonstrated in Section~\ref{char hi}, all our samples are dominated by gas with an insufficient \HI\ column to be self-shielding. Even our most conservative estimates of the potential impact of a subsample of self-shielded systems show them to be sub-dominant (Section~\ref{llincomp}).

\begin{figure}
\begin{center}
\includegraphics[angle=0, width=0.44\textwidth]{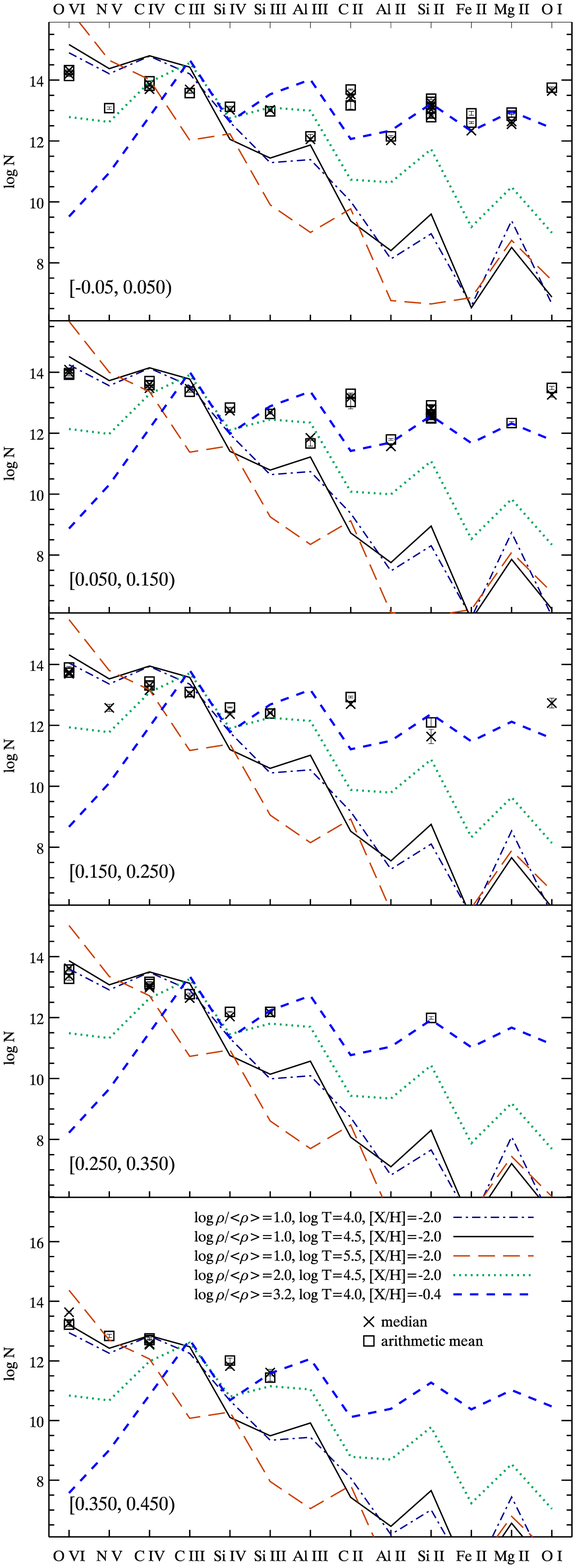}
\end{center}
\caption{The column densities of metal species measured in order of decreasing ionization potential for all our composite spectra. Metal line columns are measured using all available lines. Model curves are displayed assuming gas is optically thin, \HI\ columns shown in Table \ref{hicolumns}, a solar abundance pattern, and metallicities scaled to produce broad agreement where possible without overproducing absorption. Lines are measured using full line profiles.}
\label{stackproj_fullline}
\end{figure}

The points in Figure~\ref{stackproj_fullline} show all our metal column density measurements for full line profiles. Measurements are grouped by species and sorted from left to right in order of decreasing ionization potential. From top to bottom, the panels display increasing selected \lya\ flux (decreasing absorption). Ordered in this manner it is clear that lower ionization species are mainly present in our strongest three \lya\ absorption samples, and among these more lines are present in the stronger \lya\ samples. 

We also show five model curves produced using {\sc CLOUDY}  version 08.00 \citep{1998PASP..110..761F} assuming a solar pattern of elemental abundances and a quasar+galaxy UV background \citep{2001cghr.confE..64H}. In each panel, the \HI\ column required by the model is taken from the measured \HI\ column listed in Table~\ref{hicolumns} for that \lya\ sample. 
We do not attempt to fit these measurements in detail, as such a fit would require a full exploration of the parameter space of all the relevant effects described above. Ultimately, matched analyses of simulations that reproduce our selection and stacking procedure will be required to investigate all the information content of our composite spectra. We will take this approach in a future work. Here we limit ourselves to the overall trends arising from the physical conditions of the absorbing gas, which are clear in a visual examination.

We do not require individual gas clouds to possess all species observed. It is a natural consequence of any spectral stacking approach that inhomogeneity in the selection will give rise to multiple populations in the composite spectrum. Furthermore, it is likely that multiple phases are present even assuming a pure selection algorithm.
Thus the observed signal may arise as a combination of the models shown or even a smooth transition between them. There is, however, one key discriminating requirement: absorption must not be overproduced by a given model. If a model reproduces the absorption for a particular species, it must not overproduce absorption for another species in order to be viable. Only full line profile measurements are shown here, but the line centre approach produces similar results since both the \HI\ and metal column densities are lower by approximately the same factor.

\subsubsection{High-ionization species}

In Figure~\ref{stackproj_fullline}, four curves display a range of models representing typical conditions in the \lya\ forest with a metallicity of 
[X/H]$=-2$. Physical conditions $\rho/\bar{\rho}\approx 10$ and $T=10^{4}-10^{4.5}\K$ give
a broadly acceptable match to the high-ionization lines, 
though we do not attempt to fit samples in detail.
This  applies to 
species with ionization potentials down to and including \SiIV, and perhaps also including intermediate lines such as \SiIII. This result is to be expected as there is a extensive discussion in the literature regarding the presence of these high-ionization lines in association with the \lya\ forest, with the exception of \NV\ which is an observationally challenging species to observe at high-redshift (\citealt{2003ApJ...596..768S,2004ApJ...602...38A,2008ApJ...689..851A}). Measurements of \NV\ and \OVI\ in 2 of our strongest absorption samples show interesting signs of lower N/O than our cold gas models predict, potentially indicating some contribution from warmer $T\sim10^{5.5}$ gas. High temperature gas, $T=10^{5.5}\K$, underpredicts \CIII\ and \SiIV\ absorption, which could be interpreted as indicating that most of our high-ionization lines come from cooler photoionized gas, however it may also be possible to reproduce these measurements with a combination of $T=10^{5.5}\K$ gas and cooler high density gas.

\subsubsection{Low-ionization species}

The presence of low-ionization potential species in the data is more difficult to explain. Most of these lines went largely undetected in the \lya\ forest at high-redshift until our previous stacking analysis (P10). All of our models of typical IGM conditions underproduce lower ionization potential lines by 2--6  orders of magnitude, so an alternative approach to explaining the presence of these lines is required. The blue dashed line in Figure~\ref{stackproj_fullline} is our model that most effectively reproduces these lines. This model has a high overdensity of $\rho / \bar{\rho}=1600$, the approximate minimum temperature for photoheated gas, $T=10^4\K$, and a near-solar metallicity of [X/H]$=-0.4$. The metallicity has been scaled to produce agreement. 
\AlII, \SiII, \FeII\ and \MgII\ are well reproduced (despite using a simple solar abundance pattern). \OI\ and \CII\ are underproduced but these values may be described by a separate population or phase of gas and this is discussed below. The only substantial conflict with this model is the overproduction of \AlIII\ it provides with respect to the observations. 

Given these high densities, our ionization models and the measured \HI\ column density, we may infer the size of absorbing clouds in the optically thin regime by using
\begin{equation}
L= N_{\rm H} / n_{\rm H}=N_{\HIss}/(f_\HIss n_{\rm H})
\label{scalelength}
\end{equation}
where $N_{\rm H}$ is the total hydrogen column density, $f_\HIss$ is the neutral fraction and $ n_{\rm H}$ is the total number density of hydrogen atoms at overdensity $\rho /\bar{\rho}$.
We perform this calculation for the line centre case, which is less affected by clustered absorption;
the full line measurement yields an absorber scale 0.5 dex higher due to the higher \HI\ column density. In the strongest \lya\ absorption selection sample our most consistent model indicates 30 pc scale gas clumping. 

The next two strongest \lya\ selection samples are consistent with the same model overdensities because the measured metal and \HI\ column densities scale together. This is consistent with lower purity of CGM selection (as we argued in Section~\ref{selectiongalaxies}). Applying equation~(\ref{scalelength}) to these samples yields clumping scales 0.4 dex and 0.85 dex smaller due to their reduced  \HI\ columns. However, given their reduced purity we conclude that this constitutes a unreliable measure compared to the 30 pc clumping scale we infer from the strongest absorption sample. Despite the simplicity of our models, it is clear that observing low-ionization metal absorption with our relatively low inferred \HI\ columns requires strong small-scale clumping.

\begin{figure}
\begin{center}
\includegraphics[angle=0, width=0.48\textwidth]{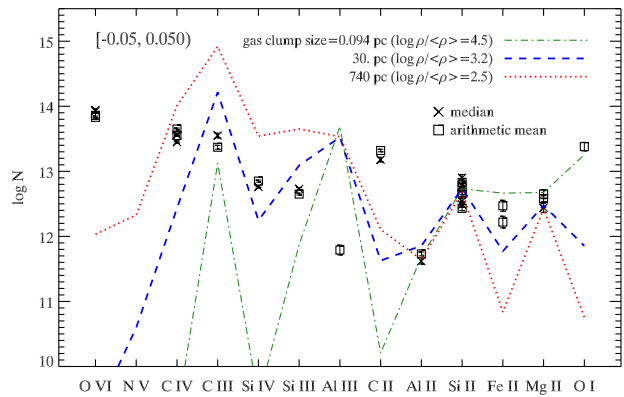}
\end{center}
\caption{As for Figure~\ref{stackproj_fullline}, highlighting variations about the model that most effectively reproduces the lowest ionization lines in the lowest flux sample when measured using only the central wavelength bin. The blue dashed line corresponds to the same model shown in Figure~\ref{stackproj_fullline}, corresponding to a gas clumping scale of 30 pc. Higher and lower density models represent gas clumping scale of 0.094 pc for [X/H]=0.3 and 740 pc for [X/H]=-0.4 respectively. $T=10^4\K$ in all cases. }
\label{stackproj_lowion}
\end{figure}

Figure~\ref{stackproj_lowion} shows details of the range of allowed gas cloud sizes. All models are for a gas temperature of $10^4\K$. The low and high density models have  [X/H]=-0.4 and 0.3, respectively, and these values are tuned for agreement with measured \SiII\ columns. The blue dashed line is the fiducial model for our observed low-ionization species. The red dotted line represents a model with lower density and so larger clumping scales, which is ruled out by its overproduction of \SiIII, \SiIV\ and \CIII\ by approximately an order of magnitude. Hence, we can rule out models with gas on scales $\ga 740$ pc as a plausible model for giving rise to the low-ionization species signal in the data.

One may also increase gas density as shown in the green dash-dotted curve in Figure~\ref{stackproj_lowion}, which assumes super-solar metallicity. As with our most consistent model and all other low-ionization models explored here,  \AlIII\ is strongly overproduced. This is the only model examined that is capable of reproducing the observed column of \OI.  
When one assumes a solar abundance pattern, there is a conflict between our measured \CII\ column densities and the relatively low columns of the low-ionization lines \SiII\ and \MgII. The \CII\ column density is at least an order of magnitude higher than that produced by any of the simple models that are consistent with these other lines.

\section{Discussion}
\label{discussion}

\subsection{The potential impact of Lyman limit systems on the composite spectra}
\label{llincomp}

In Section \ref{lya selection} we demonstrated that our selected systems, particularly those with $F<0.25$, are associated with strong, intrinsically blended \Lya\ forest lines which are indistinguishable at BOSS resolution.  We showed, that given maximal selection,  systems with $N_{\rm HI} > 10^{17.5} \cm^{-2}$ could at most give rise to 3.7\%
of our selected sample with a noise-free flux of  $F<0.25$, while systems with $N_{\rm HI} > 10^{16.9} \cm^{-2}$ potentially give rise to up to 6.1\%. We neglect the impact of such systems in our interpretation using the optically thin approximation. While this population is limited, if the metal lines associated with these systems are sufficiently strong they might have a significant impact on our composite spectra and so our interpretation.

We can assess the potential impact of this effect using the above maximum incidence rate of of $N_{\rm HI} > 10^{17.5} \cm^{-2}$ systems and the median composite spectrum in figure 9 of \citealt{2013ApJ...775...78F} (F13 hereafter). Since the strongest absorber sample represents our most efficient selection of true $F<0.25$ wavelength bins it provides the most convenient comparison with the above incidence rates. If one assumes that 3.7\% of our sample has metal lines of the strength shown in the F13 composite and that our resultant metal signal exclusively came from them, this signal would be diluted by a factor of $1/0.037= 27$. Our resultant composite would appear qualitatively like the figure of F13 but with metal line flux decrements a factor of 27 weaker. All metal lines in our strongest composite are at least a factor 4 stronger than this projected maximal contribution from systems with  $N_{\rm HI}> 10^{17} \cm^{-2}$, and so we can rule out a scenario where they dominate the signal of any of our metal lines even in this conservative case.

Note that super Lyman limit systems (or sub-DLAs) and DLAs are included in the stacked spectrum of F13. For a direct comparison with our measurements the DLAs would have to be omitted from their stacked spectrum. The excess metals associated with these DLA render the above assessment yet more conservative. Also the resolution of the F13 spectra is approximately double ours and so if the lines in this composite are narrower than ours, the minimum flux of lines will be higher and so the above calculation of their potential impact is overestimated.

The above argument does not include the impact of partial Lyman limit systems ($10^{16.9} \cm^{-2} < N_{\rm HI} < 10^{17.5} \cm^{-2}$). We may include them with the additional conservative assumptions that the strength of their associated metals is also well characterised by figure 9 of F13 and that they are also maximally selected. The maximum potential selection of partial Lyman limit systems, Lyman limit systems, super Lyman limit systems and DLAs (i.e. $N_{\rm HI} > 10^{16.9} \cm^{-2}$) is 6.1\%. This corresponds to dilution by a factor of 16 in our sample. Again comparing with our strongest absorption sample, no metal line has a greater than 50\% maximal contribution by such systems. 

One final caveat must be stated. The above maximal contributions to metal lines by high Lyman limit opacity systems are based the use of a simple arithmetic mean in constructing a composite spectrum. In this work, we use the median and the arithmetic mean discarding the highest 3\% and the lowest 3\% fluxes. Both these statistics are more resistant to the introduction of small populations of outliers in the stack of spectra than the simple arithmetic mean. This adds another layer of conservatism to the assessment of the impact of these systems.

While we cannot clearly estimate the contribution of these high Lyman limit opacity systems, we can confidently infer that they do not dominate any of our metal line absorption measurements, given the series of conservative assumptions above. Our conclusion that much of the absorption arises in optically thin gas and requires both high density and high metallicity to explain it is not altered.

\subsection{Central bin, full profile and high-resolution measurements}

We measure \HI\ and metal absorption using both the full observed absorption profile and the smallest representative portion of our spectrum associated with the selected absorbers - our 2-pixel-wide selection bin. These spectral bins are $138\kms$ wide, slightly smaller than the BOSS resolution FWHM ($150\kms$). This provides two column density integration scales, one narrow and the other inclusive. The difference in column density is approximately 0.5 dex for our whole sample. While these individual line measurements differ, the uniformity of the global shift in column density makes the choice between them inconsequential when compared with our models; i.e. if the integration scale is consistent between \HI\ and metal lines, the measurements are robust.

These large differences based on integration scale should sound a note of caution for comparisons with individual lines in high-resolution spectra. 
All our measured column densities are integrated values and the individual, resolved lines that make up this signal may be of dramatically lower column density, potentially rendering them unobservable in individual spectra.

\subsection{Limitations of our absorber size constraints}

We have neglected to consider the impact of impure CGM selection or covering fractions of low-ionization species below unity. Both these effects are present in the data and lead to underestimates of the column of low-ionization species in the systems where they are present. A relative strengthening of low-ionization lines would force models to smaller clumping scales to maintain consistency. We will investigate this effect in a future publication by constraining the absorber population size and strength in the full stack of spectra at each line centre. We will also further explore the full range of \HI\ absorption we select.

Our simple ionization models are applied assuming that the gas observed is optically thin as indicated by our results in Section~\ref{char hi}. It is possible that ionising radiation can be attenuated by the small but significant Lyman limit opacity that we see. If we conservatively assume that the UV background as a whole is attenuated by this level (rather than just blueward of the Lyman limit) and apply the highest opacity shown seen in Figure~\ref{finalhicolumns}, we find a negligible change from the results presented here. 

Changes in the UV background intensity can be treated as a simple scaling relation. Our results assumed a fixed UV background intensity at $z=2.7$ taken from \citet{2013arXiv1307.2259B}. When this intensity is increased or decreased by 0.1 dex (in line with the stated errors), this is equivalent to an equal increase or decrease in density for the ionization models. Hence such errors correspond to an equal modifications to the absorber size. This is a small effect compared to our current range of plausible absorber sizes. We also assume ionization equilibrium in this preliminary interpretation, but non-equilibrium effects may have a significant role in determining the characteristics of the gas observed \citep{2013MNRAS.434.1043O}. 

We provide only a preliminary exploration of strong effects seen in our composite spectra by comparison with a limited set of models. We do not consider modifications to the abundance profile of elements measured, but measurements from multiple species of the same element allow us to limit the importance of this effect. In particular, observations of only \SiII, \SiIII\ and \SiIV\ are sufficient to support the constraints on absorber sizes in the CGM, since reproducing \SiII\ columns without overproducing our \SiIII\ and \SiIV\ columns forces models to small absorber sizes.

Our models assume a quasar+galaxy UV background model, however, much of our results arise as a consequence of probing circumgalactic regions. Therefore, a potential source of systematic error may arise where galaxies generate a local excess in the galaxy component of this radiation field. In general, a higher amplitude radiation field would require higher gas density to compensate.

\subsection{Potential for further IGM/CGM constraints}

We detect signs of [C/$\alpha] > 1$, and this is not the first time that super solar C/$\alpha$ has been observed at high-redshift. \citet{2008MNRAS.385.2011P}, \citet{2010ApJ...721....1P} and \citet{2011MNRAS.417.1534C} find evidence of [C/$\alpha$] as high as 0.5 dex above solar in metal poor DLAs.  Our results show a discrepancy 0.5 dex higher still than these DLA observations and for much higher metallicities. A potential explanation for these observations of high [C/$\alpha$] in DLAs is a significant contribution by explosions of Pop III stars \citet{2011ApJ...730L..14K}. 

In addition, our observations of \FeII\ may allow exploration of deviations from solar $\alpha$/Fe. We see no significant deviations that are distinct from the allowed range of gas physical conditions thus far; however, a full species-by-species exploration of parameter space with modifications to the UV background, abundance patterns and physical conditions would place valuable constraints on the abundance pattern of gas in the CGM at high-redshift.

The ultimate goal of these measurements is improving our understanding of galaxy formation and evolution. The most effective way of achieving this is by direct comparison with hydrodynamic simulations. The column densities for \HI\ and several metal lines are broadly in line with those predicted for CGM regions by recent hydrodynamic simulations \citep{2013MNRAS.430.1548H,2013ApJ...765...89S},
 but a detailed comparison is necessary for the subset of sight lines through simulated CGM regions that meet our \lya\ selection functions and total \HI\ columns.
 Our measurements allow a direct comparison to simulations, but they all do so with some simplifications. The optimal way to explore this problem without loss of information content is to compare the composite spectra with composites produced from simulations and we will conduct this analysis in a future work (M. Mortonson et al. in prep). We enable the simulation community to utilise these measurements in their entirety by providing our composite spectra online \footnote{http://icg.port.ac.uk/stable/pierim/compositelyabossdr9.html}.

\subsection{Comparison with other observations}

In the above analysis we have assumed that high-ionization lines arise in gas with typical IGM physical conditions rather than a circumgalactic origin, even for systems which we infer arise in CGM regions (Section~\ref{selectiongalaxies}). We assume that we do select CGM regions in a minority of cases or that there is a nearby excess in IGM structure that we fail to resolve. However, recent results by \citet{2011Sci...334..952T} and \citet{2013ApJ...767...49M} show low, high and intermediate ionization potential species are observed to be kinematically similar despite certainly probing different phases. Although they are at low-redshift, these results are consistent with our observations and are suggestive of poor mixing of hot and cold, metal rich and metal poor, dense and rare gas as part of a bulk flow.

\citet{2006ApJ...637..648S} inferred near-solar metallicities and high-densities consistent with our results from three high-redshift galaxies with a line of sight in their circumgalactic regions.
\citet{2007MNRAS.379.1169S} explored the properties of several strong \CIV\ lines and performed ionization 
calculations using upper limits on column densities of various other species. The physical conditions derived indicated clumping scales of $\sim 100$ pc,  solar or higher metallicities and densities of around $n_H \sim 10^{-3.5} {\rm \cm^{-3}}$ (or $ \rho / {\bar \rho} \sim 50$). The low \HI\ columns probed ($\la 10^{14}\cm^{-2}$) and moderate densities indicate that these systems are likely intergalactic in origin and not verifiably proximate to galaxies. The results are primarily a measurement of mixing of metals on large scales. More recently P10 found evidence of 10 pc clumping from composite spectra of \Lya\ absorbers using SDSS-II spectra. These results suggested near solar metallicities, high densities and low-ionization species making them broadly indicative of circumgalactic regions; an inference that we verify here. 

\citet{2010ApJ...717..289S} detected sub-kpc clumping based on observations of LBGs as background light sources near LBGs. They also argue for an increase in absorber sizes with increasing galactic impact parameter.

Observations of \MgII\ absorbers have shown signs of clumping scales $\sim10^{-2}$pc by comparison of incidence rates and beam sizes along the line of sight to gamma-ray bursts and quasars \citep{2007Ap&SS.312..325F,2007ApJ...659L..99H}.  There remains, however, considerable uncertainty, particularly with respect to quasar beam sizes. Indications of sub-parsec structure in \MgII\ absorbers have been further reinforced by the results of \citet{2013MNRAS.434..163H} based on time-variability of absorbers in SDSS quasar spectra.

At low-redshift a variety of measurements have been performed. \citet{2002ApJ...575..697T} studied a system in the Virgo cluster showing substantial low-ionization lines (but a lack of high-ionization lines) and concluded a sub-kpc thickness of the gas that gave rise to them. \citet{2006MNRAS.367..139A} explored absorbers along the line of sight to a $z=0.37$ quasar and found metallicities $0.5\,{\rm dex}$ below solar and absorber size of a few kpc. A study of  \NeVIII\ and \OVI\ toward a $z=1.0$ quasar \citep{2013ApJ...767...49M} found systems with metallicities of $Z \ge 0.3 Z_{\sun}$ and sizes $\sim 1$kpc.  Supersolar neon absorption has also been seen around our galaxy \citep{2002ApJ...573..157N}.

\citet{2002ApJ...565..743R} found indications of clumping on $\sim 10$pc scales in a sub-set of their weak \MgII\ absorbers that are relatively iron rich at $z\sim1$; however, \citet{2007ApJ...666...64L} found that the incidence of such weak \MgII\ systems is much reduced at redshifts as high as $z=2.4$, the lower redshift limit of our study.

There are striking similarities between our strongest \lya\ absorption sample and the lowest \HI\ column densities of \citet{2013ApJ...770..138L} investigated at $z\la1$. These \HI\ columns are slightly higher than ours  ($\ga 10^{16.2} {\rm cm^{-2}}$ for their strongest component). They designate such systems Lyman limit systems, while formally they are only marginally partial Lyman limit systems. They report that these partial Lyman limit systems are circumgalactic in origin, and in this sample they find a bimodality in metallicity with peaks at [X/H]~$\simeq -1.6$ and -0.3. They interpret these results as evidence of both clumpy metal rich outflows and cold metal poor accretion streams. Recent metallicity measurements of components of the sub-damped system at $z=2.4$ nearby a Lyman break-selected galaxy \citep{2013ApJ...776L..18C} show similarly high and low metallicities, which suggests that this bimodality may also be present at higher redshifts. While we cannot detect this bimodality in our composite spectra, we are broadly consistent with both these phases and indeed do not require additional phases than these two broadly defined categories.

\section{Summary}

We present composite spectra of \lya\ forest absorbers selected by flux  in $138\kms$ wide bins in the BOSS DR9 sample of quasar spectra. This binning scale represents two pixels of the SDSS wavelength solution and a little under the typical resolution of the BOSS spectrograph ($150\kms$). We limit ourselves to intervening absorbers (selected within 1041--1185\,\AA\ in the quasar rest-frame) at redshifts $2.4<z<3.1$. We examine five samples with increasing flux centred at $F=0.0$, 0.1, 0.2, 0.3 and 0.4. 

Tests on high-resolution spectra show that our three strongest absorbers samples would, in the absence of noise, probe CGM regions (300 pkpc transverse, $\pm 300\kms$ line of sight) $\approx 60$\% of the time (Figure~\ref{nearlbgs}). In the presence of noise, we get steadily decreasing purity of CGM selection with increasing flux (Figure~\ref{selectiontest}).

We characterise the inferred \HI\ column density contribution to this selection by measuring Lyman series lines and comparing them with a variety of single line models. We complement this with a new modified approach to stacking directed at measuring the Lyman limit opacity resulting from our selected absorbers and any associated excess \lya\ absorption. We present measurements of the column density of 26 metal lines from 7 elements and 13  ionization species, over the full line profiles (which includes all clustered absorption both from complexes and large scale structure) and for the wavelength bin selected i.e. $138\kms$ scales (which corresponds to our most exclusive measure with respect to clustering). These measurements constitute upper and lower limits on our column density integration scale; however, both sets include clustering making them difficult to compare with observations of individual lines of sight. 

We compare our measurements to simple ionization models assuming a quasar and galaxy UV background model and a solar pattern of elemental abundances.  We find that the potential contribution of Lyman limit systems to our metal line measurements is sufficiently weak to allow the use of the optically thin approximation in our ionization corrections without modifying our interpretation. All our samples show clear measurements of high-ionization parameter lines broadly in line with physical conditions associated with the IGM and a metallicity of approximately  [X/H]$=-2$. Our composite spectra corresponding to CGM regions show a number of additional lines. These are mostly low-ionization lines and we are able to reproduce them in our models only with much higher densities and metallicities. The combination of density and column density implies a characteristic length scale of these low-ionization absorbers, and we 
find greatest consistency with absorbers typically 30 pc in size with a near-solar metallicity of [X/H]$=-0.4$ . There are indications that measurements of \OI\ require yet smaller scale gas clumping. 
We also detect tentative signs of a strongly elevated C/$\alpha$ compared to solar, and note the potential to measure both $\alpha$/Fe and the shape of the UV background as part of a full search of parameter space.

We make available all our composite spectra available to allow a direct comparison between simulations and our absorption profiles.

\section*{Acknowledgments}
We would like to offer special thanks to the anonymous referee, whose constructive criticism led to significant improvement of this manuscript.
We also thank Romeel Dav\'e, Neal Katz, Benjamin Oppenheimer and Mark Fardal for the use of their SPH simulations and Benjamin Oppenheimer for the use of his CLOUDY look up tables. Also we thank Todd Tripp and Joop Schaye for their insightful comments on this work.

The research leading to these results has received funding from the European Union Seventh Framework Programme (FP7/2007-2013) under grant agreement n¡ [PIIF-GA-2011-301665]. MM was partially supported by NASA grant NNX12AG71G and CCAPP at Ohio State. DW acknowledges support of NSF Grant AST-1009505 and NASA ATP grant NNX10AJ95G

Funding for SDSS-III has been provided by the Alfred P. Sloan
Foundation, the Participating Institutions, the National Science
Foundation, and the U.S. Department of Energy Office of Science.
The SDSS-III web site is http://www.sdss3.org/.
SDSS-III is managed by the Astrophysical Research Consortium for the
Participating Institutions of the SDSS-III Collaboration including the
University of Arizona,
the Brazilian Participation Group,
Brookhaven National Laboratory,
University of Cambridge,
Carnegie Mellon University,
University of Florida,
the French Participation Group,
the German Participation Group,
Harvard University,
the Instituto de Astrofisica de Canarias,
the Michigan State/Notre Dame/JINA Participation Group,
Johns Hopkins University,
Lawrence Berkeley National Laboratory,
Max Planck Institute for Astrophysics,
Max Planck Institute for Extraterrestrial Physics,
New Mexico State University,
New York University,
Ohio State University,
Pennsylvania State University,
University of Portsmouth,
Princeton University,
the Spanish Participation Group,
University of Tokyo,
University of Utah,
Vanderbilt University,
University of Virginia
University of Washington,
and Yale University.

\appendix

\section{Metal Line Measurements}
\label{metalcat}

\begin{figure*}
\begin{center}
\includegraphics[angle=0, width=.195\textwidth]{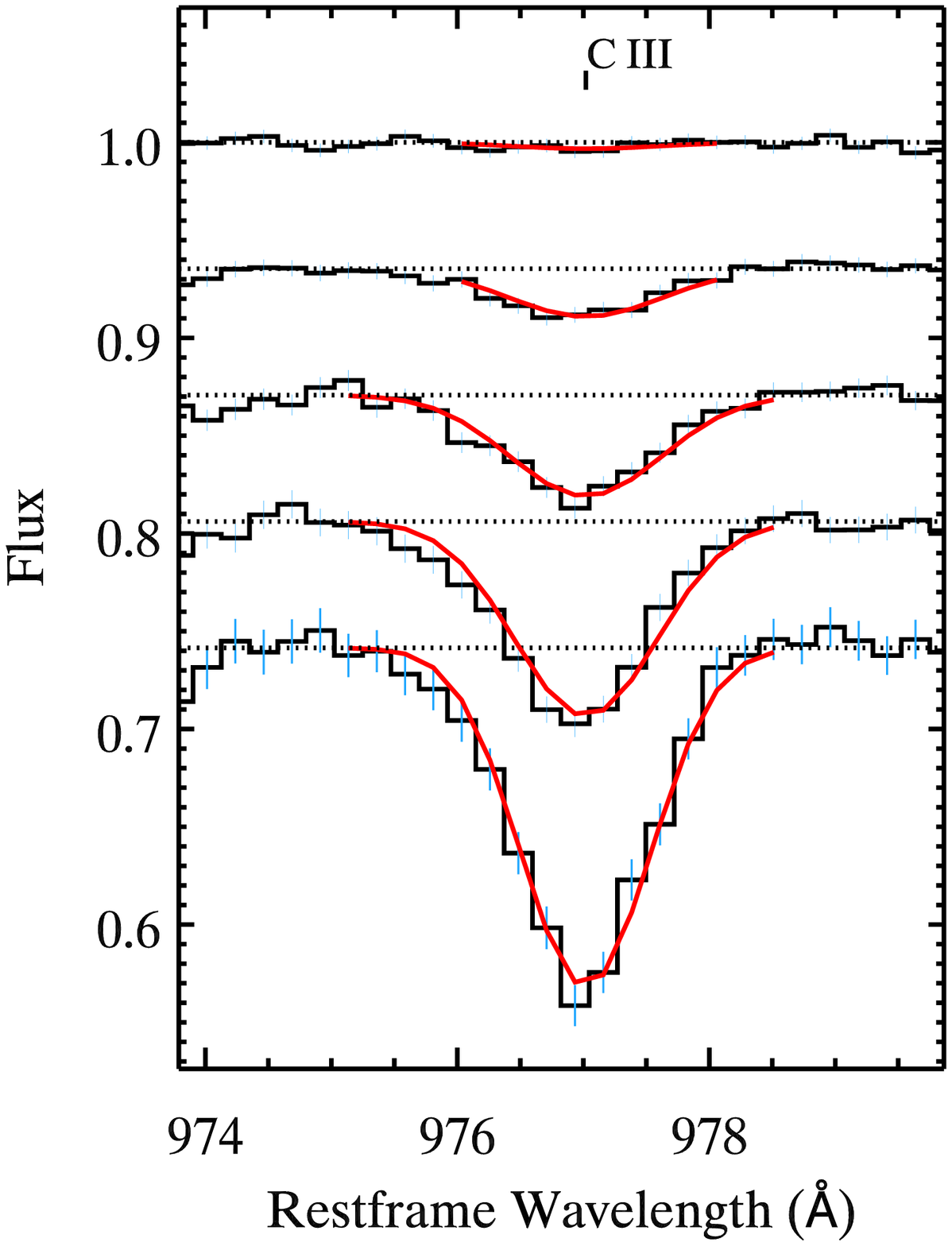}
\includegraphics[angle=0, width=.195\textwidth]{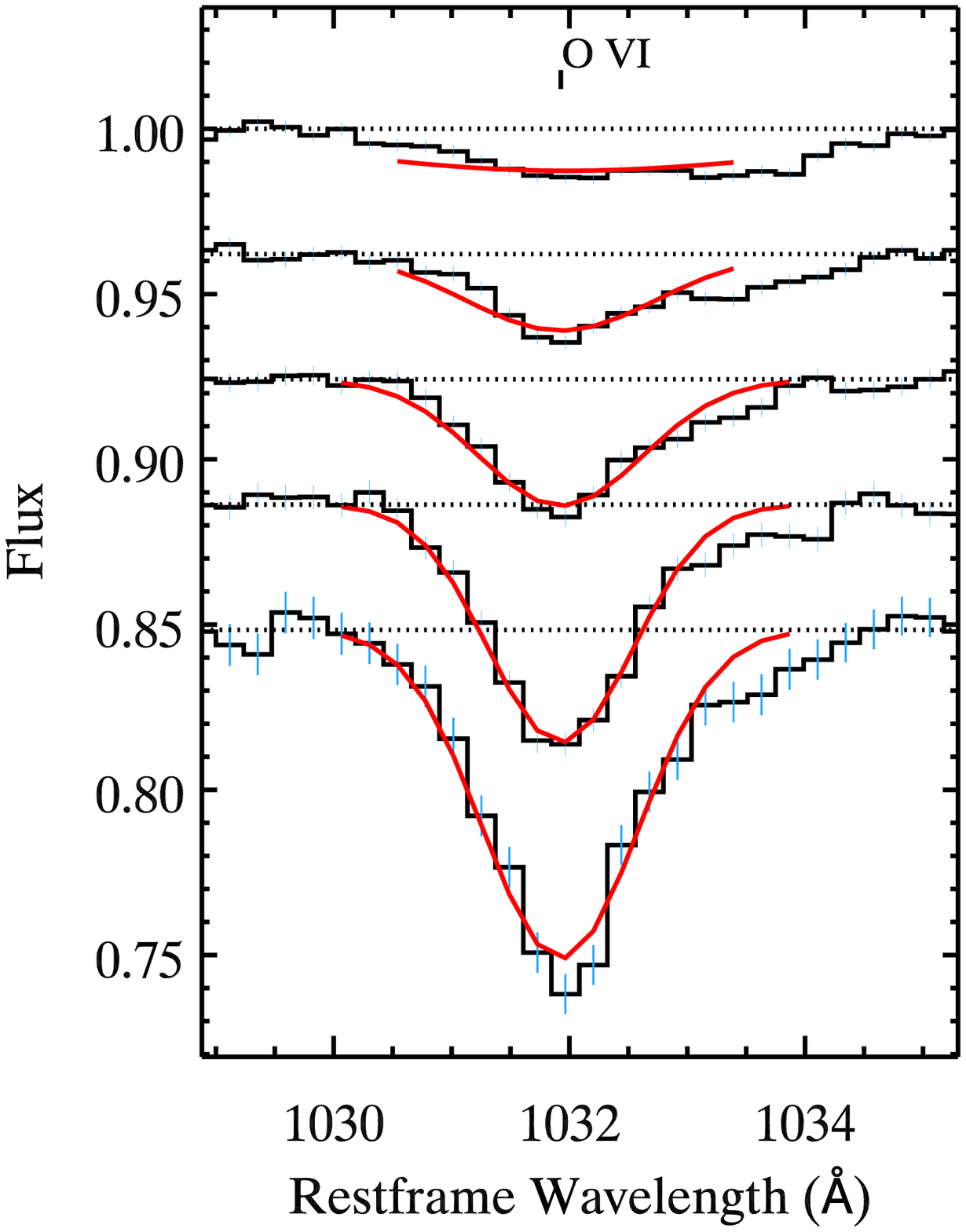}
\includegraphics[angle=0, width=.195\textwidth]{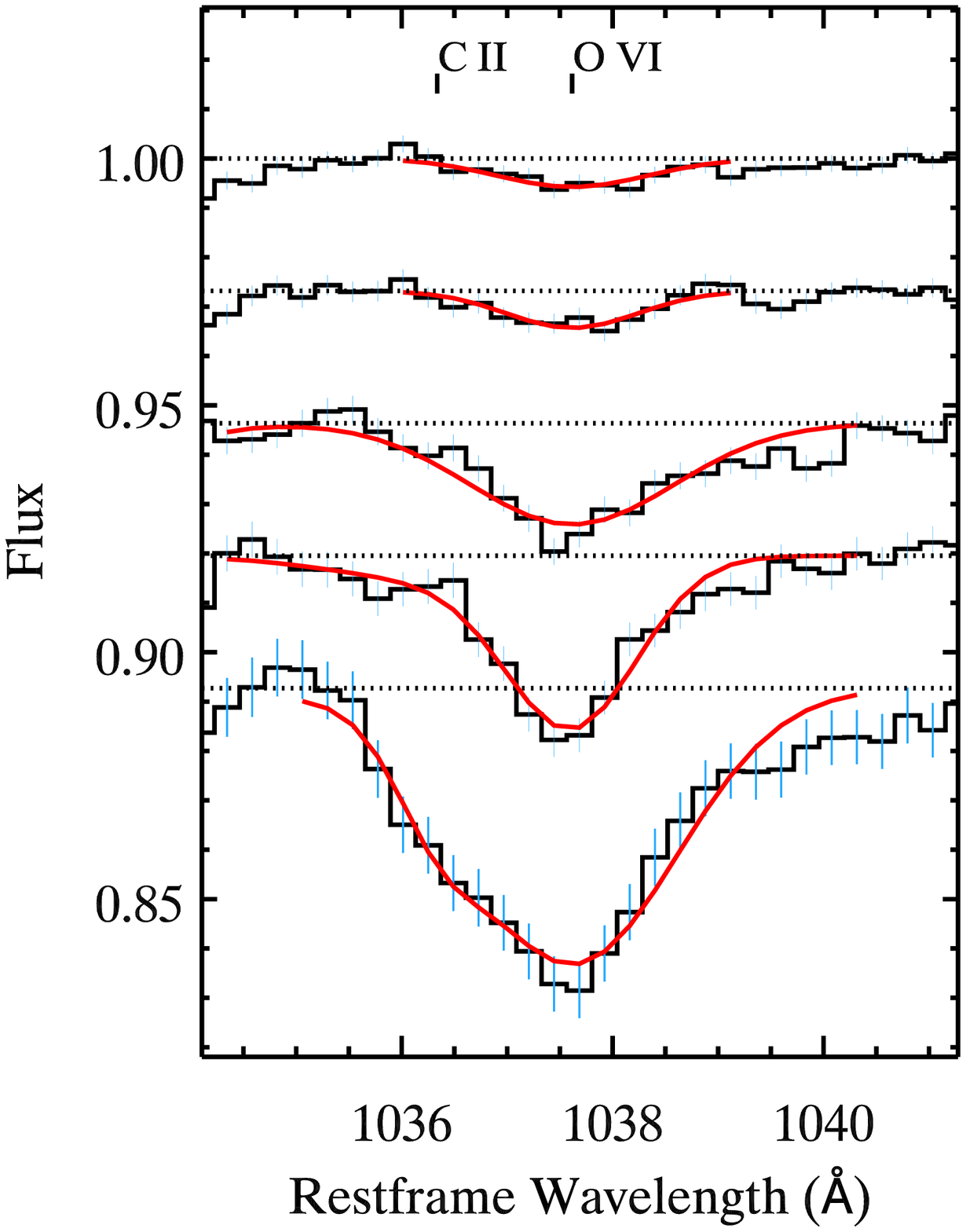}
\includegraphics[angle=0, width=.195\textwidth]{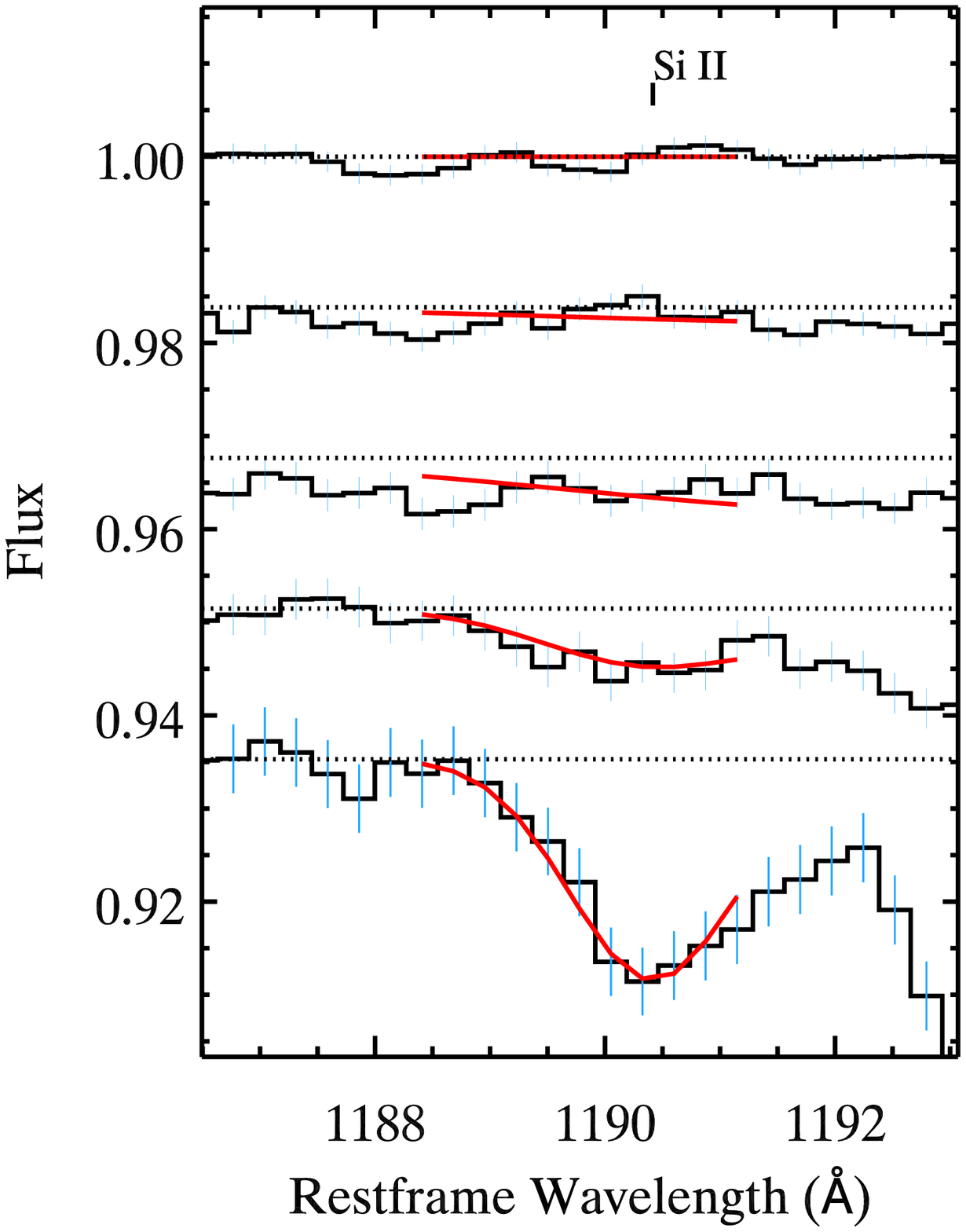}
\includegraphics[angle=0, width=.195\textwidth]{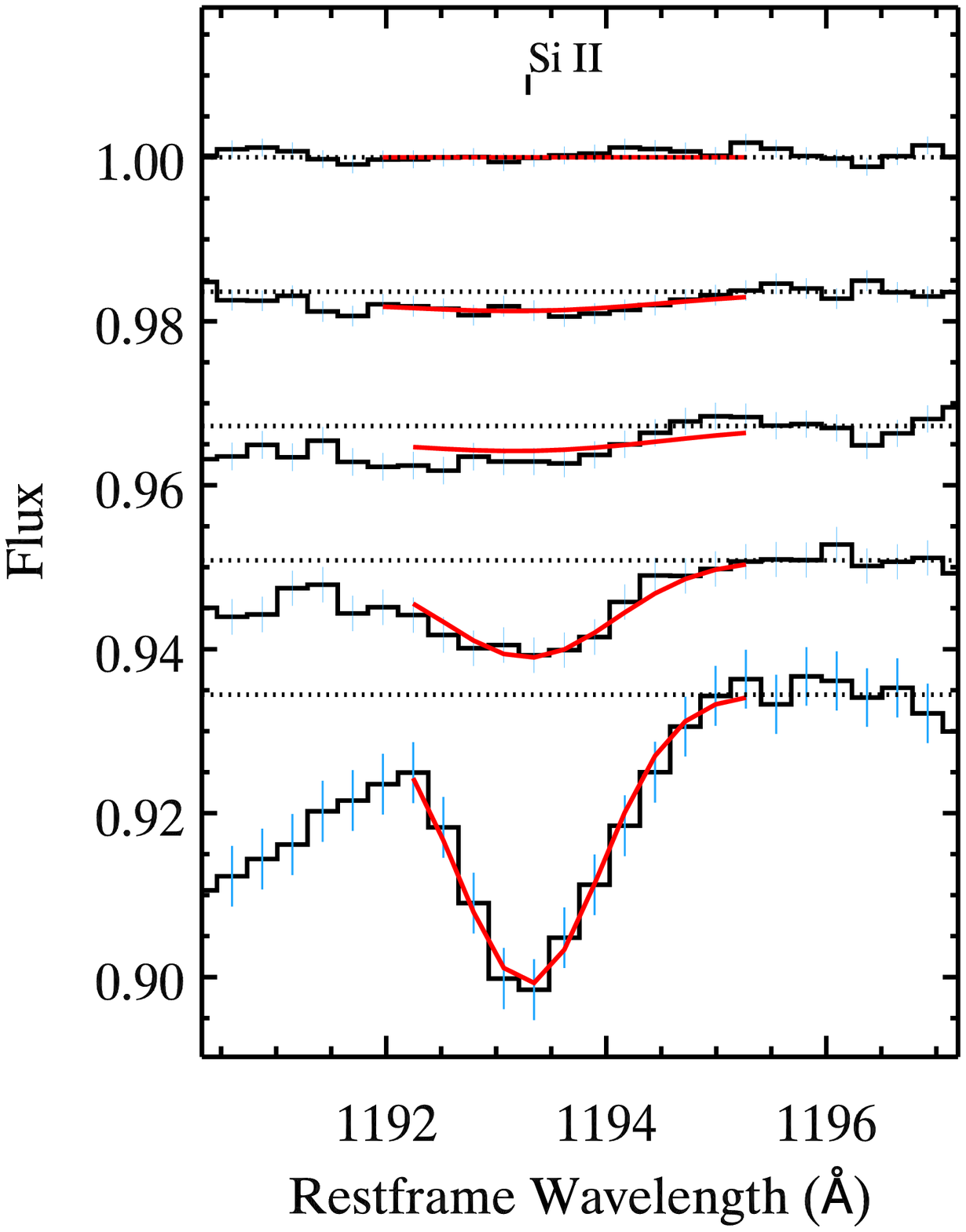}
\includegraphics[angle=0, width=.195\textwidth]{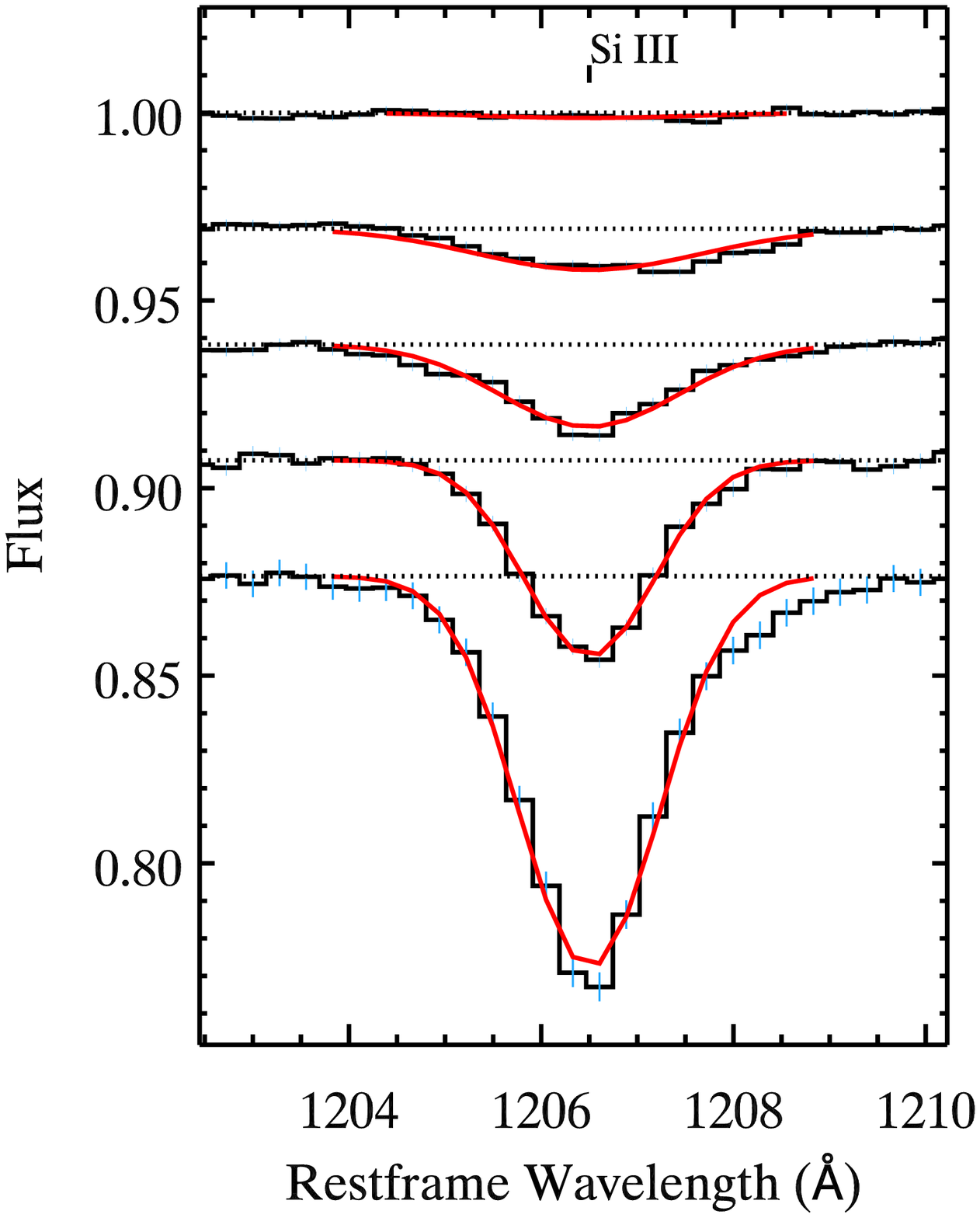}
\includegraphics[angle=0, width=.195\textwidth]{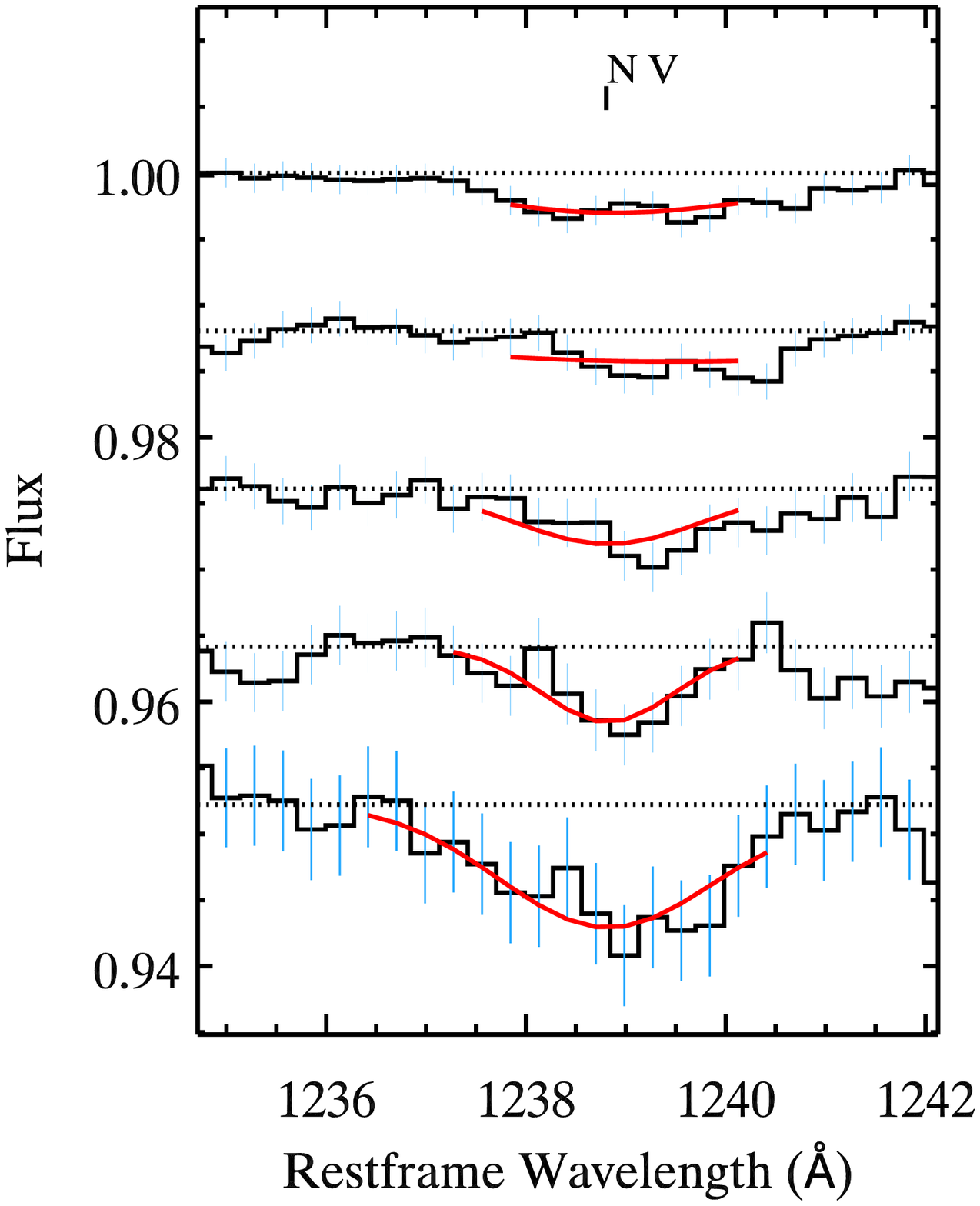}
\includegraphics[angle=0, width=.195\textwidth]{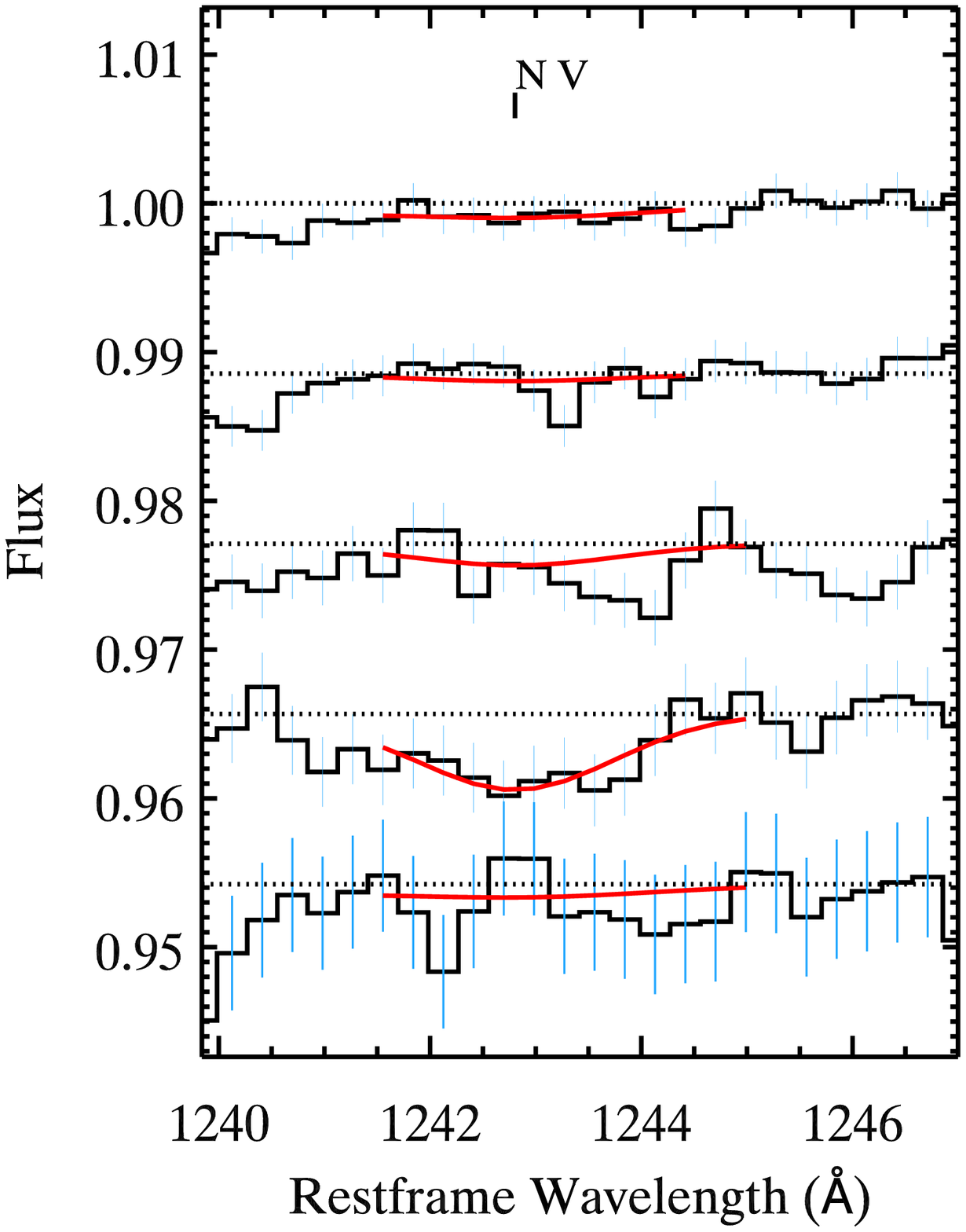}
\includegraphics[angle=0, width=.195\textwidth]{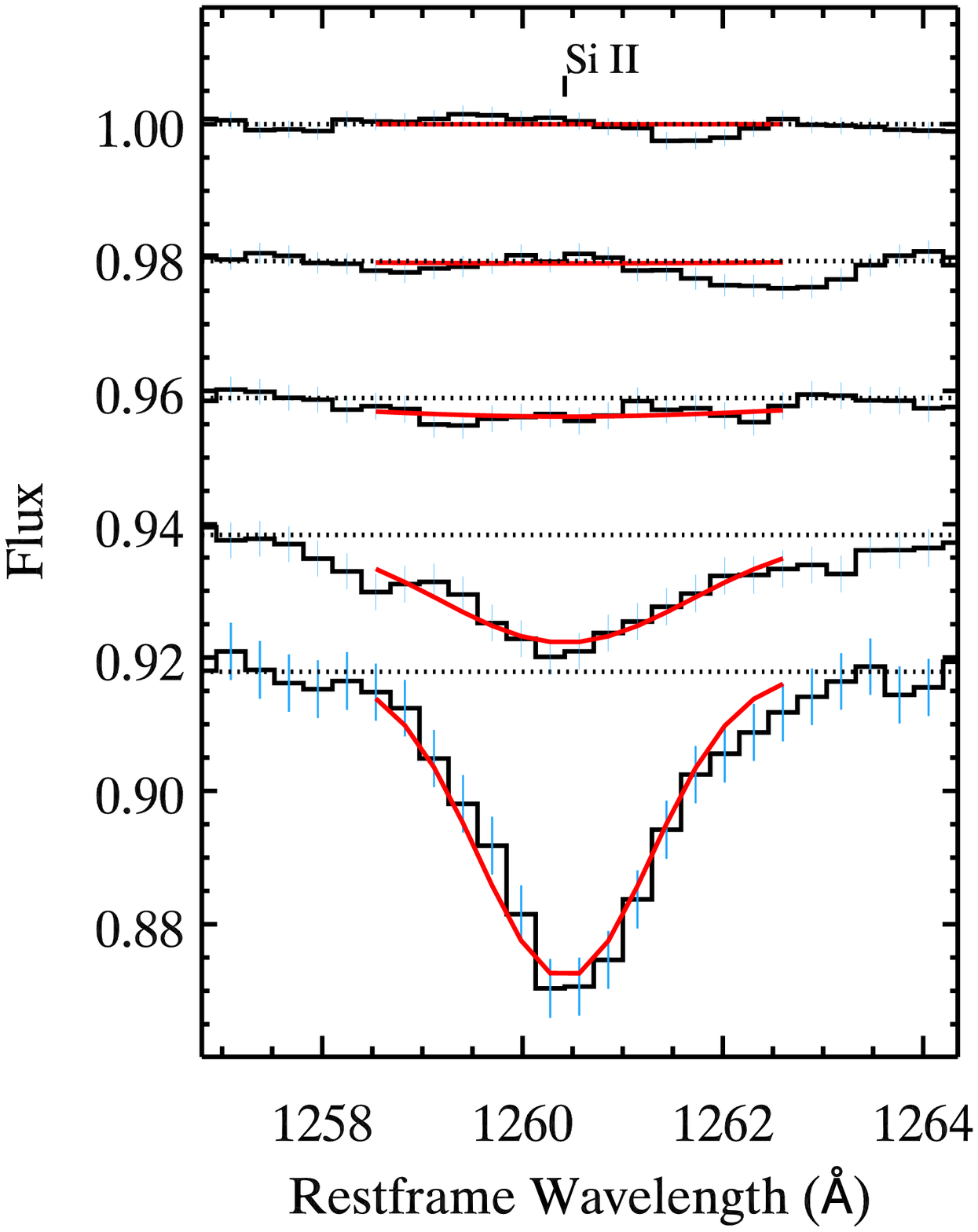}
\includegraphics[angle=0, width=.195\textwidth]{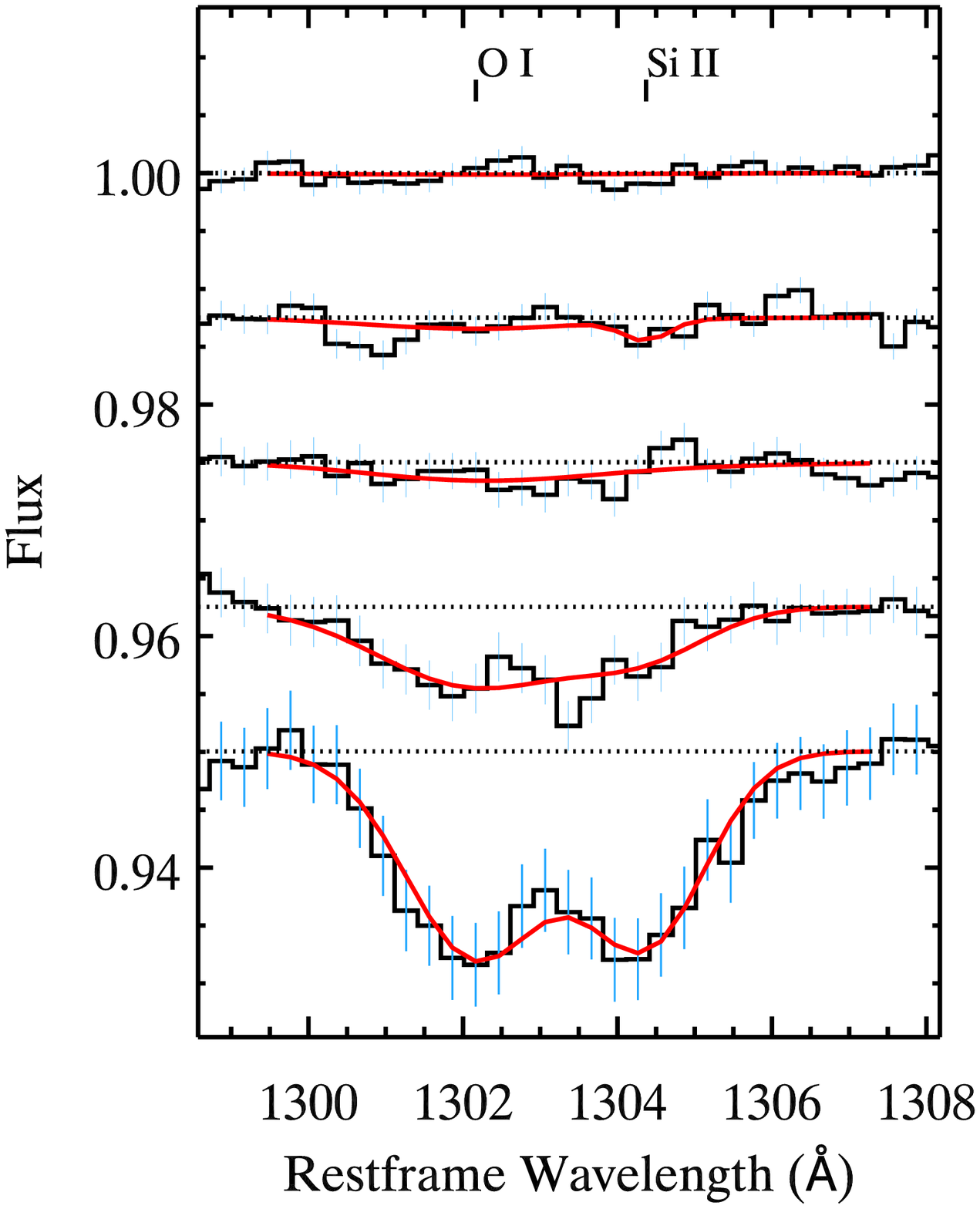}
\includegraphics[angle=0, width=.195\textwidth]{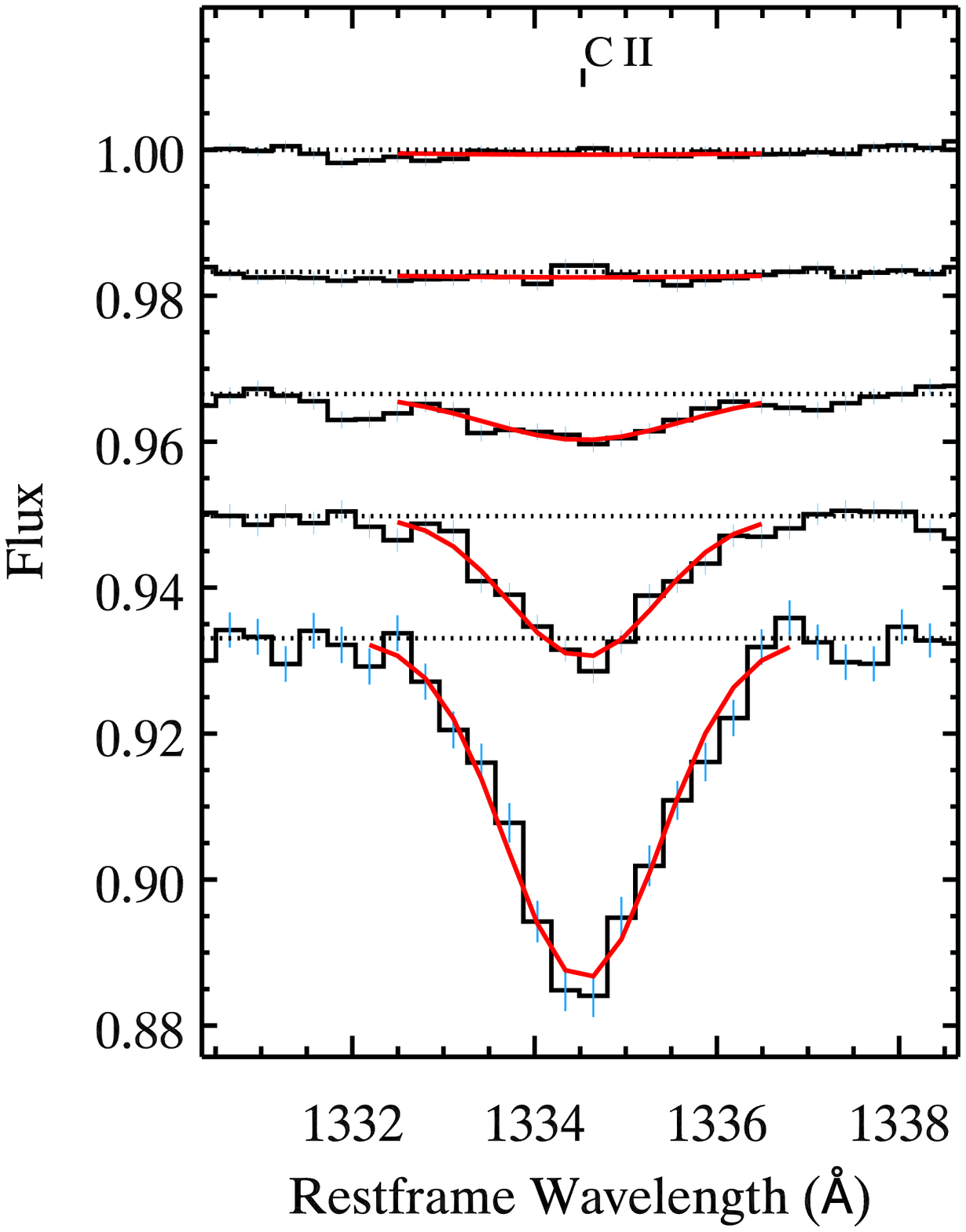}
\includegraphics[angle=0, width=.195\textwidth]{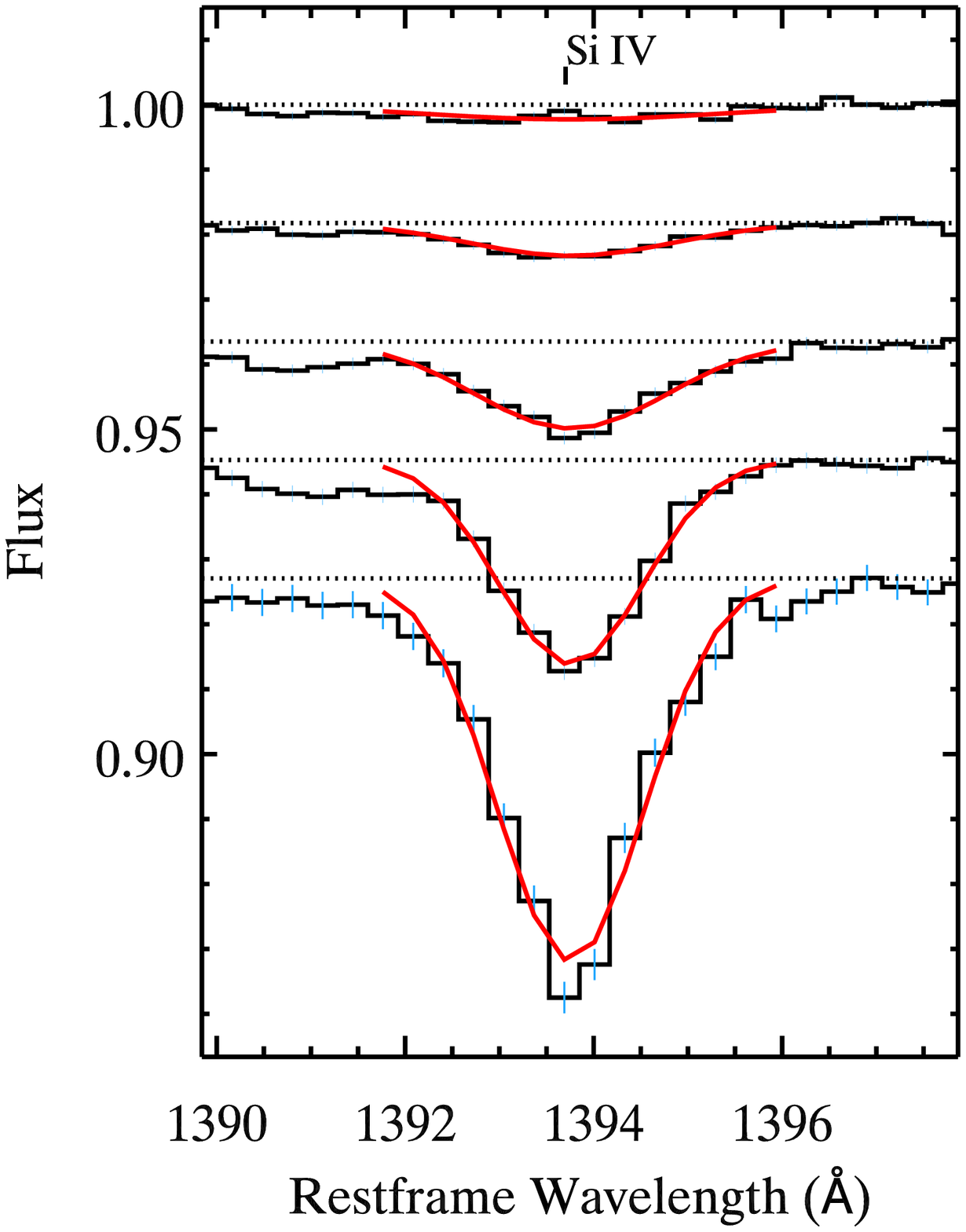}
\includegraphics[angle=0, width=.195\textwidth]{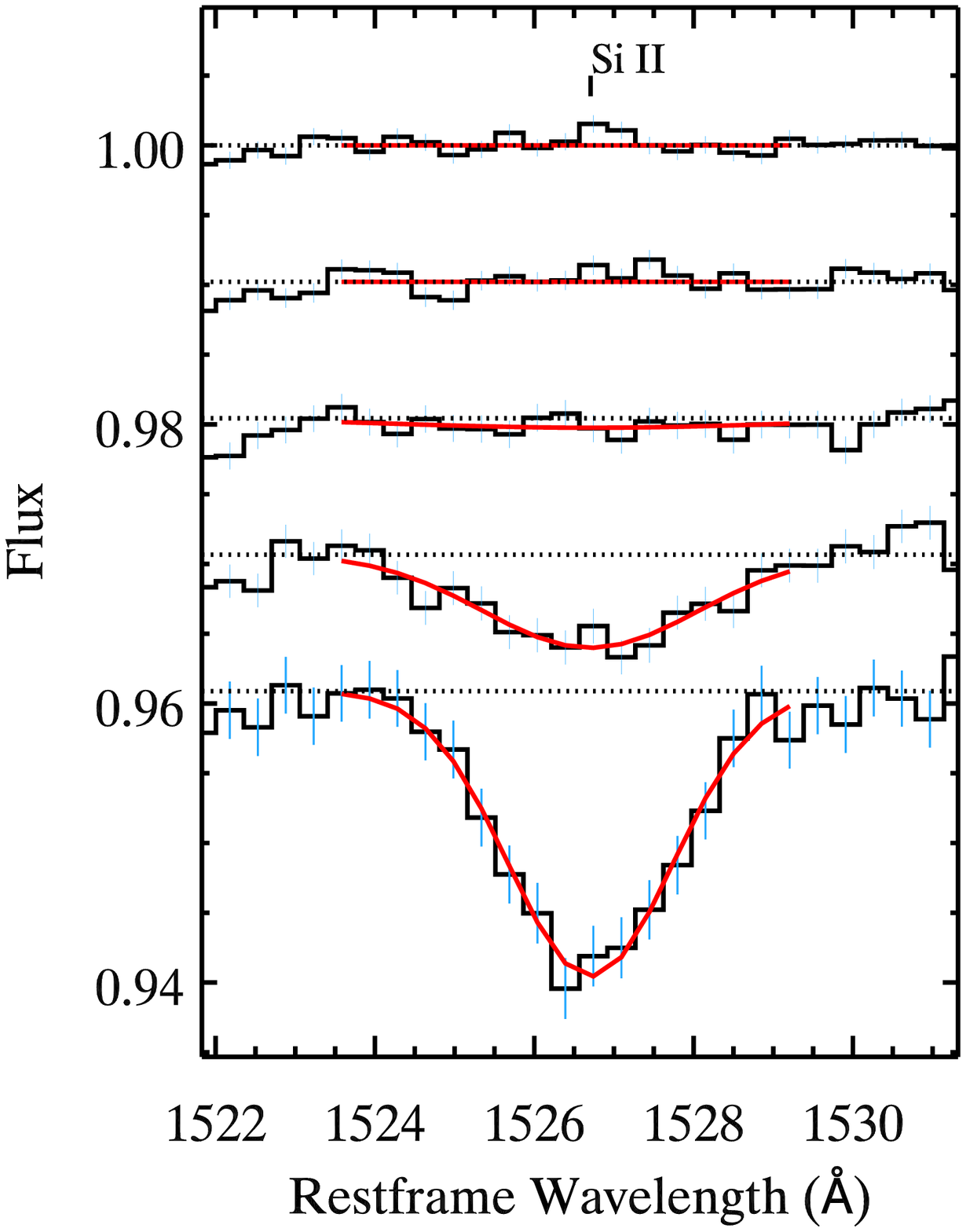}
\includegraphics[angle=0, width=.195\textwidth]{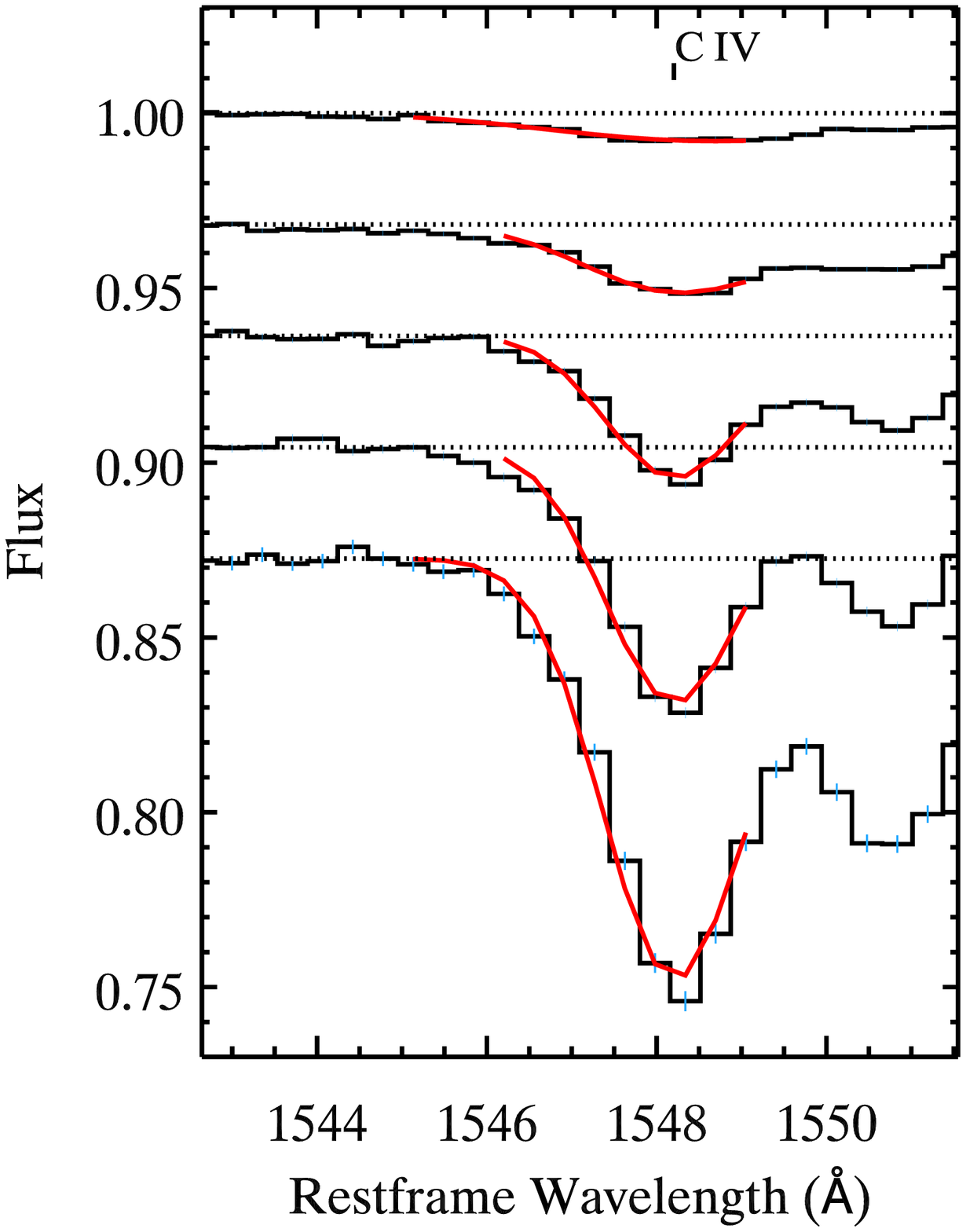}
\includegraphics[angle=0, width=.195\textwidth]{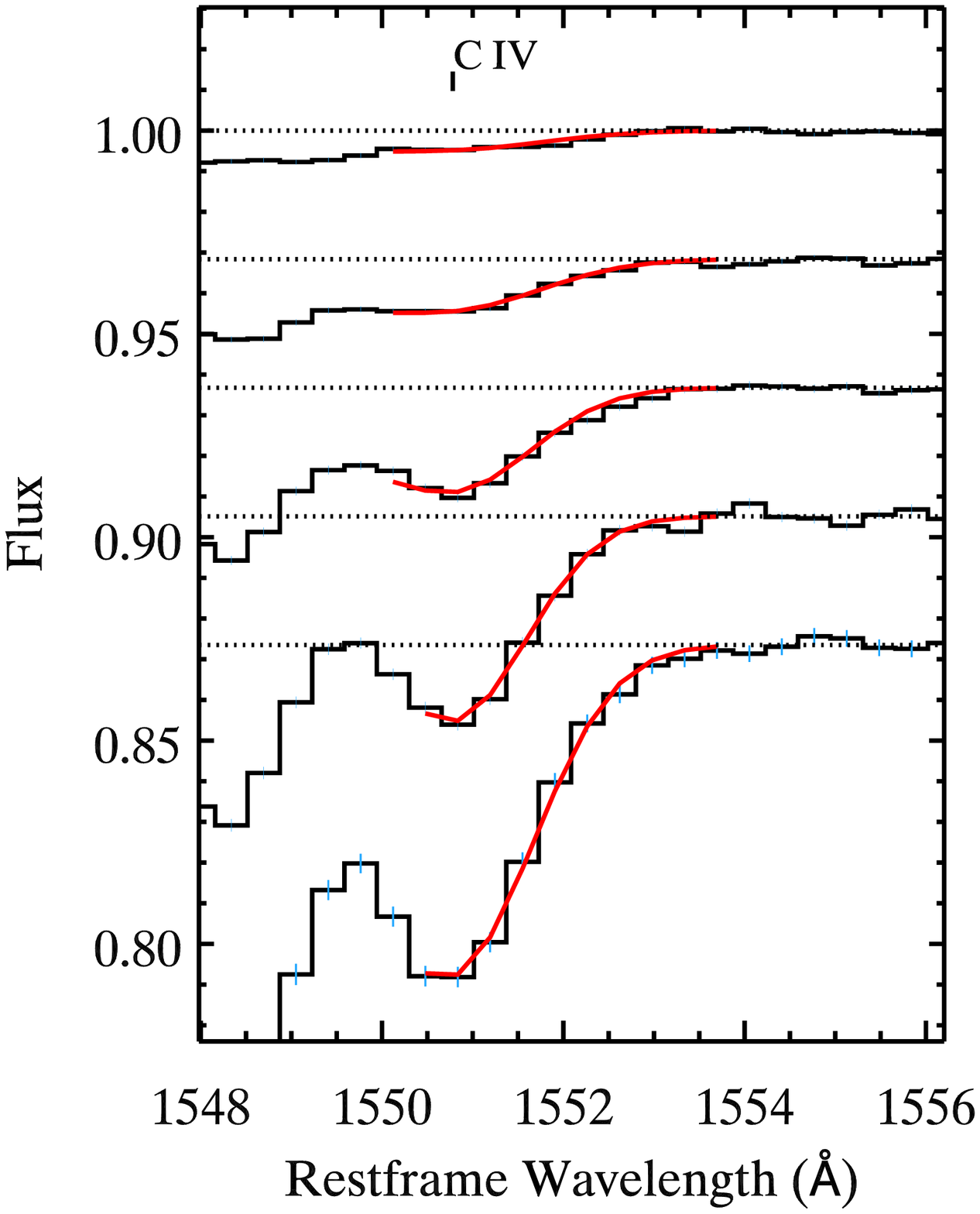}
\includegraphics[angle=0, width=.195\textwidth]{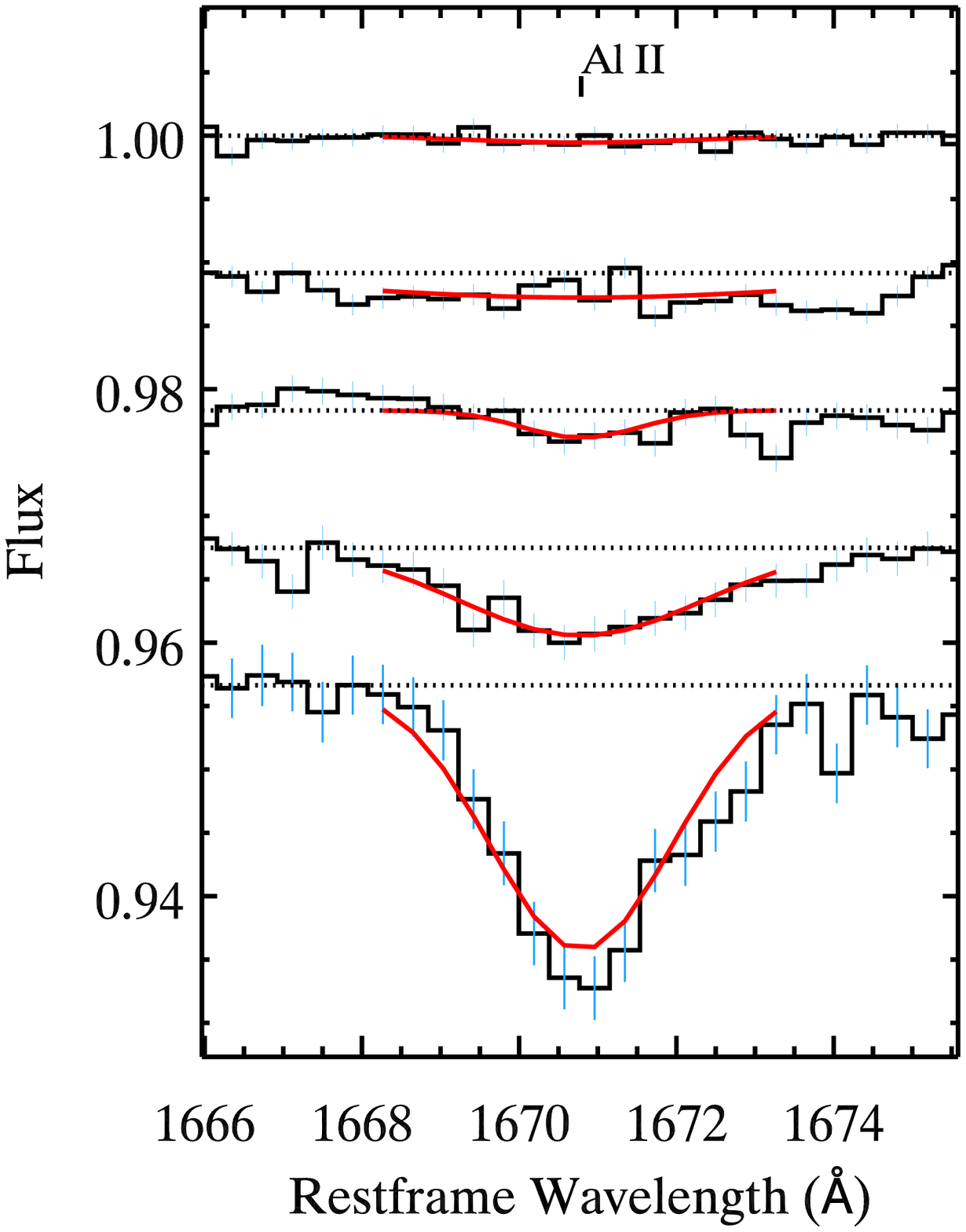}
\includegraphics[angle=0, width=.195\textwidth]{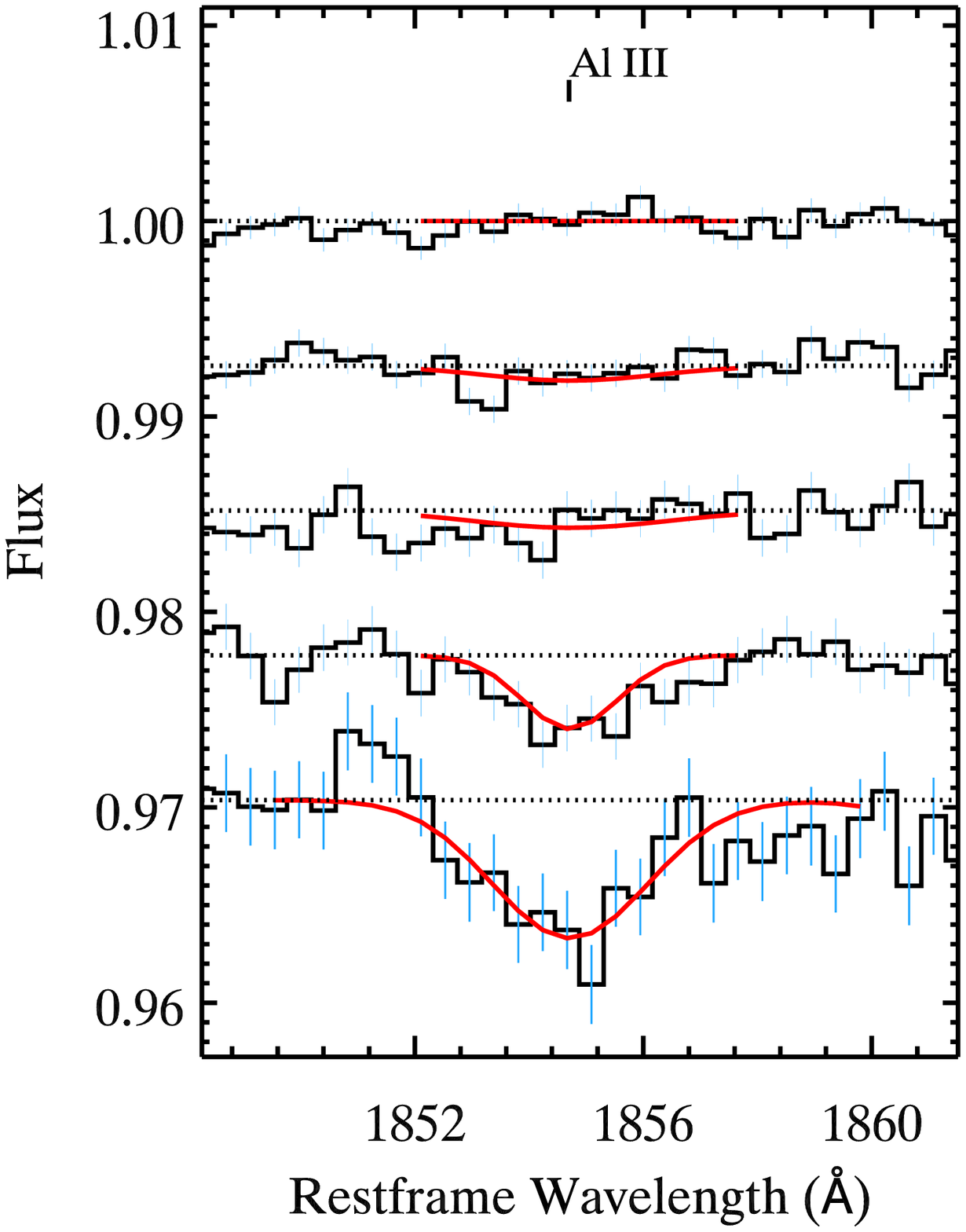}
\includegraphics[angle=0, width=.195\textwidth]{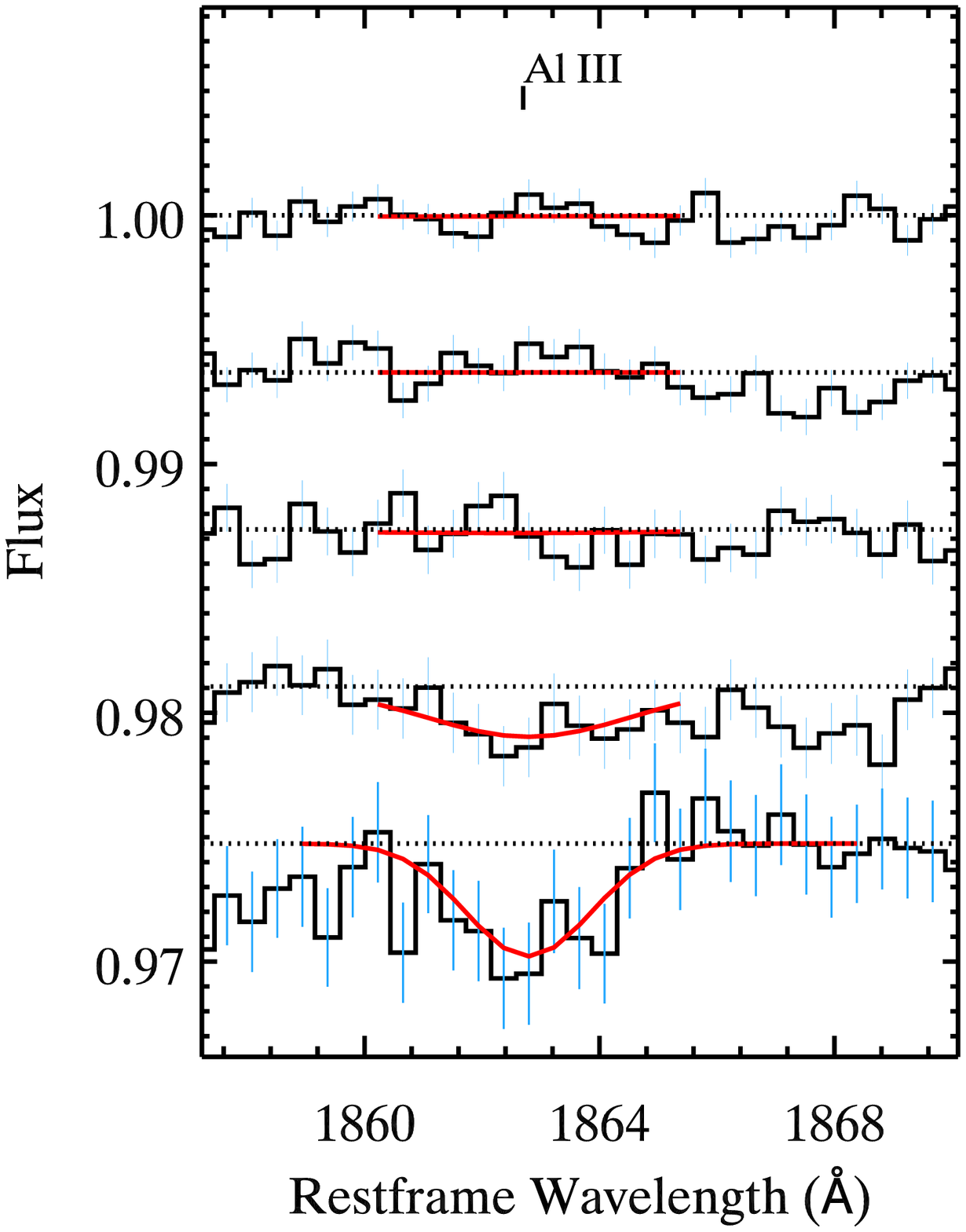}
\includegraphics[angle=0, width=.195\textwidth]{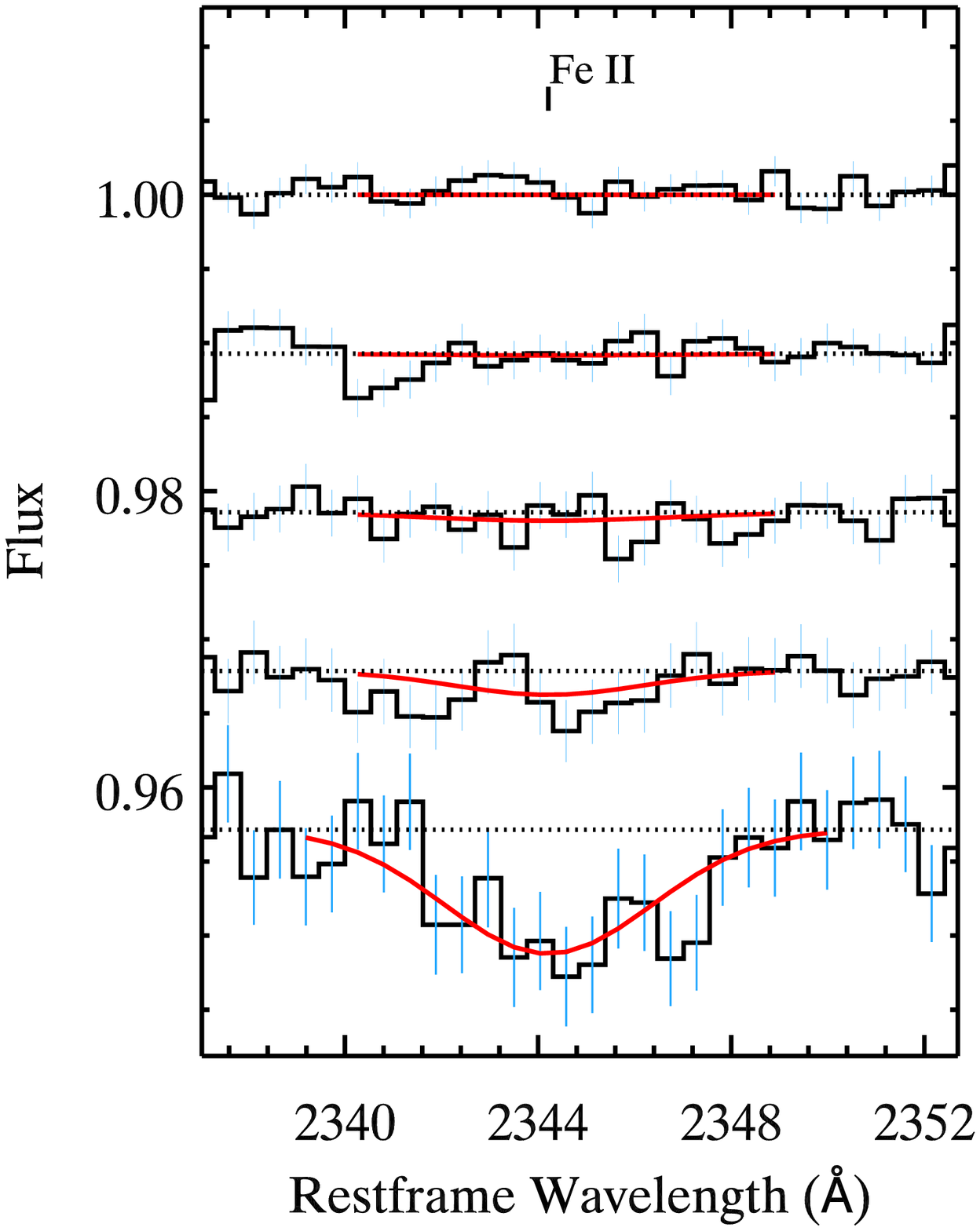}
\includegraphics[angle=0, width=.195\textwidth]{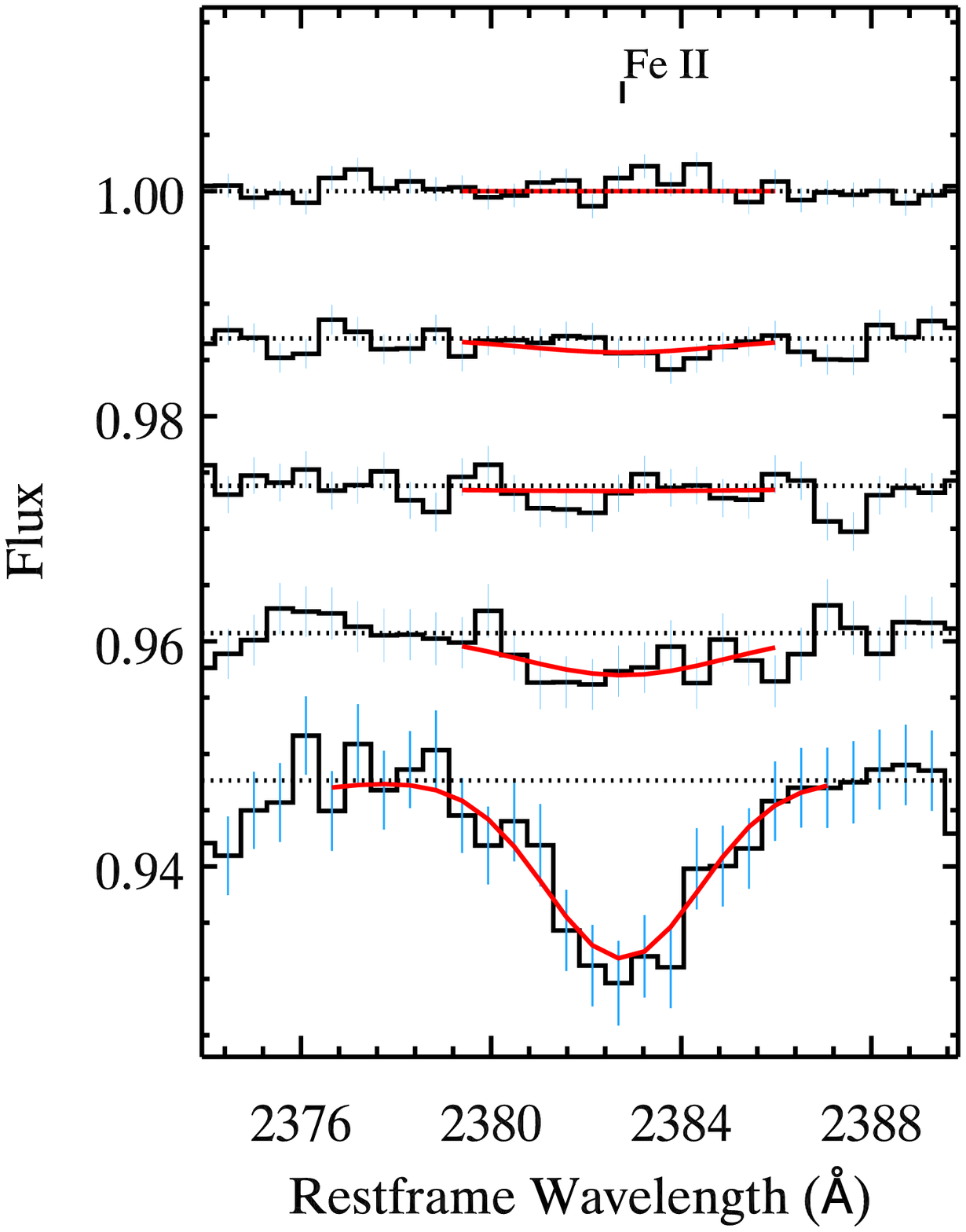}
\includegraphics[angle=0, width=.195\textwidth]{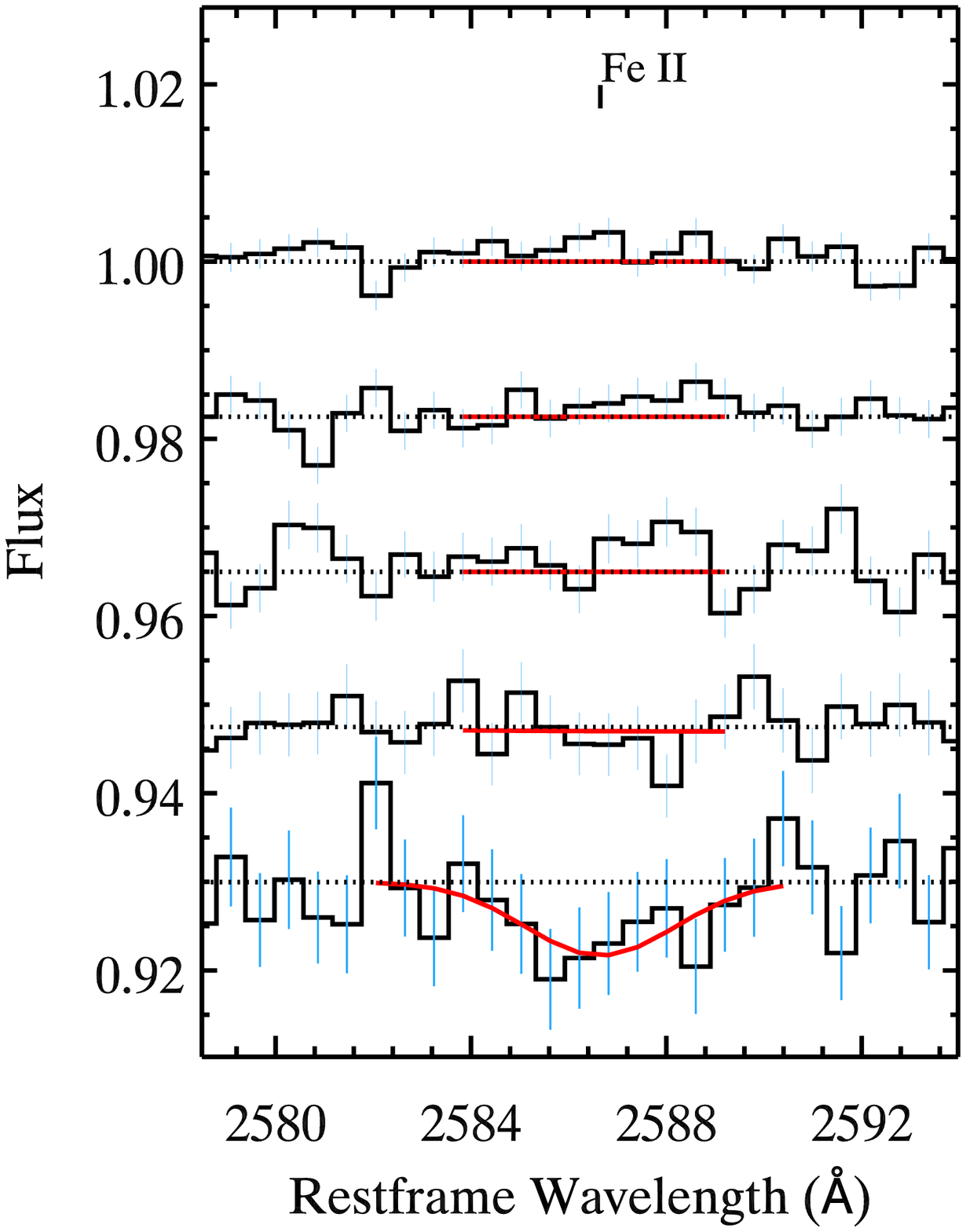}
\includegraphics[angle=0, width=.195\textwidth]{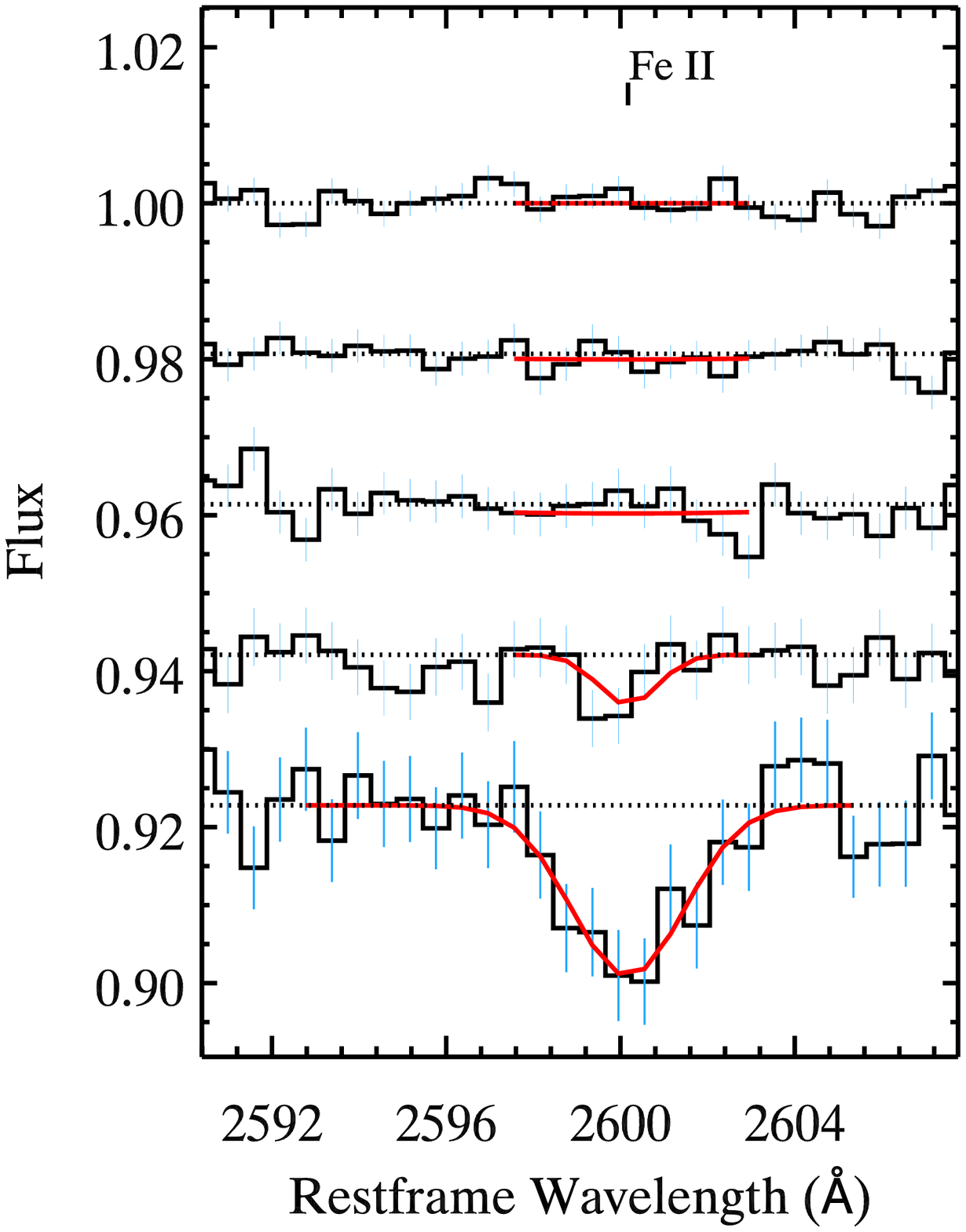}
\includegraphics[angle=0, width=.195\textwidth]{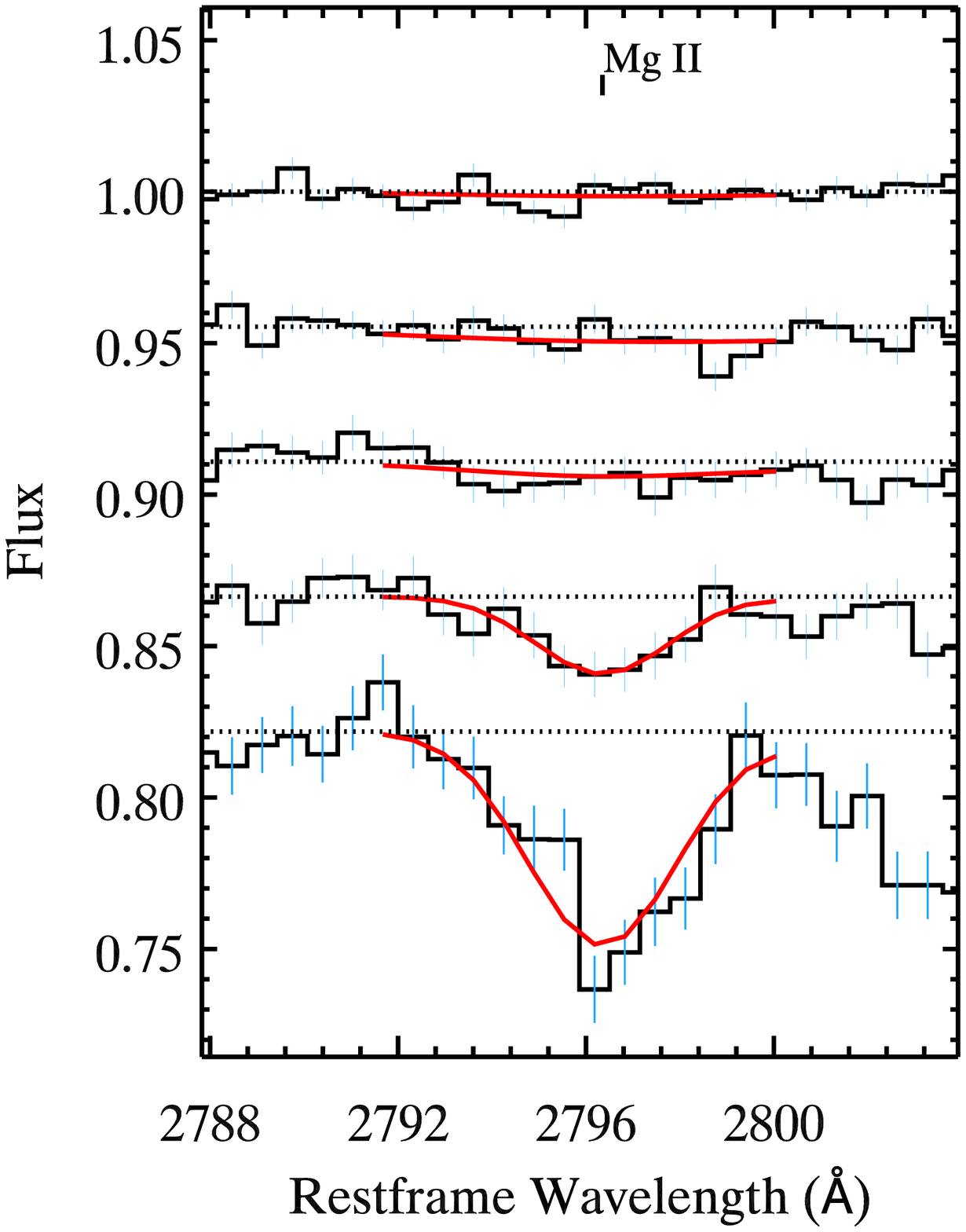}
\includegraphics[angle=0, width=.195\textwidth]{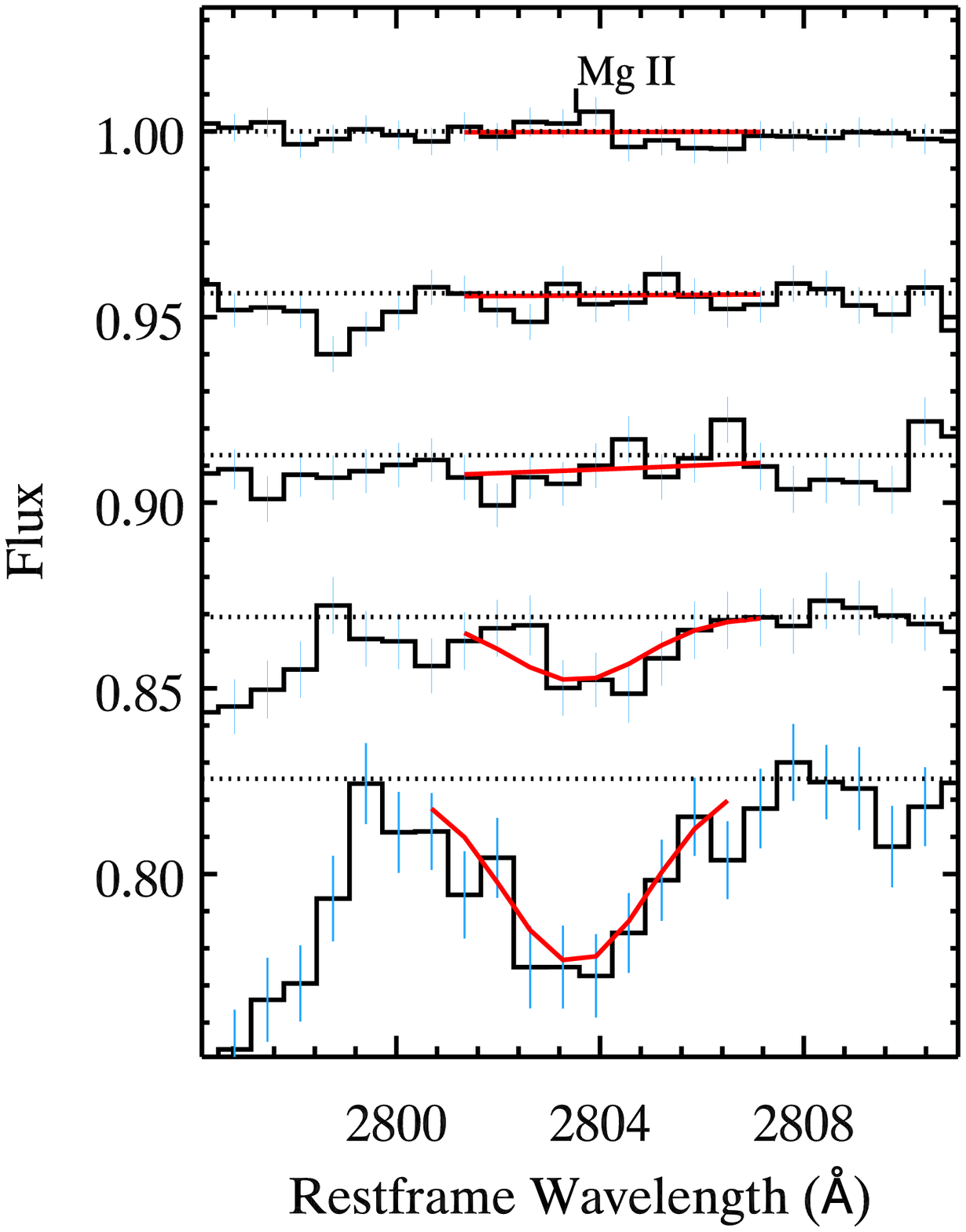}
\end{center}
\caption{Metal line profiles (with error bars shown in blue) and fits for the arithmetic mean composite spectrum. Each panel displays the profile for one line or blend of lines. In each panel the top profile corresponds to the $0.35 \le F<0.45$ case with increasingly strong selection of \lya\ absorption offset downwards. In each case the horizontal dotted line shows $F=1$. The red line shows the fit and the range of this line in wavelength corresponds to the spectral range for which the fit was performed.
}
\label{am_fits}
\end{figure*}

\begin{figure*}
\begin{center}
\includegraphics[angle=0, width=.195\textwidth]{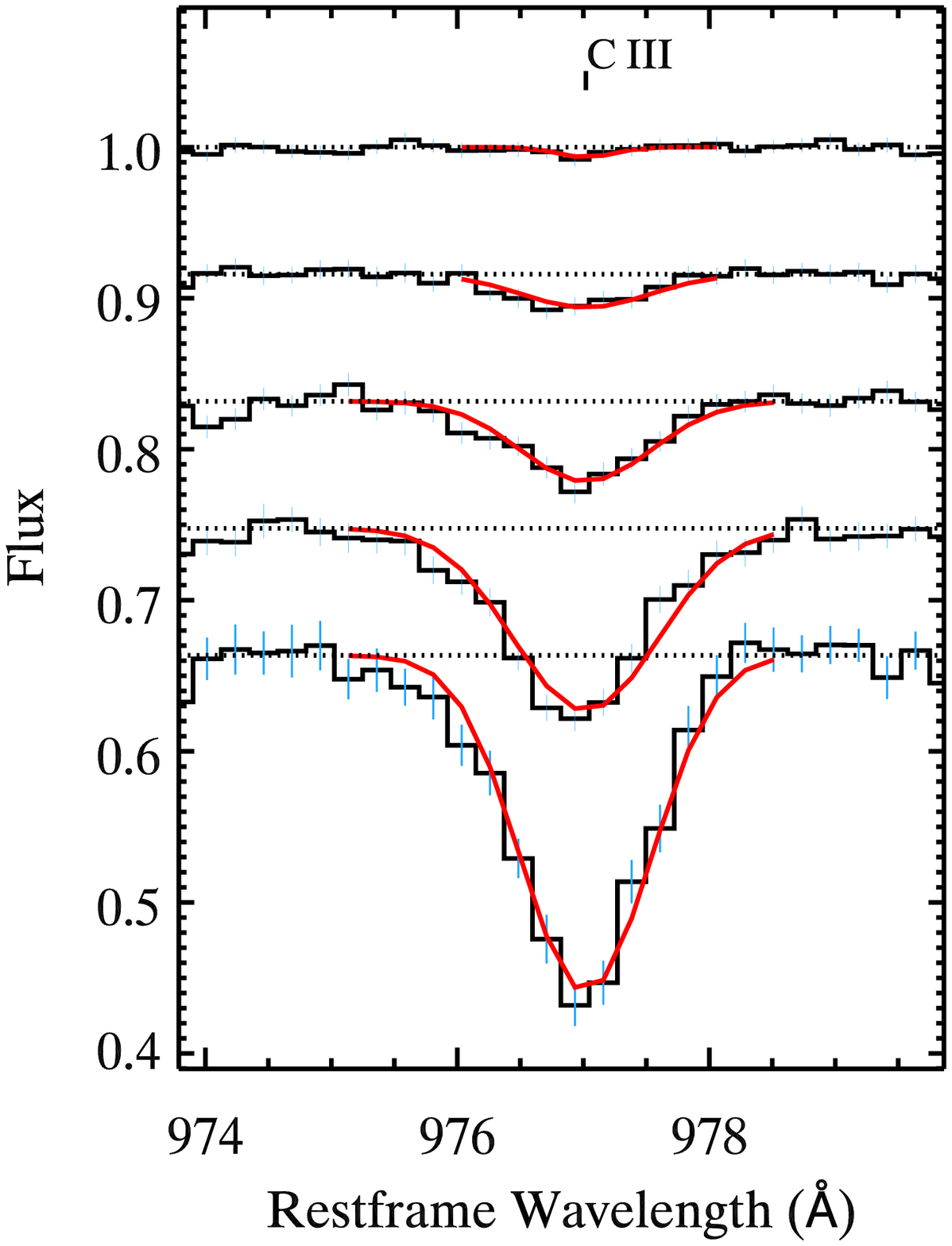}
\includegraphics[angle=0, width=.195\textwidth]{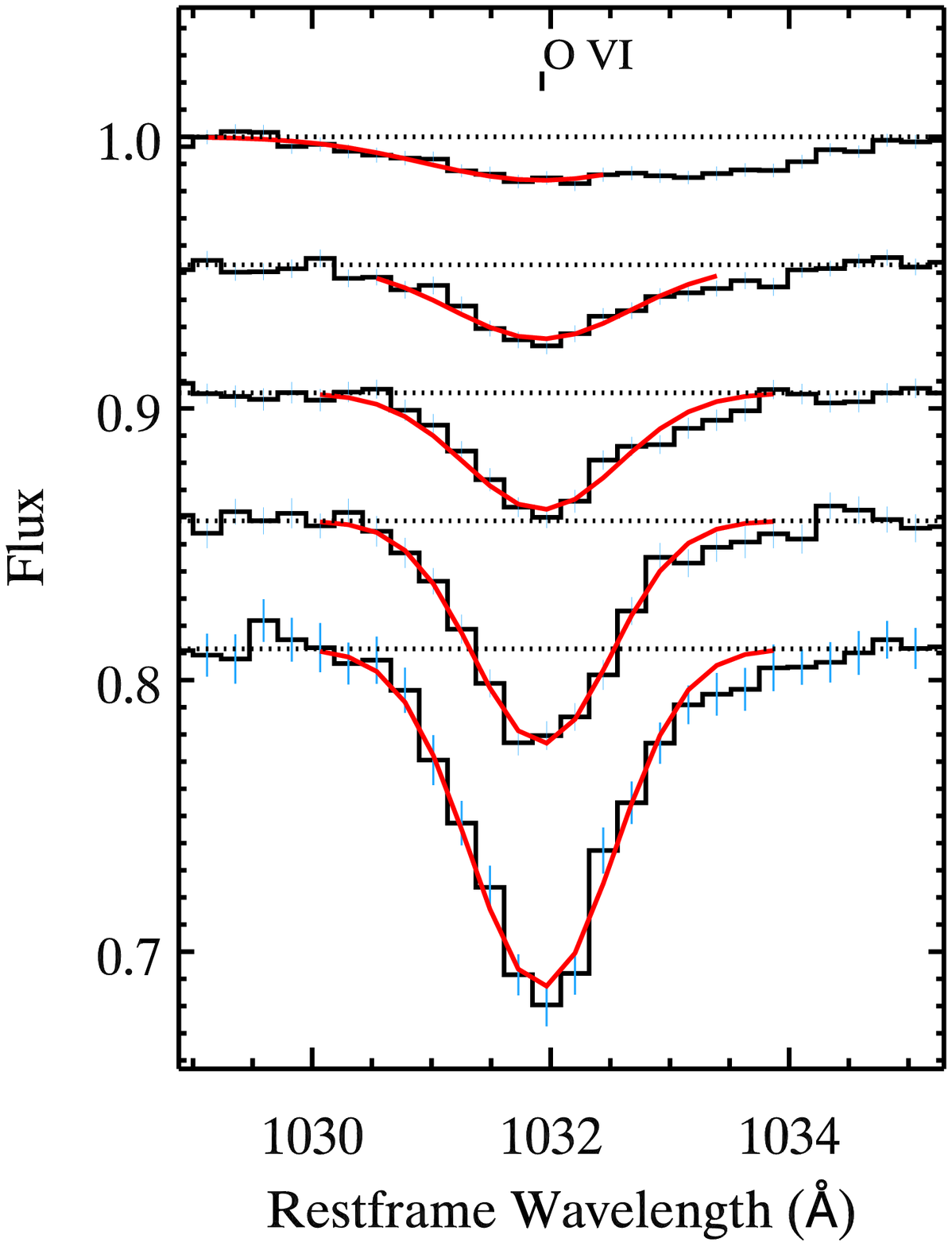}
\includegraphics[angle=0, width=.195\textwidth]{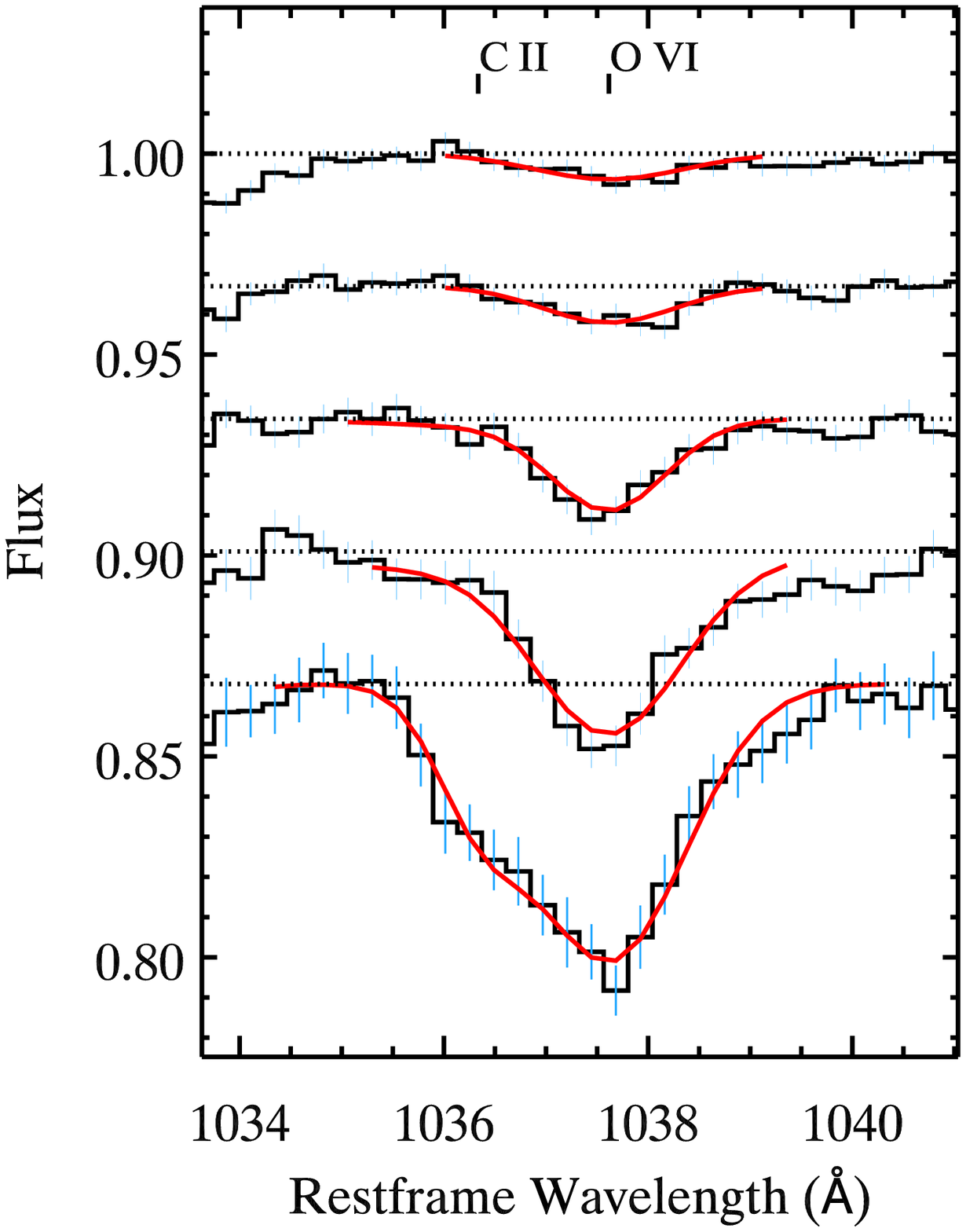}
\includegraphics[angle=0, width=.195\textwidth]{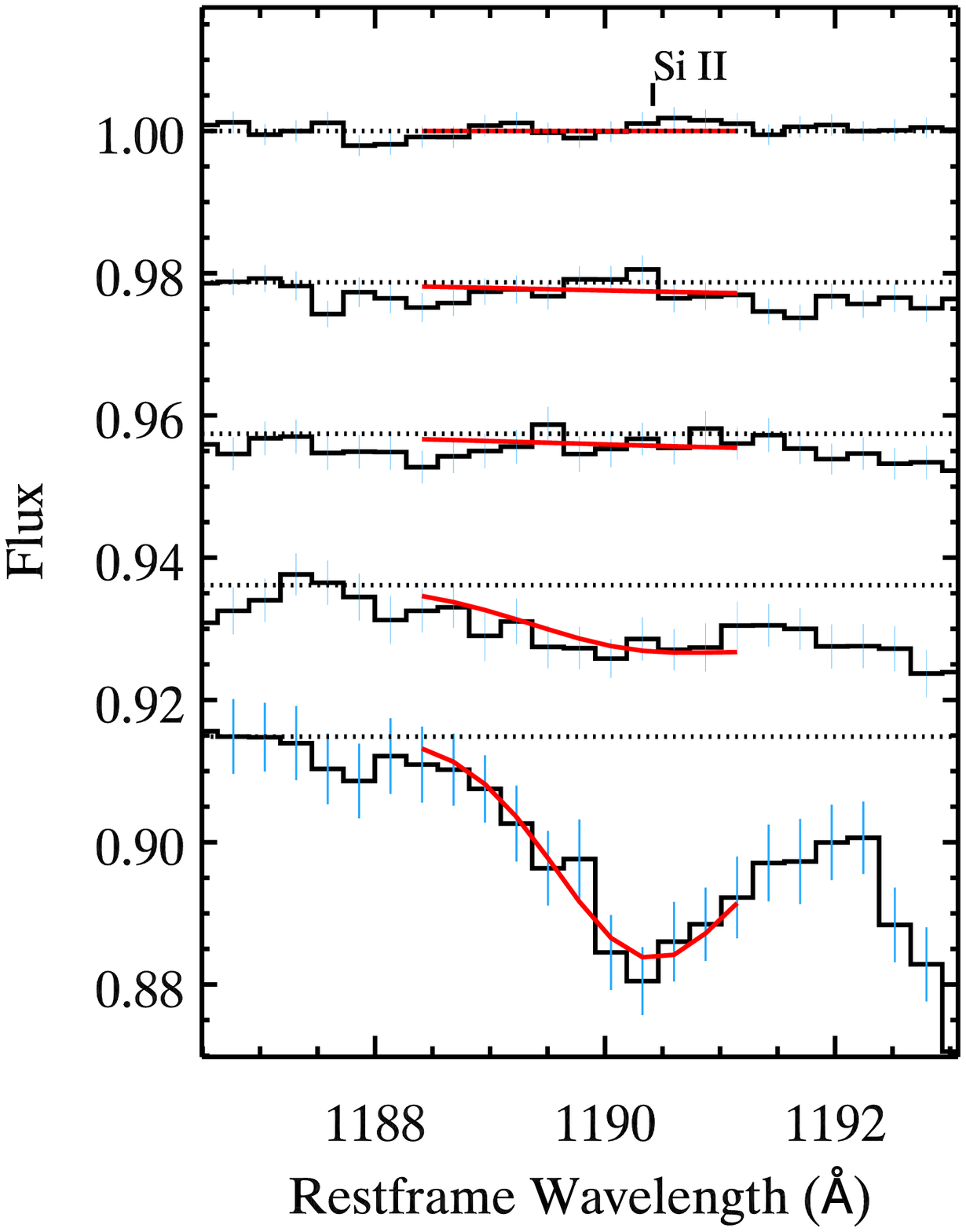}
\includegraphics[angle=0, width=.195\textwidth]{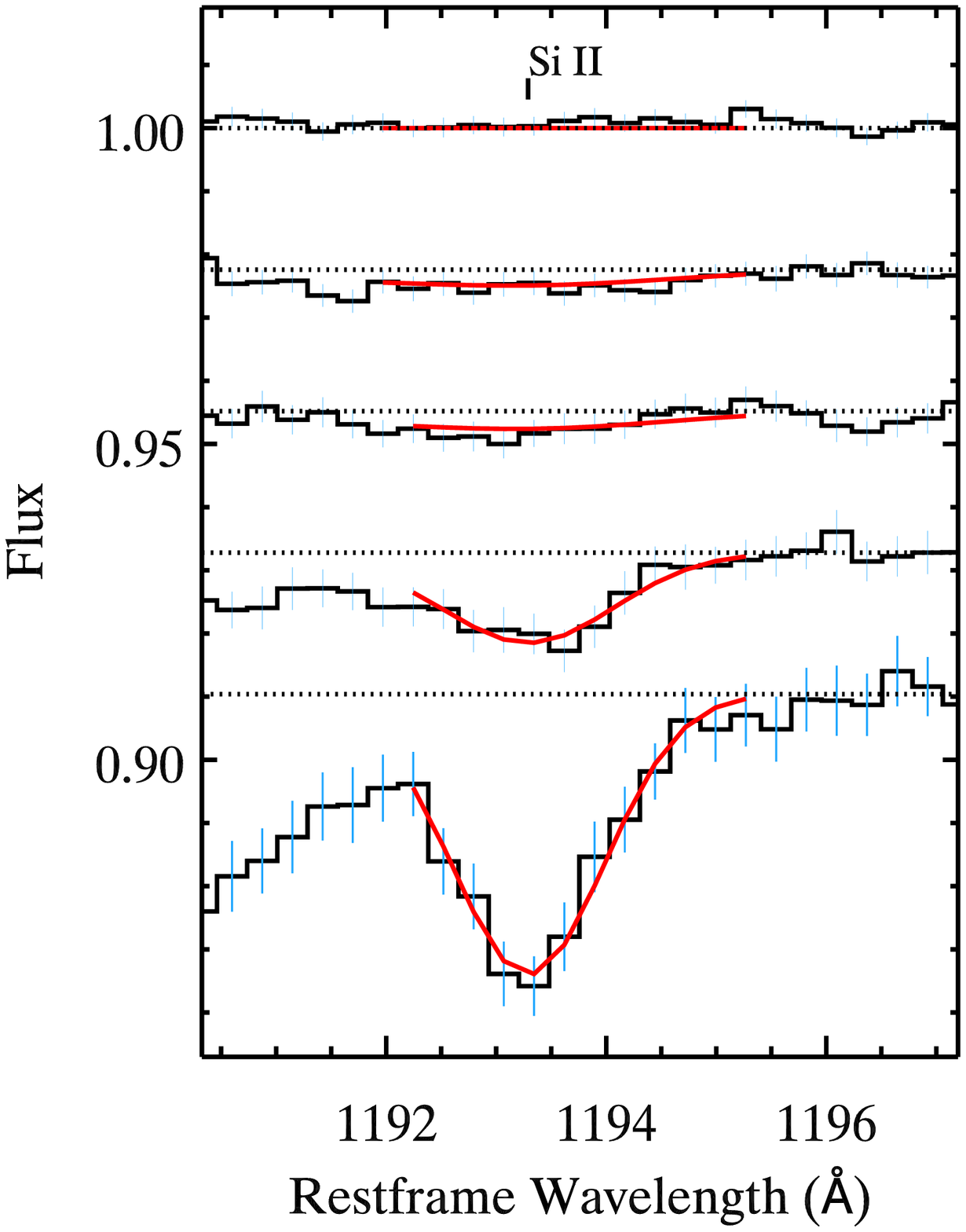}
\includegraphics[angle=0, width=.195\textwidth]{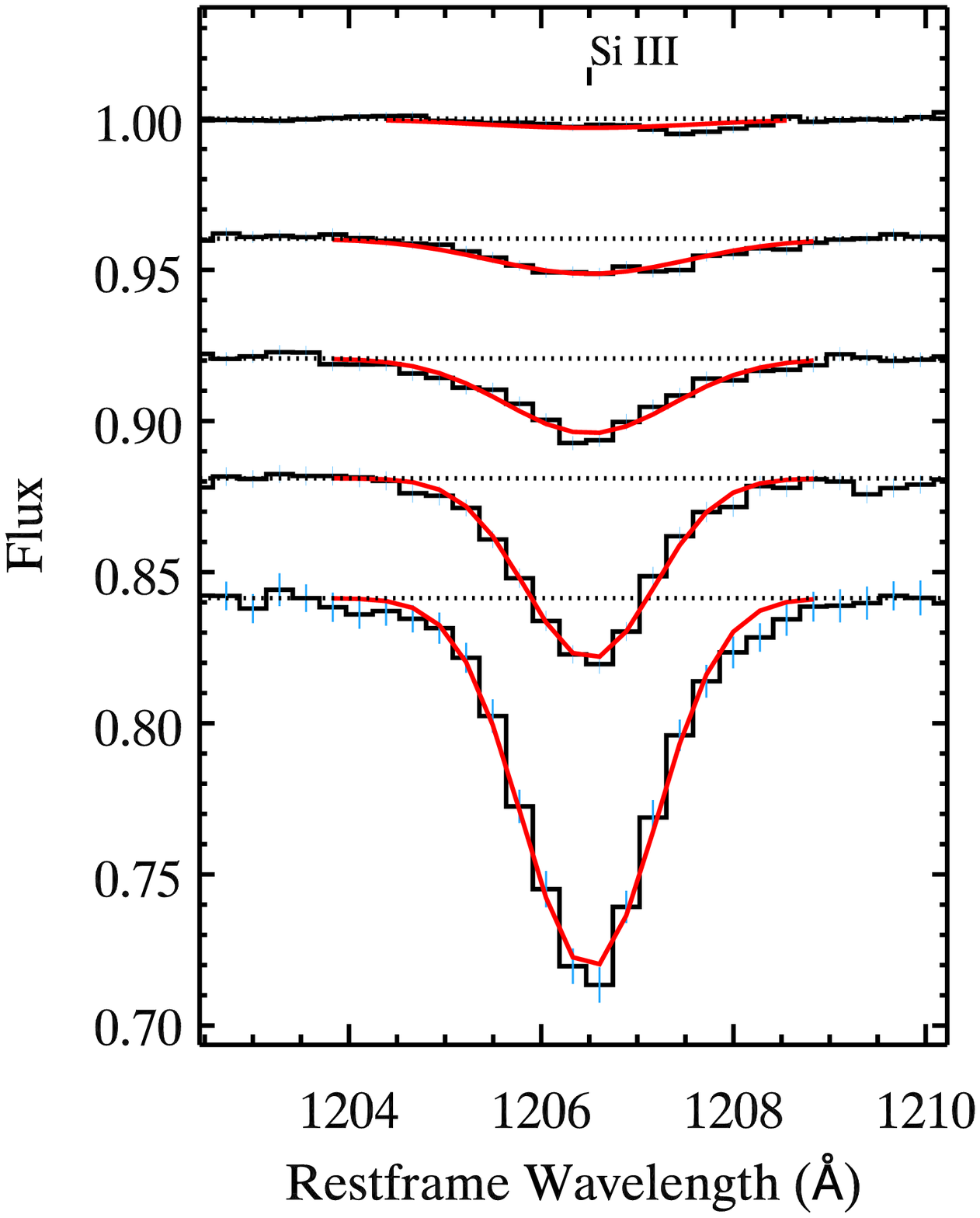}
\includegraphics[angle=0, width=.195\textwidth]{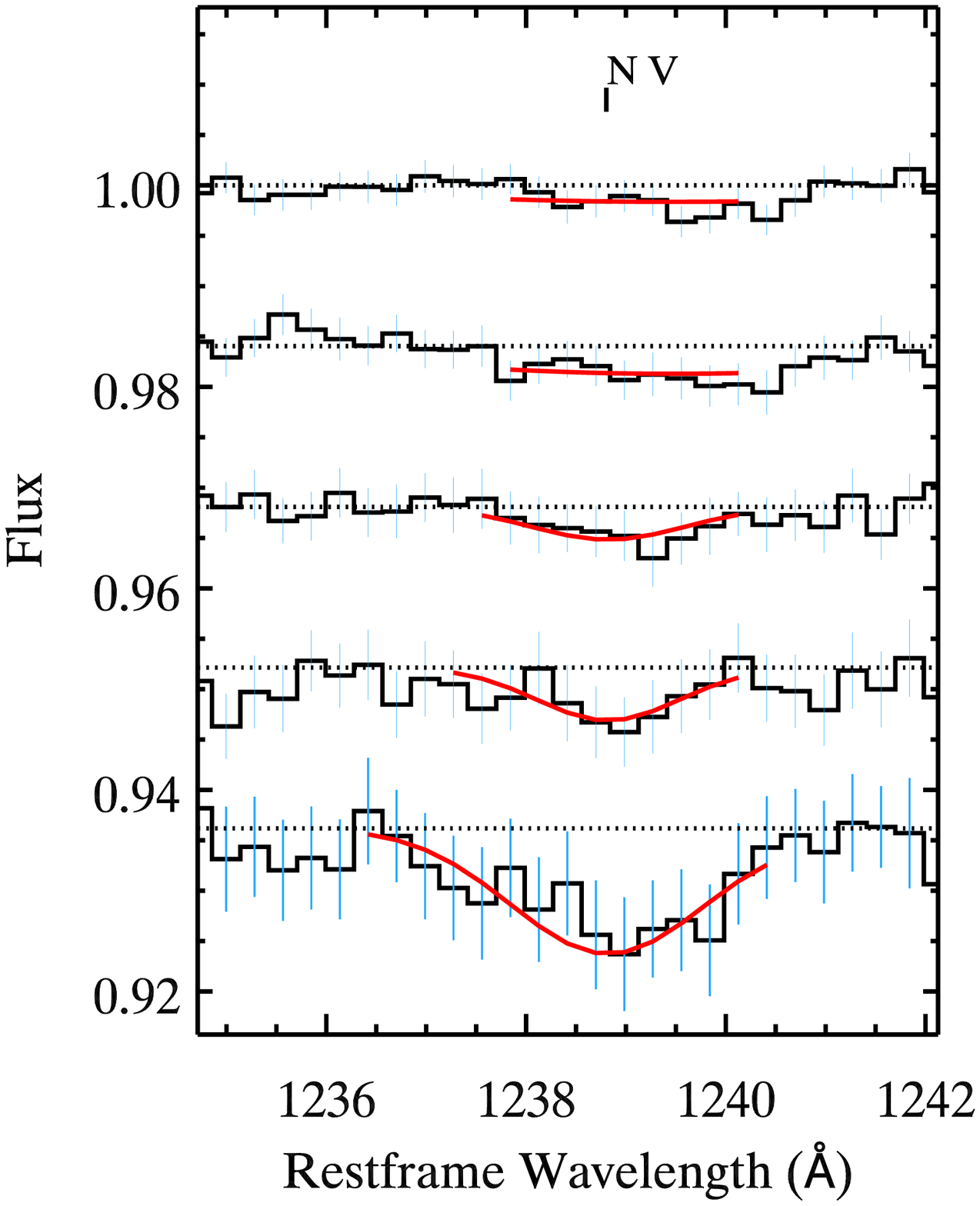}
\includegraphics[angle=0, width=.195\textwidth]{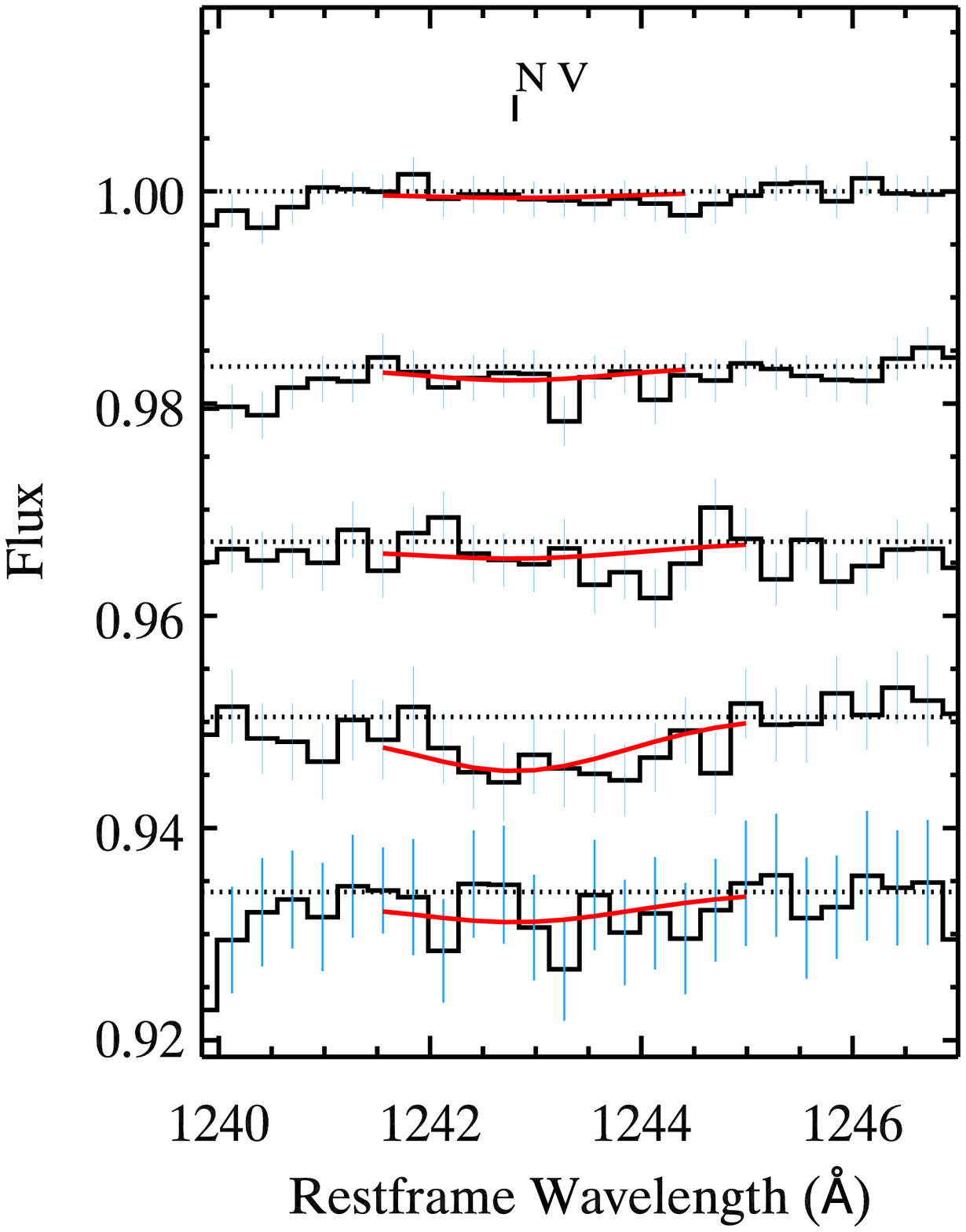}
\includegraphics[angle=0, width=.195\textwidth]{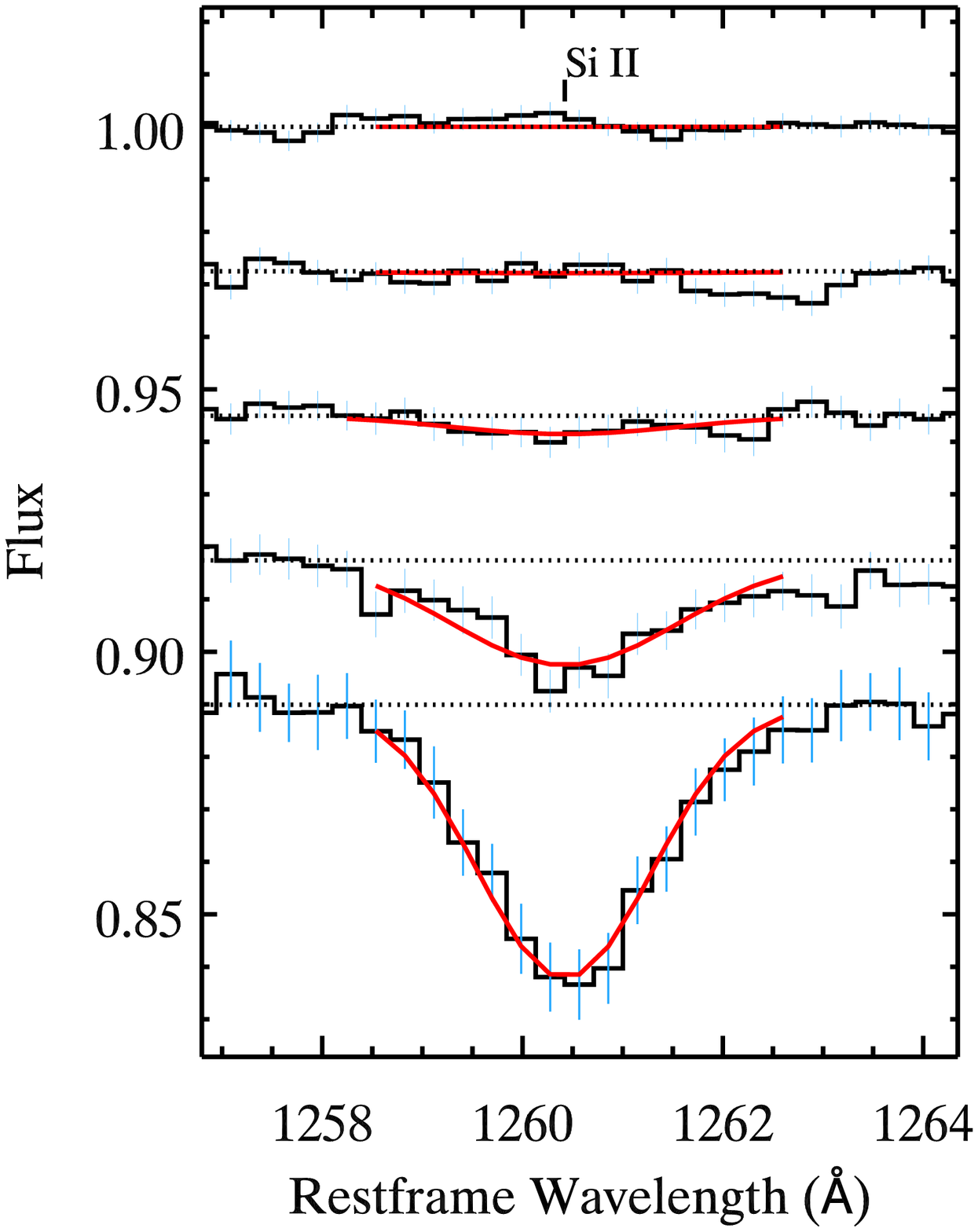}
\includegraphics[angle=0, width=.195\textwidth]{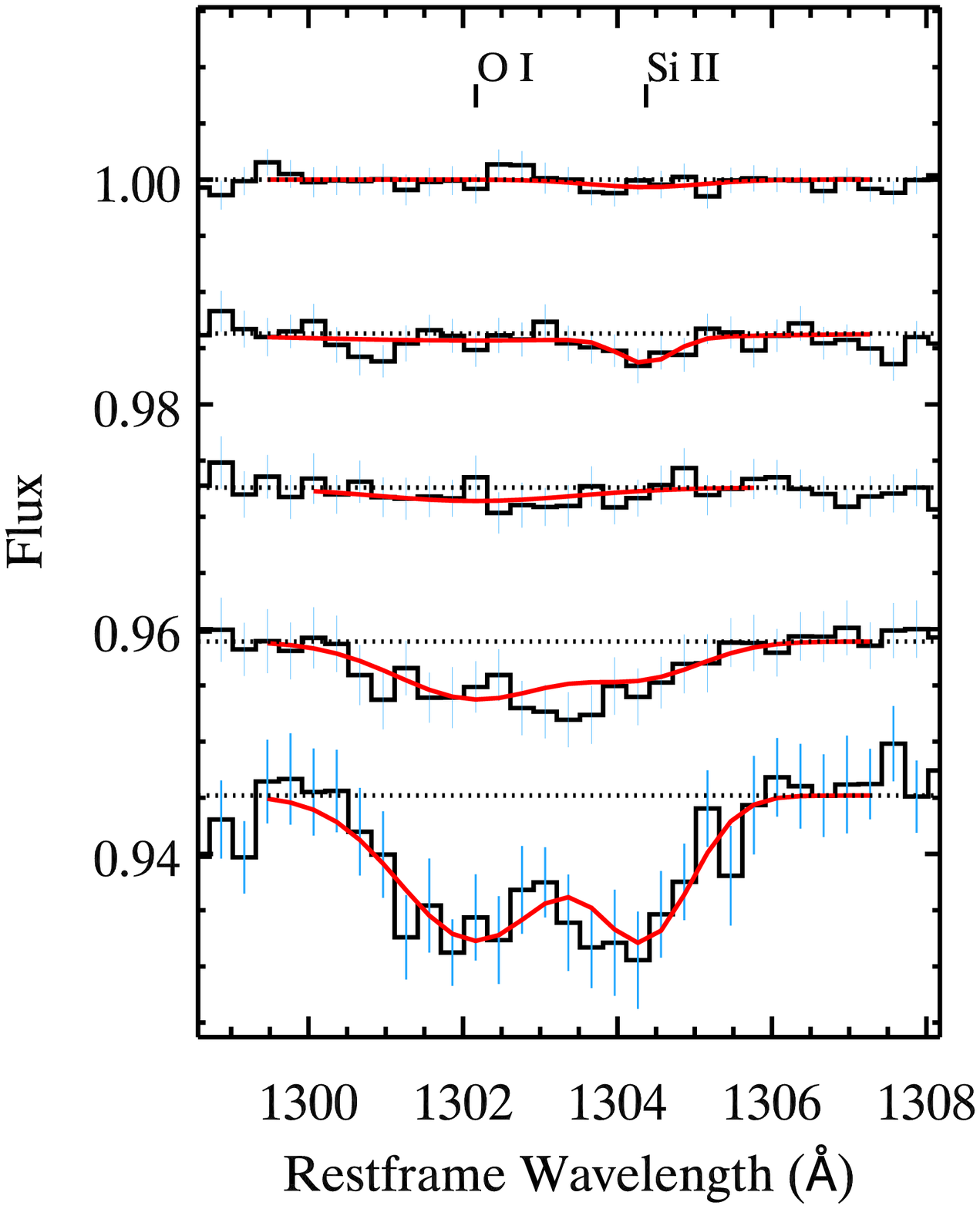}
\includegraphics[angle=0, width=.195\textwidth]{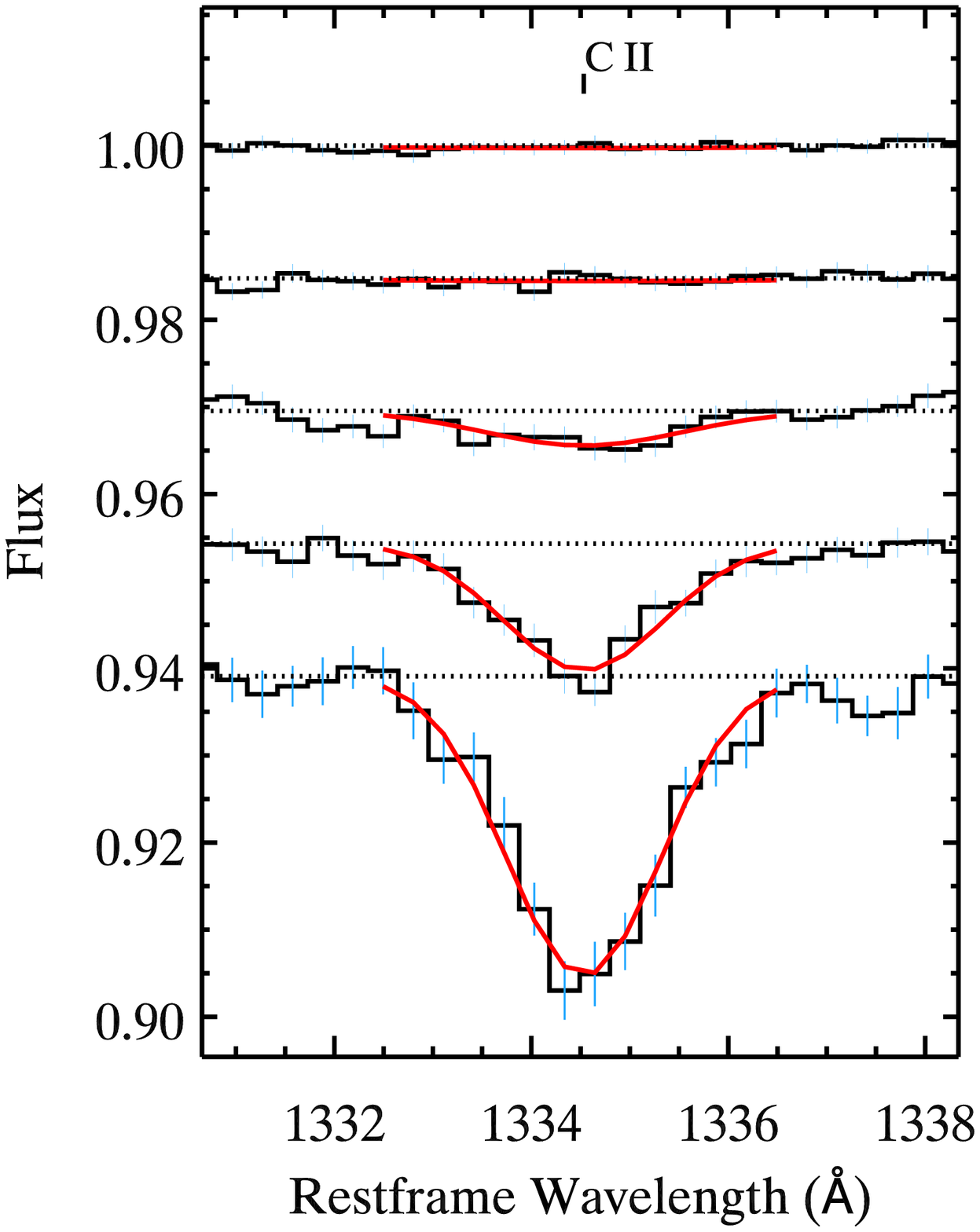}
\includegraphics[angle=0, width=.195\textwidth]{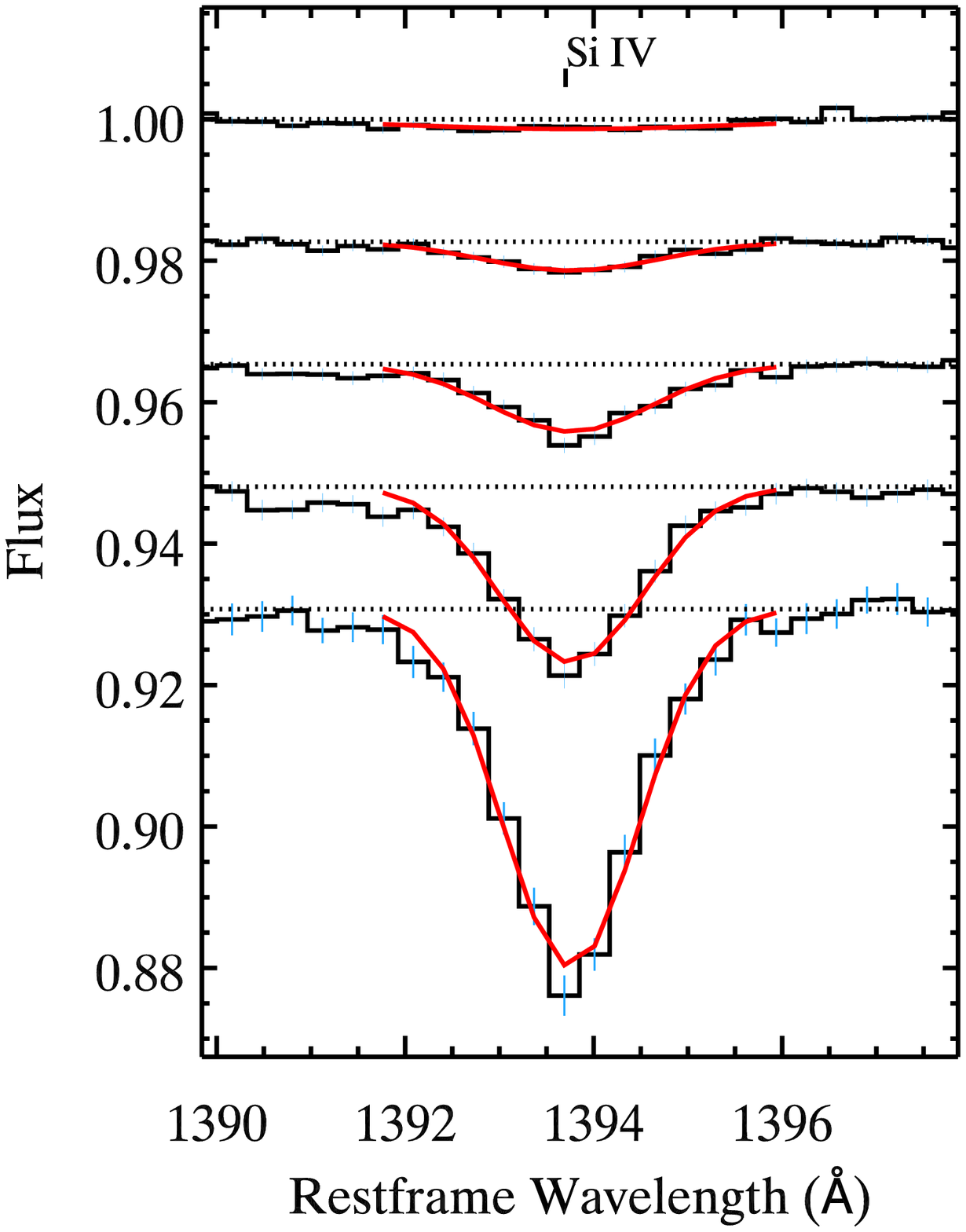}
\includegraphics[angle=0, width=.195\textwidth]{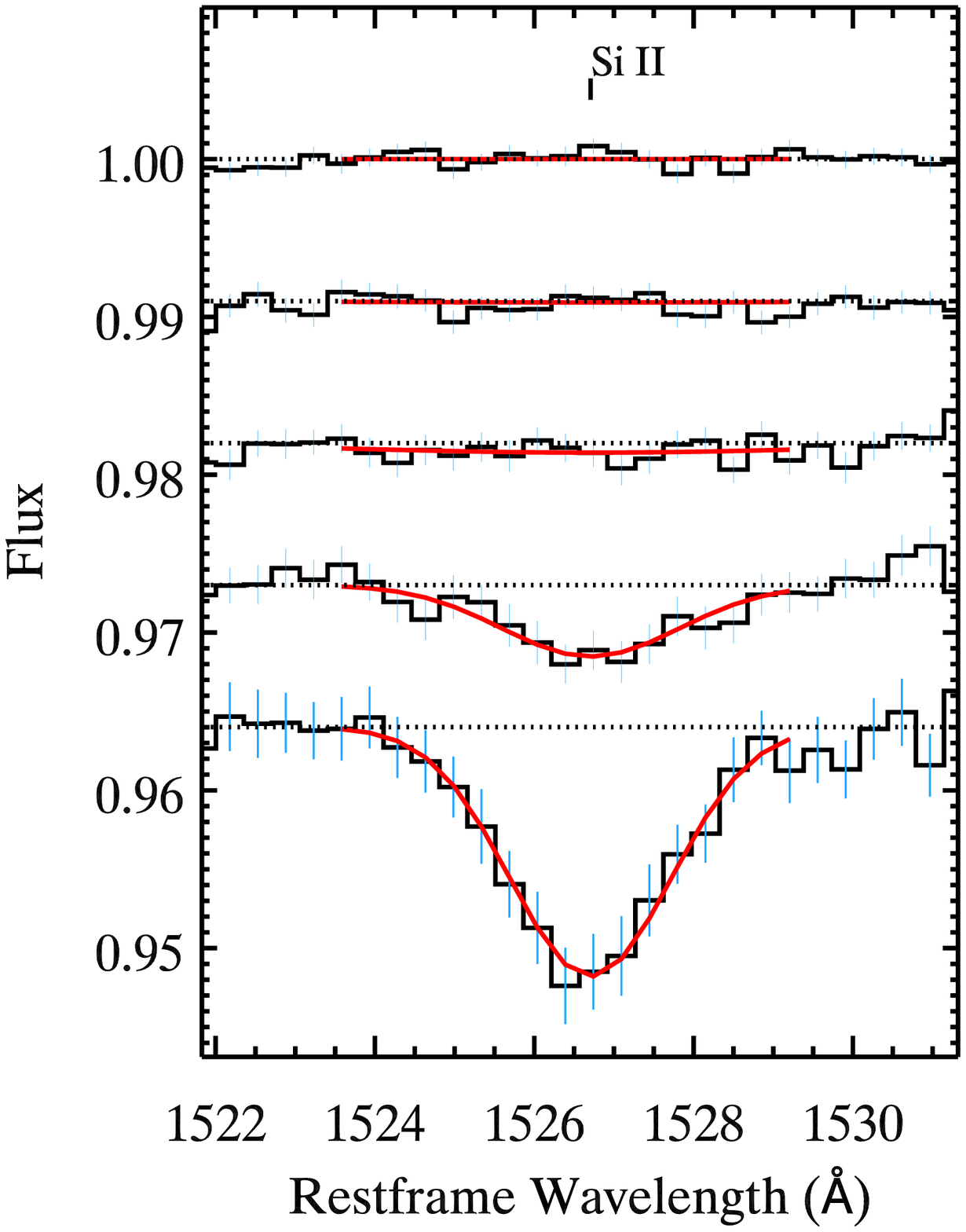}
\includegraphics[angle=0, width=.195\textwidth]{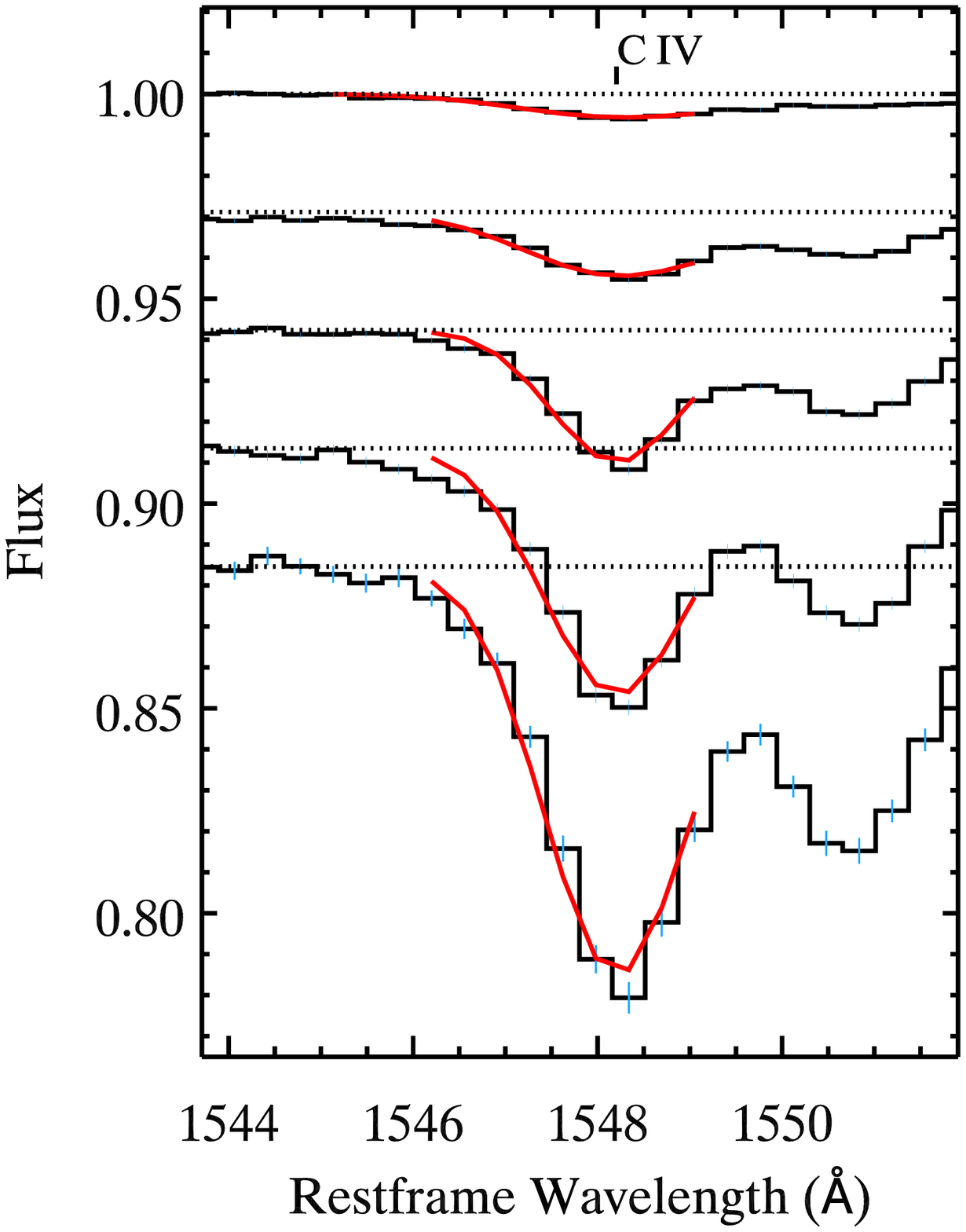}
\includegraphics[angle=0, width=.195\textwidth]{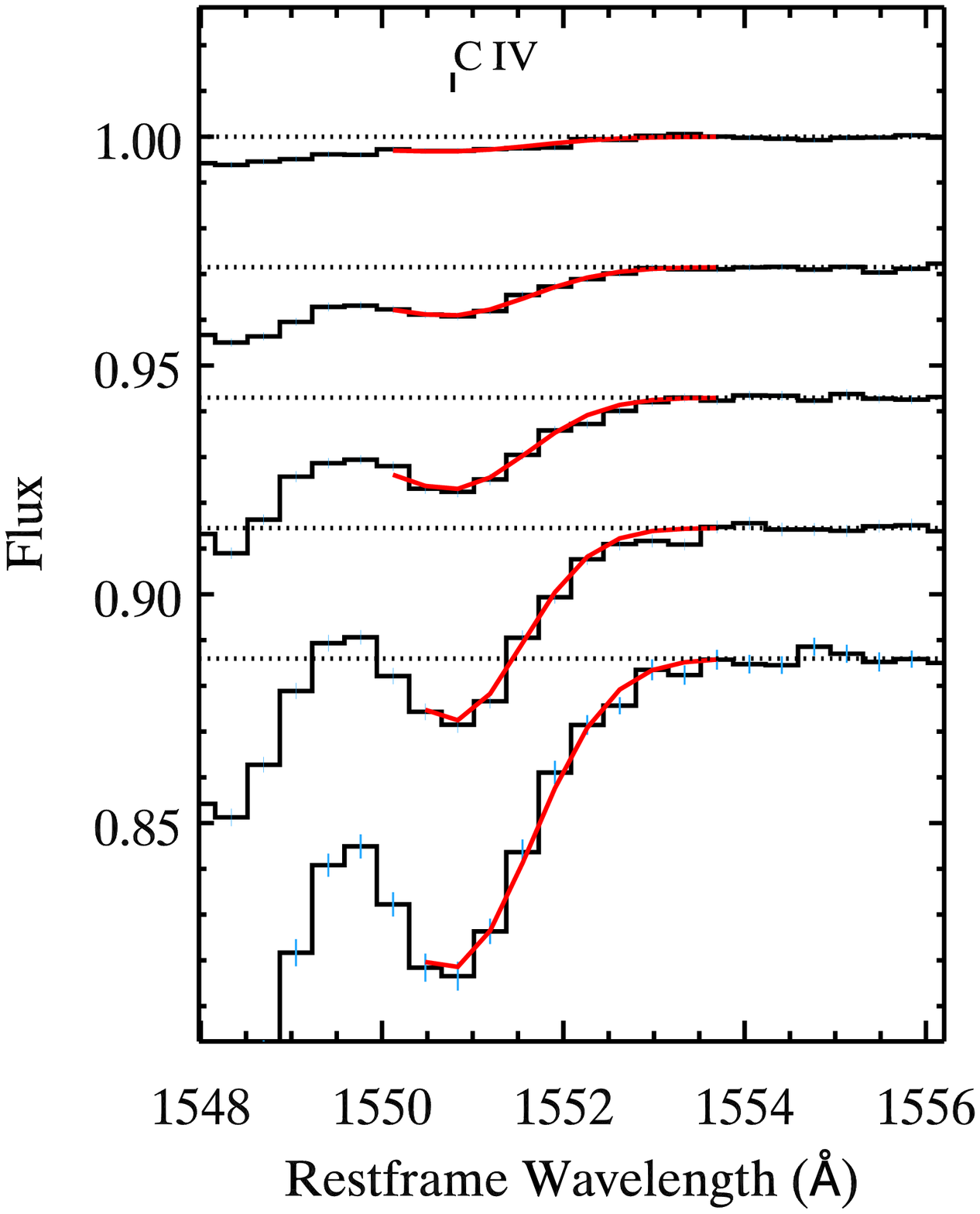}
\includegraphics[angle=0, width=.195\textwidth]{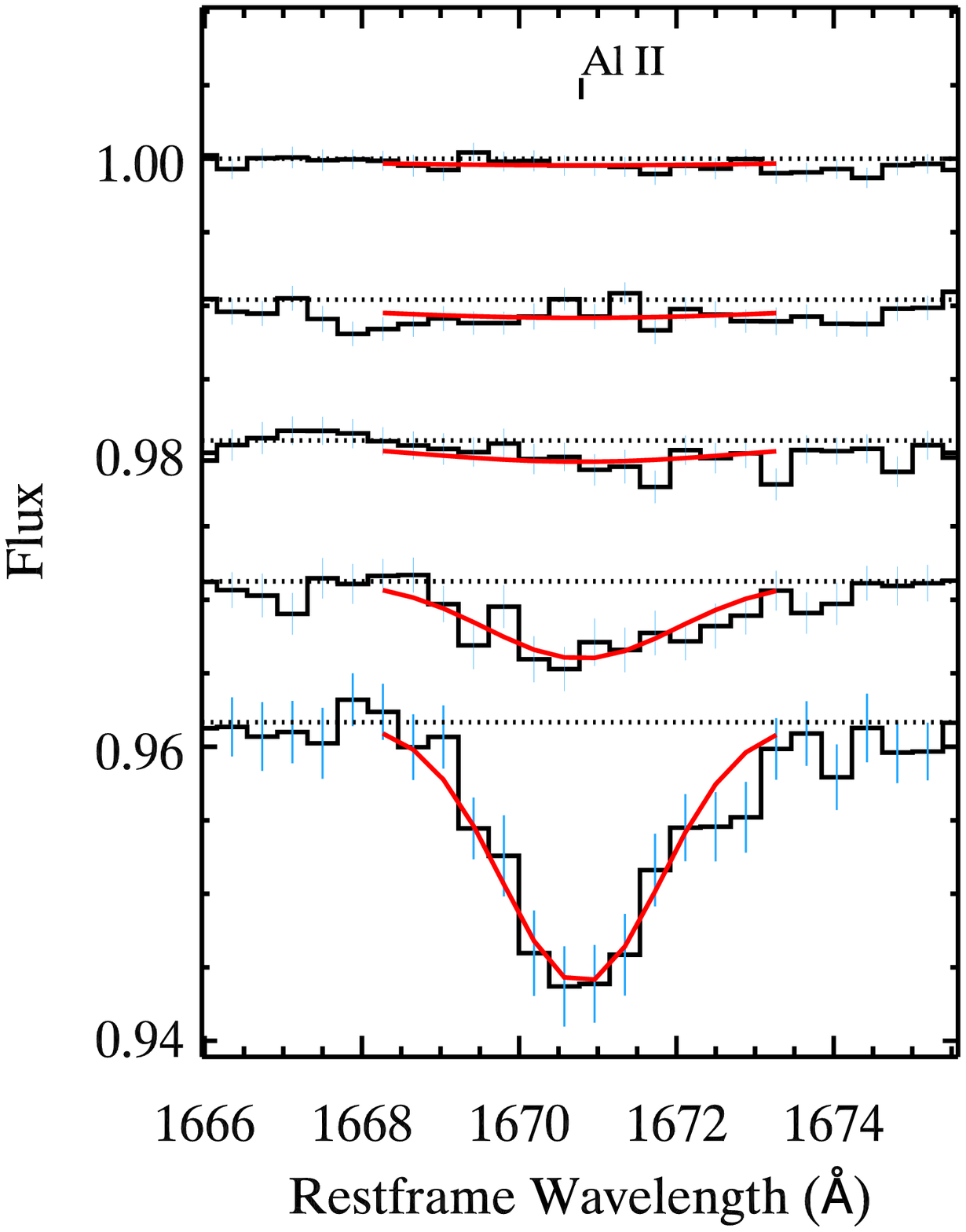}
\includegraphics[angle=0, width=.195\textwidth]{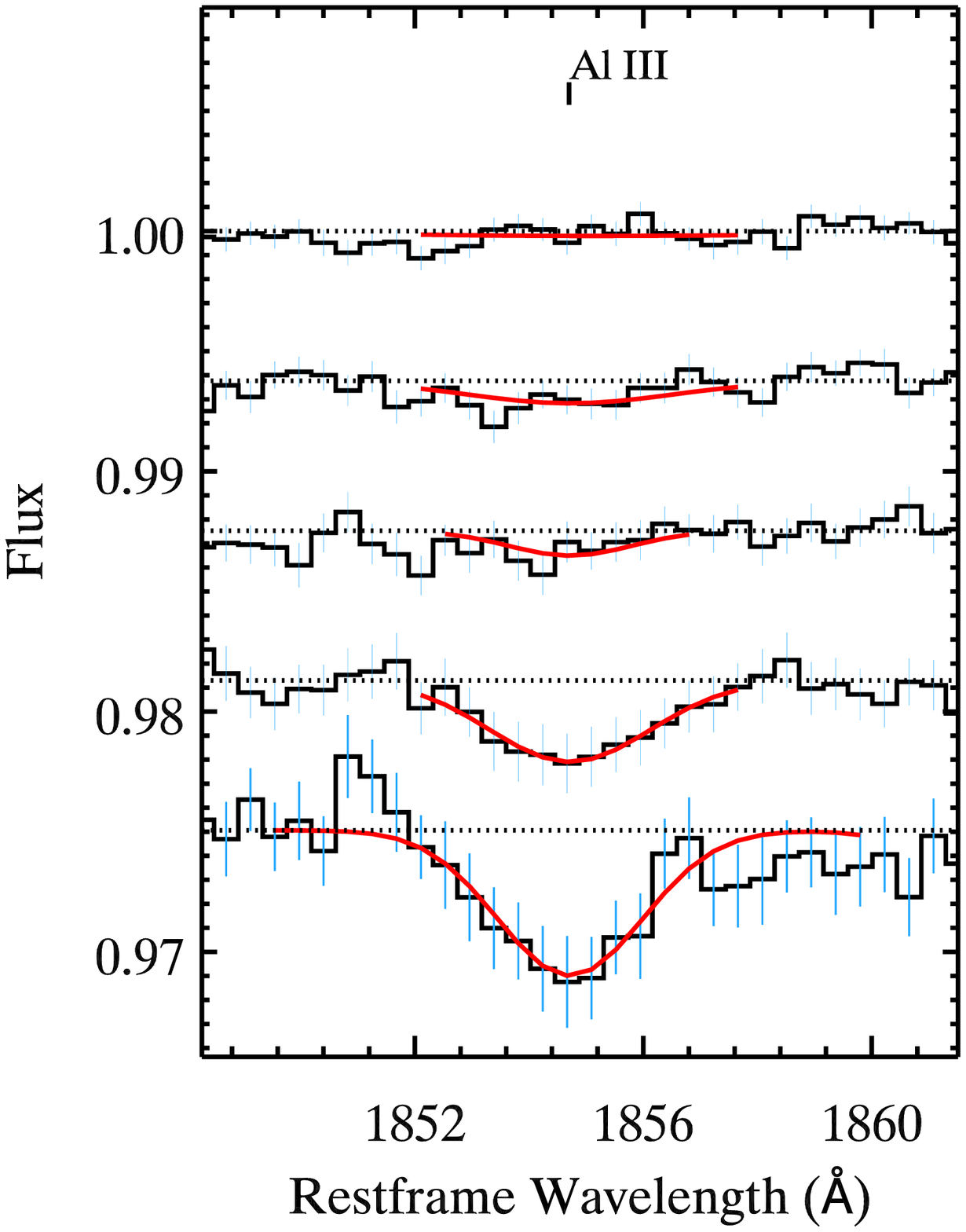}
\includegraphics[angle=0, width=.195\textwidth]{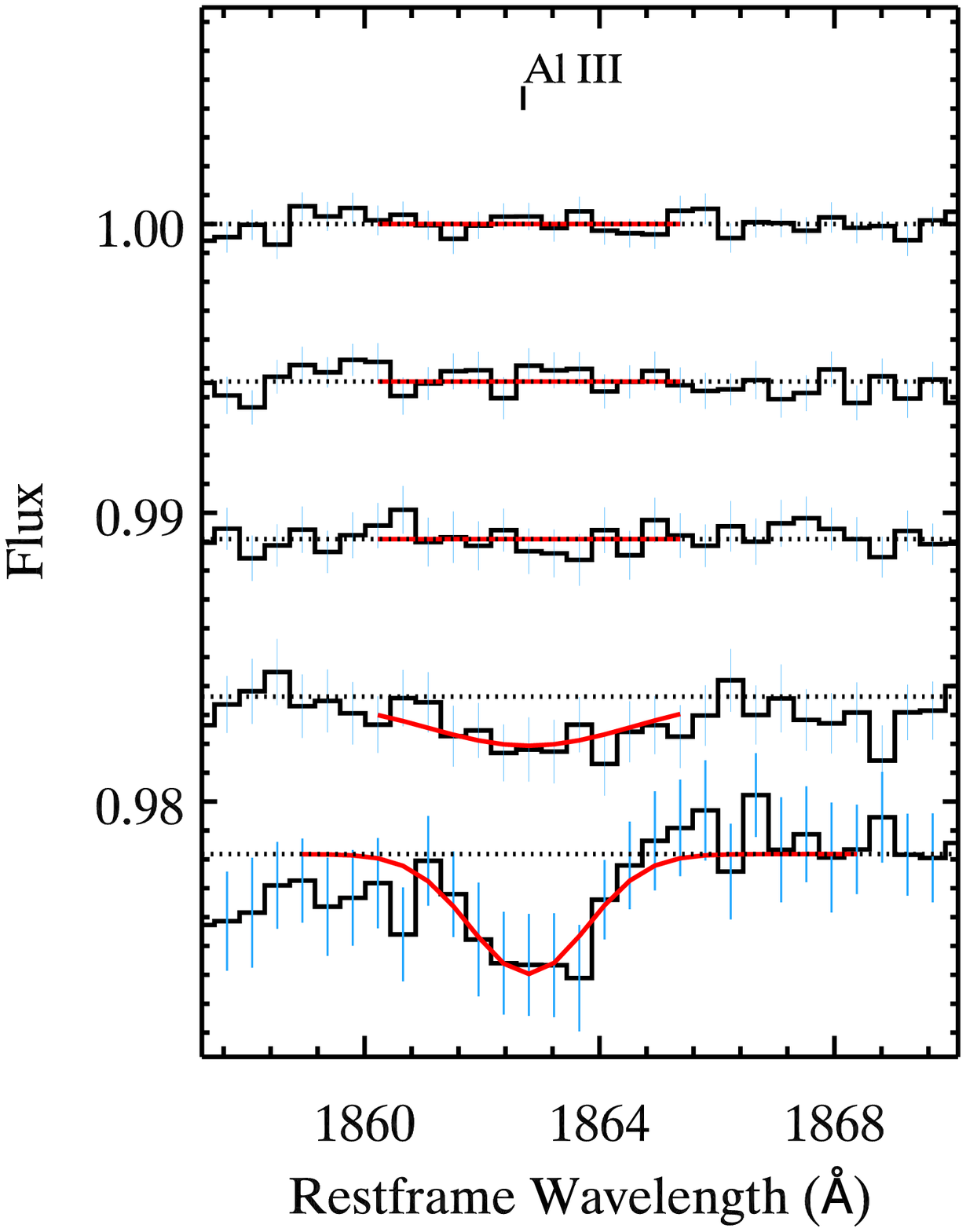}
\includegraphics[angle=0, width=.195\textwidth]{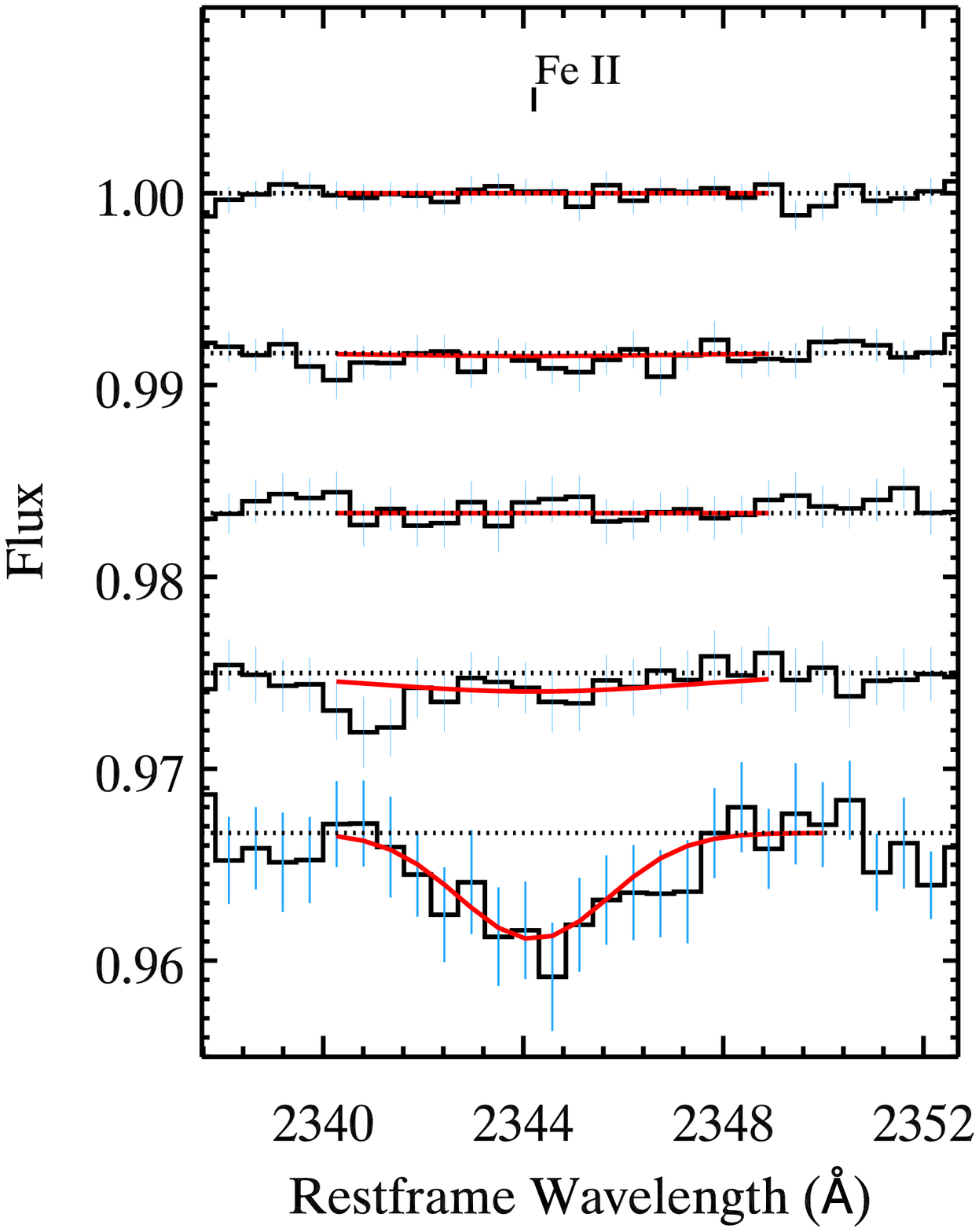}
\includegraphics[angle=0, width=.195\textwidth]{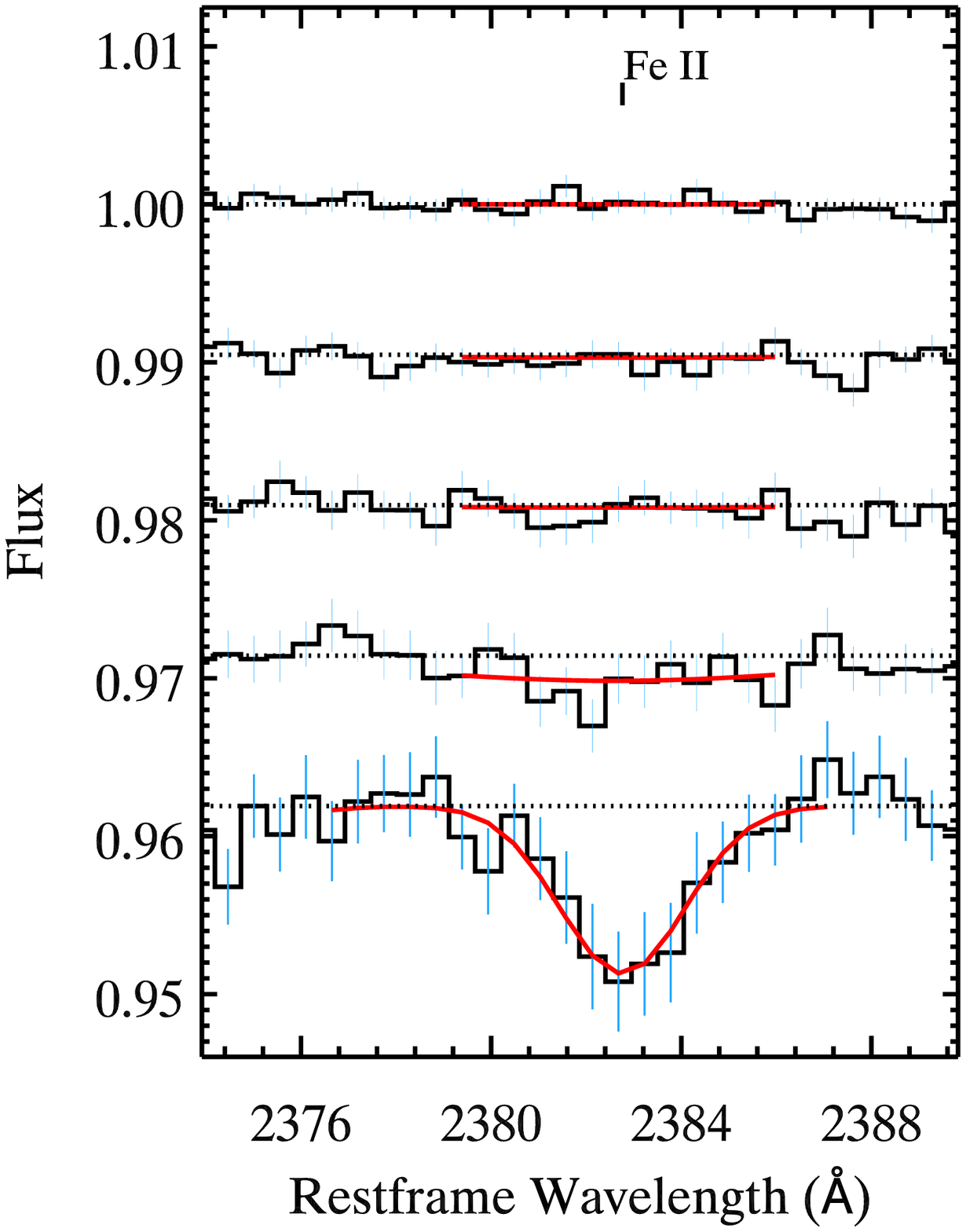}
\includegraphics[angle=0, width=.195\textwidth]{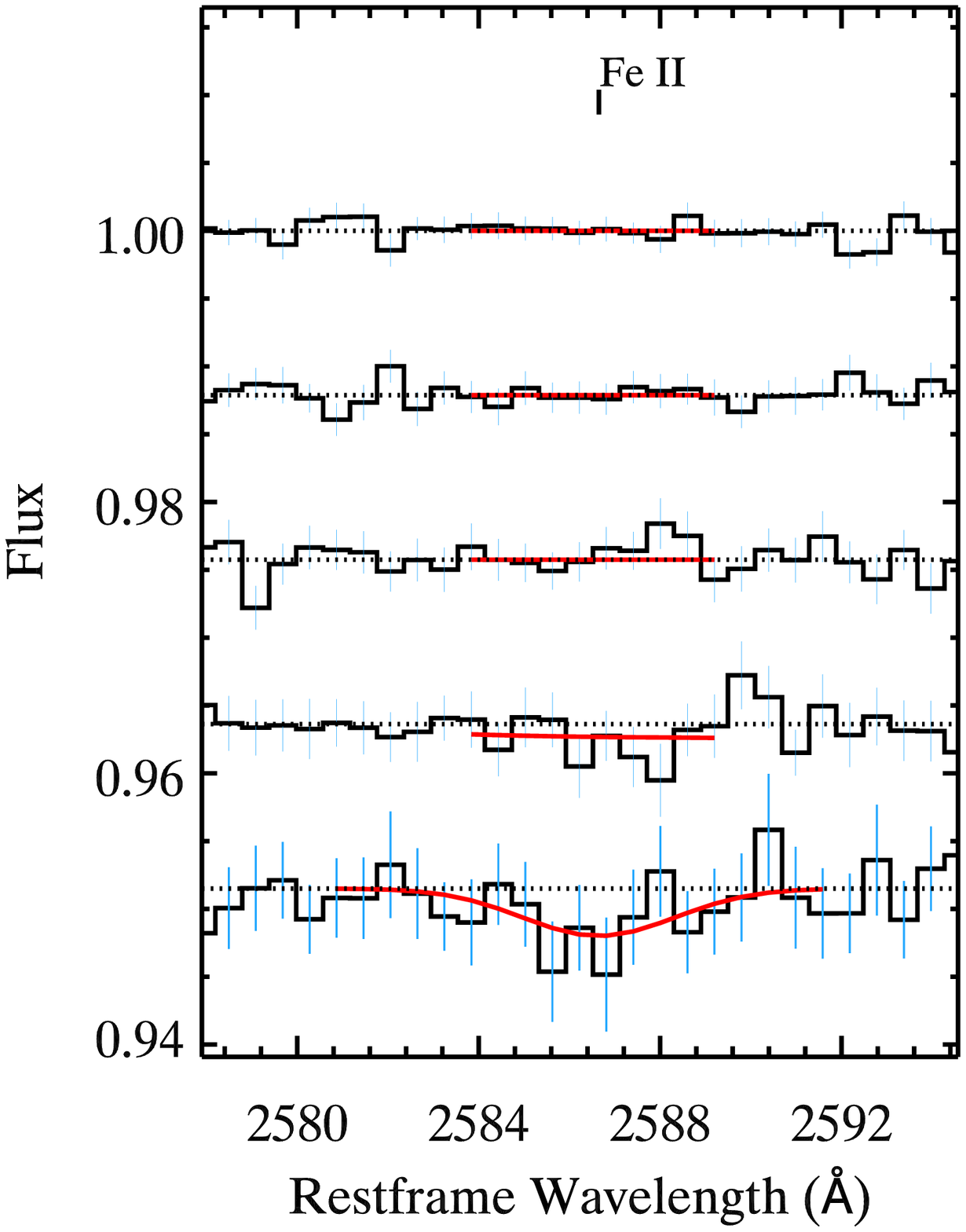}
\includegraphics[angle=0, width=.195\textwidth]{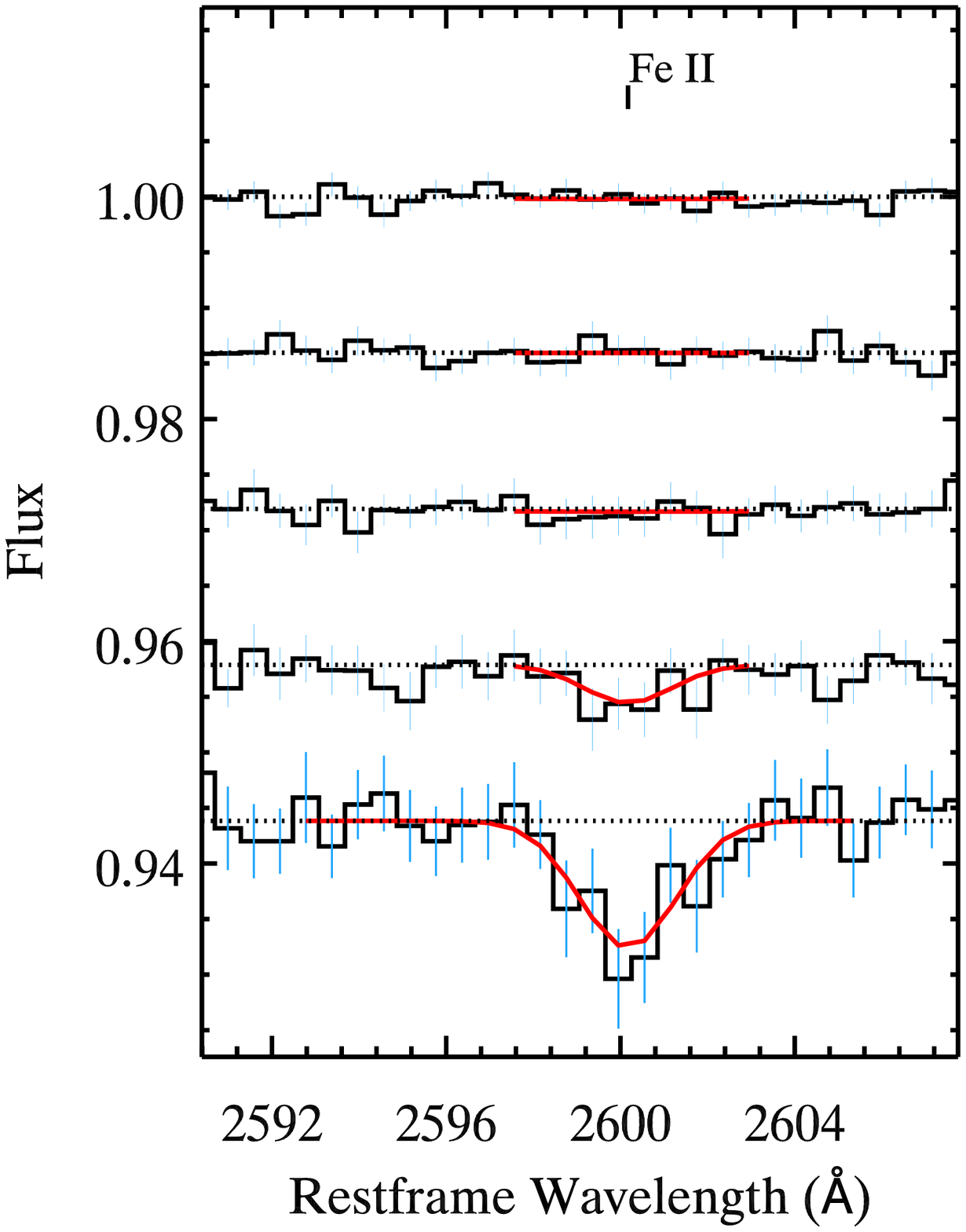}
\includegraphics[angle=0, width=.195\textwidth]{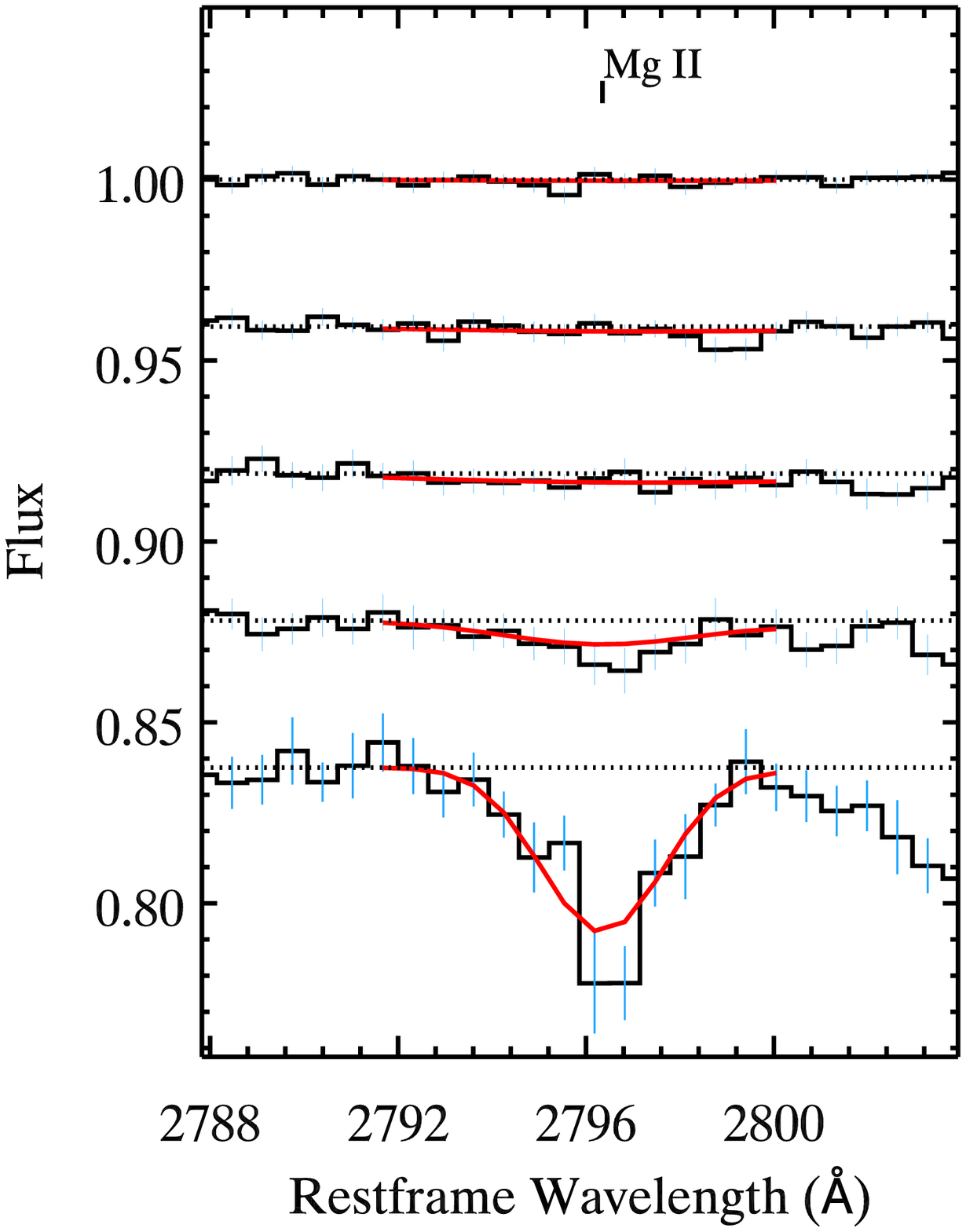}
\includegraphics[angle=0, width=.195\textwidth]{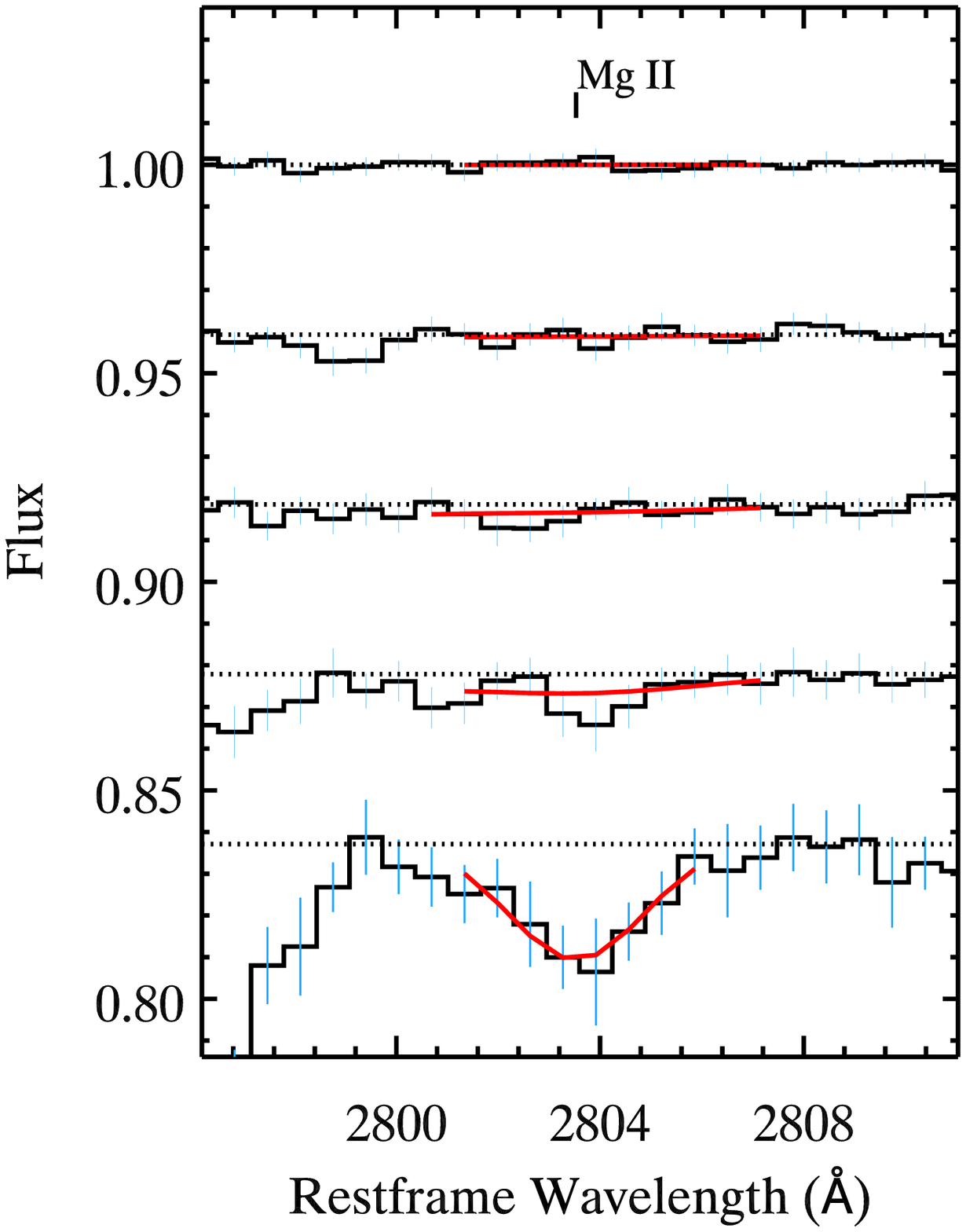}
\end{center}
\caption{As for \ref{am_fits} with profiles and fits for the median composite spectrum.  }
\label{med_fits}
\end{figure*}

\begin{table*}
 \caption{Metal lines measured from full line profiles in the median composite rest-frame spectrum of \lya\ 
absorbers. A joint fit was performed to $N_f$ and $b_f$. Values are given in units of $\cm^{-2}$ and $\kms$, respectively.}
 \label{tab_med_fullline}
 \begin{tabular}{@{}l*{12}{r}}
  \hline
Species & $\lambda (\rm {\AA})$&  \multicolumn{2}{c}{$-0.05\le F < 0.05$} &  \multicolumn{2}{c}{$0.05\le F < 0.15$}&  \multicolumn{2}{c}{$0.15\le F < 0.25$}&  \multicolumn{2}{c}{$0.25\le F < 0.35$}&  \multicolumn{2}{c}{$0.35\le F < 0.45$} \\
 & & \multicolumn{1}{c}{$\log N_f$}  & \multicolumn{1}{c}{$b_f$}  & \multicolumn{1}{c}{$\log N_f$}  & \multicolumn{1}{c}{$b_f$} & \multicolumn{1}{c}{$\log N_f$}  & \multicolumn{1}{c}{$b_f$} & \multicolumn{1}{c}{$\log N_f$}  & \multicolumn{1}{c}{$b_f$} & \multicolumn{1}{c}{$\log N_f$}  & \multicolumn{1}{c}{$b_f $} \\
\CIII\ & 977 & $13.69 \pm 0.02$ & $190 \pm 12$ & $13.45 \pm 0.02$ & $230 \pm 16$ & $13.04 \pm 0.03$ & $200 \pm 20$ & $12.64 \pm 0.05$ & $200 \pm 34$ & - & - \\
\OVI\ & 1032 & $14.2 \pm 0.01$ & $220 \pm 10$ & $13.99 \pm 0.01$ & $210 \pm 9$ & $13.75 \pm 0.03$ & $250 \pm 19$ & $13.62 \pm 0.02$ & $290 \pm 22$ & $13.5 \pm 0.02$ & $390 \pm 26$ \\
\CII\ & 1036 & $13.4 \pm 0.07$ & $150 \pm 33$ & - & - & - & - & - & - & - & - \\
\OVI\ & 1038 & $14.33 \pm 0.02$ & $290 \pm 17$ & $14.08 \pm 0.06$ & $270 \pm 32$ & $13.7 \pm 0.05$ & $210 \pm 22$ & $13.37 \pm 0.05$ & $250 \pm 38$ & $13.27 \pm 0.06$ & $280 \pm 52$ \\
\SiII\ & 1190 & $13.25 \pm 0.02$ & $280 \pm 20$ & $12.81 \pm 0.05$ & $370 \pm 74$ & - & - & - & - & - & - \\
\SiII\ & 1193 & $13.03 \pm 0.01$ & $230 \pm 10$ & $12.58 \pm 0.03$ & $260 \pm 24$ & - & - & - & - & - & - \\
\SiIII\ & 1207 & $13.02 \pm 0.01$ & $220 \pm 7$ & $12.69 \pm 0.01$ & $210 \pm 6$ & $12.41 \pm 0.02$ & $290 \pm 16$ & $12.16 \pm 0.02$ & $350 \pm 23$ & $11.61 \pm 0.11$ & $390 \pm 123$ \\
\NV\ & 1239 & - & - & - & - & - & - & - & - & - & - \\
\NV\ & 1242 & - & - & - & - & - & - & - & - & - & - \\
\SiII\ & 1260 & $12.85 \pm 0.01$ & $280 \pm 11$ & $12.53 \pm 0.03$ & $370 \pm 37$ & - & - & - & - & - & - \\
\OI\ & 1302 & $13.65 \pm 0.05$ & $300 \pm 42$ & $13.26 \pm 0.08$ & $320 \pm 71$ & - & - & - & - & - & - \\
\SiII\ & 1304 & $13.14 \pm 0.07$ & $170 \pm 35$ & $12.63 \pm 0.16$ & $230 \pm 92$ & - & - & - & - & - & - \\
\CII\ & 1335 & $13.53 \pm 0.02$ & $230 \pm 12$ & $13.17 \pm 0.03$ & $240 \pm 18$ & $12.7 \pm 0.06$ & $310 \pm 56$ & - & - & - & - \\
\SiIV\ & 1394 & $13.02 \pm 0.01$ & $200 \pm 8$ & $12.74 \pm 0.02$ & $210 \pm 10$ & $12.37 \pm 0.03$ & $250 \pm 19$ & $12.03 \pm 0.03$ & $270 \pm 23$ & $11.83 \pm 0.08$ & $510 \pm 136$ \\
\SiII\ & 1527 & $13.17 \pm 0.01$ & $260 \pm 10$ & $12.66 \pm 0.04$ & $290 \pm 36$ & - & - & - & - & - & - \\
\CIV\ & 1548 & $13.7 \pm 0.02$ & $190 \pm 11$ & $13.48 \pm 0.02$ & $190 \pm 15$ & $13.16 \pm 0.02$ & $170 \pm 12$ & $12.98 \pm 0.02$ & $260 \pm 14$ & $12.57 \pm 0.02$ & $280 \pm 17$ \\
\CIV\ & 1551 & $13.86 \pm 0.01$ & $210 \pm 9$ & $13.61 \pm 0.02$ & $190 \pm 10$ & $13.31 \pm 0.02$ & $200 \pm 12$ & $13.04 \pm 0.01$ & $210 \pm 8$ & $12.54 \pm 0.03$ & $230 \pm 30$ \\
\AlII\ & 1671 & $12.02 \pm 0.03$ & $240 \pm 20$ & $11.57 \pm 0.05$ & $290 \pm 44$ & - & - & - & - & - & - \\
\AlIII\ & 1855 & $12.06 \pm 0.05$ & $280 \pm 41$ & $11.85 \pm 0.02$ & $310 \pm 20$ & - & - & - & - & - & - \\
\AlIII\ & 1863 & - & - & - & - & - & - & - & - & - & - \\
\FeII\ & 2344 & - & - & - & - & - & - & - & - & - & - \\
\FeII\ & 2383 & $12.34 \pm 0.04$ & $220 \pm 26$ & - & - & - & - & - & - & - & - \\
\FeII\ & 2587 & - & - & - & - & - & - & - & - & - & - \\
\FeII\ & 2600 & - & - & - & - & - & - & - & - & - & - \\
\MgII\ & 2796 & $12.55 \pm 0.05$ & $170 \pm 28$ & - & - & - & - & - & - & - & - \\
\MgII\ & 2804 & $12.64 \pm 0.04$ & $180 \pm 22$ & - & - & - & - & - & - & - & - \\
\hline
\end{tabular}
\end{table*}
\begin{table*}
 \caption{Metal lines measured from full line profiles in the arithmetic mean composite rest-frame spectrum of \lya\ 
absorbers. A joint fit was performed to $N_f$ and $b_f$. Values are given in units of $\cm^{-2}$ and $\kms$, respectively.}
 \label{ab_am_fullline}
 \begin{tabular}{@{}l*{12}{r}}
  \hline
Species & $\lambda (\rm {\AA})$&  \multicolumn{2}{c}{$-0.05\le F < 0.05$} &  \multicolumn{2}{c}{$0.05\le F < 0.15$}&  \multicolumn{2}{c}{$0.15\le F < 0.25$}&  \multicolumn{2}{c}{$0.25\le F < 0.35$}&  \multicolumn{2}{c}{$0.35\le F < 0.45$} \\
  & & \multicolumn{1}{c}{$\log N_f$}  & \multicolumn{1}{c}{$b_f$}  & \multicolumn{1}{c}{$\log N_f$}  & \multicolumn{1}{c}{$b_f$} & \multicolumn{1}{c}{$\log N_f$}  & \multicolumn{1}{c}{$b_f$} & \multicolumn{1}{c}{$\log N_f$}  & \multicolumn{1}{c}{$b_f$} & \multicolumn{1}{c}{$\log N_f$}  & \multicolumn{1}{c}{$b_f $} \\
\CIII\ & 977 & $13.57 \pm 0.01$ & $200 \pm 8$ & $13.36 \pm 0.02$ & $220 \pm 14$ & $13.09 \pm 0.02$ & $240 \pm 18$ & $12.77 \pm 0.03$ & $250 \pm 25$ & - & - \\
\OVI\ & 1032 & $14.14 \pm 0.02$ & $250 \pm 16$ & $13.96 \pm 0.02$ & $230 \pm 12$ & $13.73 \pm 0.03$ & $270 \pm 20$ & $13.58 \pm 0.05$ & $310 \pm 46$ & - & - \\
\CII\ & 1036 & $13.17 \pm 0.12$ & $150 \pm 54$ & $13.02 \pm 0.22$ & $400 \pm 207$ & - & - & - & - & - & - \\
\OVI\ & 1038 & $14.33 \pm 0.02$ & $370 \pm 20$ & $13.92 \pm 0.04$ & $230 \pm 21$ & $13.9 \pm 0.03$ & $380 \pm 34$ & $13.27 \pm 0.05$ & $240 \pm 37$ & $13.22 \pm 0.07$ & $270 \pm 54$ \\
\SiII\ & 1190 & $13.07 \pm 0.02$ & $240 \pm 16$ & $12.6 \pm 0.05$ & $320 \pm 58$ & - & - & - & - & - & - \\
\SiII\ & 1193 & $12.9 \pm 0.01$ & $210 \pm 8$ & $12.51 \pm 0.02$ & $260 \pm 20$ & $12.1 \pm 0.15$ & $430 \pm 241$ & $11.99 \pm 0.05$ & $440 \pm 80$ & - & - \\
\SiIII\ & 1207 & $12.97 \pm 0.01$ & $230 \pm 9$ & $12.63 \pm 0.01$ & $220 \pm 4$ & $12.39 \pm 0.02$ & $310 \pm 16$ & $12.19 \pm 0.03$ & $400 \pm 39$ & $11.44 \pm 0.12$ & $410 \pm 145$ \\
\NV\ & 1239 & $13.08 \pm 0.04$ & $360 \pm 44$ & - & - & - & - & - & - & $12.84 \pm 0.05$ & $450 \pm 82$ \\
\NV\ & 1242 & - & - & - & - & - & - & - & - & - & - \\
\SiII\ & 1260 & $12.78 \pm 0.02$ & $270 \pm 14$ & $12.49 \pm 0.02$ & $410 \pm 29$ & - & - & - & - & - & - \\
\OI\ & 1302 & $13.75 \pm 0.03$ & $280 \pm 21$ & $13.49 \pm 0.06$ & $400 \pm 62$ & - & - & - & - & - & - \\
\SiII\ & 1304 & $13.39 \pm 0.03$ & $230 \pm 21$ & $12.76 \pm 0.13$ & $250 \pm 80$ & - & - & - & - & - & - \\
\CII\ & 1335 & $13.69 \pm 0.02$ & $240 \pm 13$ & $13.29 \pm 0.02$ & $240 \pm 15$ & $12.93 \pm 0.03$ & $330 \pm 30$ & - & - & - & - \\
\SiIV\ & 1394 & $13.12 \pm 0.02$ & $210 \pm 11$ & $12.84 \pm 0.01$ & $210 \pm 10$ & $12.59 \pm 0.02$ & $290 \pm 13$ & $12.19 \pm 0.02$ & $310 \pm 15$ & $12.01 \pm 0.08$ & $480 \pm 126$ \\
\SiII\ & 1527 & $13.29 \pm 0.01$ & $270 \pm 11$ & $12.92 \pm 0.03$ & $360 \pm 31$ & - & - & - & - & - & - \\
\CIV\ & 1548 & $13.81 \pm 0.01$ & $200 \pm 10$ & $13.58 \pm 0.02$ & $200 \pm 11$ & $13.31 \pm 0.02$ & $200 \pm 11$ & $13.1 \pm 0.02$ & $280 \pm 17$ & $12.7 \pm 0.02$ & $340 \pm 28$ \\
\CIV\ & 1551 & $13.96 \pm 0.01$ & $220 \pm 6$ & $13.71 \pm 0.01$ & $200 \pm 8$ & $13.44 \pm 0.02$ & $220 \pm 13$ & $13.17 \pm 0.01$ & $250 \pm 14$ & $12.76 \pm 0.03$ & $260 \pm 34$ \\
\AlII\ & 1671 & $12.15 \pm 0.03$ & $280 \pm 25$ & $11.79 \pm 0.03$ & $380 \pm 33$ & - & - & - & - & - & - \\
\AlIII\ & 1855 & $12.15 \pm 0.06$ & $300 \pm 53$ & $11.66 \pm 0.09$ & $160 \pm 53$ & - & - & - & - & - & - \\
\AlIII\ & 1863 & - & - & - & - & - & - & - & - & - & - \\
\FeII\ & 2344 & $12.91 \pm 0.06$ & $380 \pm 67$ & - & - & - & - & - & - & - & - \\
\FeII\ & 2383 & $12.6 \pm 0.03$ & $270 \pm 26$ & - & - & - & - & - & - & - & - \\
\FeII\ & 2587 & - & - & - & - & - & - & - & - & - & - \\
\FeII\ & 2600 & - & - & - & - & - & - & - & - & - & - \\
\MgII\ & 2796 & $12.83 \pm 0.05$ & $220 \pm 30$ & $12.33 \pm 0.05$ & $190 \pm 32$ & - & - & - & - & - & - \\
\MgII\ & 2804 & $12.93 \pm 0.05$ & $200 \pm 32$ & - & - & - & - & - & - & - & - \\

\hline
\end{tabular}
\end{table*}
\begin{table*}
 \caption{Metal lines measured from the central bin in the median composite rest-frame spectrum of \lya\ 
absorbers. A fit to $N_c$ was performed and values of $b_c$ taken from Lyman series measurements. Values are given in units of $\cm^{-2}$ and $\kms$, respectively.}
 \label{tab_med_onebin}
 \begin{tabular}{@{}l*{12}{r}}
  \hline
Species & $\lambda (\rm {\AA})$&  \multicolumn{2}{c}{$-0.05\le F < 0.05$} &  \multicolumn{2}{c}{$0.05\le F < 0.15$}&  \multicolumn{2}{c}{$0.15\le F < 0.25$}&  \multicolumn{2}{c}{$0.25\le F < 0.35$}&  \multicolumn{2}{c}{$0.35\le F < 0.45$} \\
 & & \multicolumn{1}{c}{$\log N_c$}  & \multicolumn{1}{c}{$b_c$}  & \multicolumn{1}{c}{$\log N_c$}  & \multicolumn{1}{c}{$b_c$} & \multicolumn{1}{c}{$\log N_c$}  & \multicolumn{1}{c}{$b_c$} & \multicolumn{1}{c}{$\log N_c$}  & \multicolumn{1}{c}{$b_c$} & \multicolumn{1}{c}{$\log N_c$}  & \multicolumn{1}{c}{$b_c $} \\
\CIII\ & 977 & $13.55 \pm 0.04$ & $25$ & $13.23 \pm 0.04$ & $15$ & $12.77 \pm 0.06$ & $15$ & - & - & - & - \\
\OVI\ & 1032 & $13.94 \pm 0.03$ & $25$ & $13.74 \pm 0.03$ & $15$ & $13.39 \pm 0.03$ & $15$ & $13.17 \pm 0.04$ & $15$ & $12.9 \pm 0.05$ & $15$ \\
\CII\ & 1036 & - & - & - & - & - & - & - & - & - & - \\
\OVI\ & 1038 & $13.93 \pm 0.03$ & $25$ & $13.75 \pm 0.04$ & $15$ & $13.4 \pm 0.05$ & $15$ & - & - & - & - \\
\SiII\ & 1190 & $12.9 \pm 0.05$ & $25$ & - & - & - & - & - & - & - & - \\
\SiII\ & 1193 & $12.78 \pm 0.03$ & $25$ & $12.13 \pm 0.11$ & $15$ & - & - & - & - & - & - \\
\SiIII\ & 1207 & $12.73 \pm 0.02$ & $25$ & $12.38 \pm 0.02$ & $15$ & $11.97 \pm 0.03$ & $15$ & $11.56 \pm 0.06$ & $15$ & - & - \\
\NV\ & 1239 & - & - & - & - & - & - & - & - & - & - \\
\NV\ & 1242 & - & - & - & - & - & - & - & - & - & - \\
\SiII\ & 1260 & $12.49 \pm 0.05$ & $25$ & $12.07 \pm 0.06$ & $15$ & - & - & - & - & - & - \\
\OI\ & 1302 & - & - & - & - & - & - & - & - & - & - \\
\SiII\ & 1304 & - & - & - & - & - & - & - & - & - & - \\
\CII\ & 1335 & $13.18 \pm 0.04$ & $25$ & $12.84 \pm 0.04$ & $15$ & - & - & - & - & - & - \\
\SiIV\ & 1394 & $12.76 \pm 0.02$ & $25$ & $12.4 \pm 0.03$ & $15$ & $12.05 \pm 0.03$ & $15$ & $11.62 \pm 0.1$ & $15$ & - & - \\
\SiII\ & 1527 & $12.54 \pm 0.05$ & $25$ & $11.78 \pm 0.14$ & $15$ & - & - & - & - & - & - \\
\CIV\ & 1548 & $13.45 \pm 0.02$ & $25$ & $13.22 \pm 0.01$ & $15$ & $12.9 \pm 0.01$ & $15$ & $12.58 \pm 0.02$ & $15$ & $11.97 \pm 0.08$ & $15$ \\
\CIV\ & 1551 & $13.56 \pm 0.02$ & $25$ & $13.34 \pm 0.02$ & $15$ & $13 \pm 0.02$ & $15$ & $12.71 \pm 0.03$ & $15$ & $11.59 \pm 0.18$ & $15$ \\
\AlII\ & 1671 & $11.62 \pm 0.05$ & $25$ & - & - & - & - & - & - & - & - \\
\AlIII\ & 1855 & - & - & - & - & - & - & - & - & - & - \\
\AlIII\ & 1863 & - & - & - & - & - & - & - & - & - & - \\
\FeII\ & 2344 & - & - & - & - & - & - & - & - & - & - \\
\FeII\ & 2383 & - & - & - & - & - & - & - & - & - & - \\
\FeII\ & 2587 & - & - & - & - & - & - & - & - & - & - \\
\FeII\ & 2600 & - & - & - & - & - & - & - & - & - & - \\
\MgII\ & 2796 & $12.44 \pm 0.07$ & $25$ & - & - & - & - & - & - & - & - \\
\MgII\ & 2804 & - & - & - & - & - & - & - & - & - & - \\
\hline
\end{tabular}
\end{table*}
\begin{table*}
 \caption{
Metal lines measured from the central bin in the  arithmetic mean  rest-frame spectrum of \lya\ 
absorbers. A fit to $N_c$ was performed and values of $b_c$ taken from Lyman series measurements. Values are given in units of $\cm^{-2}$ and $\kms$, respectively.}
 \label{tab_am_onebin}
 \begin{tabular}{@{}l*{12}{r}}
  \hline
Species & $\lambda (\rm {\AA})$&  \multicolumn{2}{c}{$-0.05\le F < 0.05$} &  \multicolumn{2}{c}{$0.05\le F < 0.15$}&  \multicolumn{2}{c}{$0.15\le F < 0.25$}&  \multicolumn{2}{c}{$0.25\le F < 0.35$}&  \multicolumn{2}{c}{$0.35\le F < 0.45$} \\
 & & \multicolumn{1}{c}{$\log N_f$}  & \multicolumn{1}{c}{$b_f$}  & \multicolumn{1}{c}{$\log N_f$}  & \multicolumn{1}{c}{$b_f$} & \multicolumn{1}{c}{$\log N_f$}  & \multicolumn{1}{c}{$b_f$} & \multicolumn{1}{c}{$\log N_f$}  & \multicolumn{1}{c}{$b_f$} & \multicolumn{1}{c}{$\log N_f$}  & \multicolumn{1}{c}{$b_f $} \\
\CIII\ & 977 & $13.37 \pm 0.03$ & $25$ & $13.11 \pm 0.03$ & $15$ & $12.75 \pm 0.04$ & $15$ & $12.32 \pm 0.06$ & $15$ & - & - \\
\OVI\ & 1032 & $13.83 \pm 0.02$ & $25$ & $13.67 \pm 0.02$ & $15$ & $13.35 \pm 0.03$ & $15$ & $13.13 \pm 0.03$ & $15$ & $12.86 \pm 0.04$ & $15$ \\
\CII\ & 1036 & - & - & - & - & - & - & - & - & - & - \\
\OVI\ & 1038 & $13.86 \pm 0.03$ & $25$ & $13.61 \pm 0.04$ & $15$ & $13.41 \pm 0.04$ & $15$ & - & - & - & - \\
\SiII\ & 1190 & $12.76 \pm 0.05$ & $25$ & - & - & - & - & - & - & - & - \\
\SiII\ & 1193 & $12.67 \pm 0.03$ & $25$ & $12.06 \pm 0.08$ & $15$ & - & - & - & - & - & - \\
\SiIII\ & 1207 & $12.65 \pm 0.02$ & $25$ & $12.3 \pm 0.02$ & $15$ & $11.91 \pm 0.03$ & $15$ & $11.5 \pm 0.04$ & $15$ & - & - \\
\NV\ & 1239 & - & - & - & - & - & - & - & - & - & - \\
\NV\ & 1242 & - & - & - & - & - & - & - & - & - & - \\
\SiII\ & 1260 & $12.43 \pm 0.04$ & $25$ & $11.96 \pm 0.05$ & $15$ & - & - & - & - & - & - \\
\OI\ & 1302 & $13.38 \pm 0.07$ & $25$ & - & - & - & - & - & - & - & - \\
\SiII\ & 1304 & $12.83 \pm 0.06$ & $25$ & - & - & - & - & - & - & - & - \\
\CII\ & 1335 & $13.32 \pm 0.02$ & $25$ & $12.93 \pm 0.03$ & $15$ & $12.41 \pm 0.07$ & $15$ & - & - & - & - \\
\SiIV\ & 1394 & $12.85 \pm 0.02$ & $25$ & $12.51 \pm 0.02$ & $15$ & $12.16 \pm 0.02$ & $15$ & $11.73 \pm 0.05$ & $15$ & - & - \\
\SiII\ & 1527 & $12.65 \pm 0.04$ & $25$ & $11.97 \pm 0.12$ & $15$ & - & - & - & - & - & - \\
\CIV\ & 1548 & $13.55 \pm 0.01$ & $25$ & $13.32 \pm 0.01$ & $15$ & $13.01 \pm 0.01$ & $15$ & $12.67 \pm 0.01$ & $15$ & $12.19 \pm 0.05$ & $15$ \\
\CIV\ & 1551 & $13.65 \pm 0.02$ & $25$ & $13.42 \pm 0.01$ & $15$ & $13.11 \pm 0.01$ & $15$ & $12.8 \pm 0.02$ & $15$ & $12.24 \pm 0.11$ & $15$ \\
\AlII\ & 1671 & $11.73 \pm 0.03$ & $25$ & $11.05 \pm 0.12$ & $15$ & - & - & - & - & - & - \\
\AlIII\ & 1855 & $11.79 \pm 0.08$ & $25$ & - & - & - & - & - & - & - & - \\
\AlIII\ & 1863 & - & - & - & - & - & - & - & - & - & - \\
\FeII\ & 2344 & - & - & - & - & - & - & - & - & - & - \\
\FeII\ & 2383 & $12.22 \pm 0.09$ & $25$ & - & - & - & - & - & - & - & - \\
\FeII\ & 2587 & - & - & - & - & - & - & - & - & - & - \\
\FeII\ & 2600 & $12.47 \pm 0.09$ & $25$ & - & - & - & - & - & - & - & - \\
\MgII\ & 2796 & $12.58 \pm 0.05$ & $25$ & - & - & - & - & - & - & - & - \\
\MgII\ & 2804 & $12.65 \pm 0.07$ & $25$ & - & - & - & - & - & - & - & - \\
\hline
\end{tabular}
\end{table*}

\label{lastpage}


\begin{thebibliography}{89}
\expandafter\ifx\csname natexlab\endcsname\relax\def\natexlab#1{#1}\fi

\bibitem[{{Adelberger} {et~al}\mbox{.}(2004){Adelberger}, {Steidel}, {Shapley},
  {Hunt}, {Erb}, {Reddy}, \& {Pettini}}]{2004ApJ...607..226A}
{Adelberger} K.~L., {Steidel} C.~C., {Shapley} A.~E., {Hunt} M.~P., {Erb}
  D.~K., {Reddy} N.~A., {Pettini} M., 2004, \apj, 607, 226

\bibitem[{{Adelberger} {et~al}\mbox{.}(2003){Adelberger}, {Steidel}, {Shapley},
  \& {Pettini}}]{2003ApJ...584...45A}
{Adelberger} K.~L., {Steidel} C.~C., {Shapley} A.~E., {Pettini} M., 2003, \apj,
  584, 45

\bibitem[{{Adelman-McCarthy} {et~al}\mbox{.}(2007){Adelman-McCarthy},
  {Ag{\"u}eros}, {Allam}, {Anderson}, {Anderson}, {Annis}, {Bahcall},
  {Bailer-Jones}, {Baldry}, {Barentine}, {Beers}, {Belokurov}, {Berlind},
  {Bernardi}, {Blanton}, {Bochanski}, {Boroski}, {Bramich}, {Brewington},
  {Brinchmann}, {Brinkmann}, {Brunner}, {Budav{\'a}ri}, {Carey}, {Carliles},
  {Carr}, {Castander}, {Connolly}, {Cool}, {Cunha}, {Csabai}, {Dalcanton},
  {Doi}, {Eisenstein}, {Evans}, {Evans}, {Fan}, {Finkbeiner}, {Friedman},
  {Frieman}, {Fukugita}, {Gillespie}, {Gilmore}, {Glazebrook}, {Gray},
  {Grebel}, {Gunn}, {de Haas}, {Hall}, {Harvanek}, {Hawley}, {Hayes},
  {Heckman}, {Hendry}, {Hennessy}, {Hindsley}, {Hirata}, {Hogan}, {Hogg},
  {Holtzman}, {Ichikawa}, {Ichikawa}, {Ivezi{\'c}}, {Jester}, {Johnston},
  {Jorgensen}, {Juri{\'c}}, {Kauffmann}, {Kent}, {Kleinman}, {Knapp},
  {Kniazev}, {Kron}, {Krzesinski}, {Kuropatkin}, {Lamb}, {Lampeitl}, {Lee},
  {Leger}, {Lima}, {Lin}, {Long}, {Loveday}, {Lupton}, {Mandelbaum}, {Margon},
  {Mart{\'{\i}}nez-Delgado}, {Matsubara}, {McGehee}, {McKay}, {Meiksin},
  {Munn}, {Nakajima}, {Nash}, {Neilsen}, {Newberg}, {Nichol},
  {Nieto-Santisteban}, {Nitta}, {Oyaizu}, {Okamura}, {Ostriker}, {Padmanabhan},
  {Park}, {Peoples}, {Pier}, {Pope}, {Pourbaix}, {Quinn}, {Raddick}, {Re
  Fiorentin}, {Richards}, {Richmond}, {Rix}, {Rockosi}, {Schlegel},
  {Schneider}, {Scranton}, {Seljak}, {Sheldon}, {Shimasaku}, {Silvestri},
  {Smith}, {Smol{\v c}i{\'c}}, {Snedden}, {Stebbins}, {Stoughton}, {Strauss},
  {SubbaRao}, {Suto}, {Szalay}, {Szapudi}, {Szkody}, {Tegmark}, {Thakar},
  {Tremonti}, {Tucker}, {Uomoto}, {Vanden Berk}, {Vandenberg}, {Vidrih},
  {Vogeley}, {Voges}, {Vogt}, {Weinberg}, {West}, {White}, {Wilhite}, {Yanny},
  {Yocum}, {York}, {Zehavi}, {Zibetti}, \& {Zucker}}]{2007ApJS..172..634A}
{Adelman-McCarthy} J.~K. {et~al.}, 2007, \apjs, 172, 634

\bibitem[{{Aguirre} {et~al}\mbox{.}(2008){Aguirre}, {Dow-Hygelund}, {Schaye},
  \& {Theuns}}]{2008ApJ...689..851A}
{Aguirre} A., {Dow-Hygelund} C., {Schaye} J., {Theuns} T., 2008, \apj, 689, 851

\bibitem[{{Aguirre} {et~al}\mbox{.}(2001){Aguirre}, {Hernquist}, {Schaye},
  {Weinberg}, {Katz}, \& {Gardner}}]{2001ApJ...560..599A}
{Aguirre} A., {Hernquist} L., {Schaye} J., {Weinberg} D.~H., {Katz} N.,
  {Gardner} J., 2001, \apj, 560, 599

\bibitem[{{Aguirre} {et~al}\mbox{.}(2004){Aguirre}, {Schaye}, {Kim}, {Theuns},
  {Rauch}, \& {Sargent}}]{2004ApJ...602...38A}
{Aguirre} A., {Schaye} J., {Kim} T.-S., {Theuns} T., {Rauch} M., {Sargent}
  W.~L.~W., 2004, \apj, 602, 38

\bibitem[{{Ahn} {et~al}\mbox{.}(2012){Ahn}, {Alexandroff}, {Allende Prieto},
  {Anderson}, {Anderton}, {Andrews}, {Aubourg}, {Bailey}, {Balbinot}, {Barnes},
  \& et~al.}]{2012ApJS..203...21A}
{Ahn} C.~P. {et~al.}, 2012, \apjs, 203, 21

\bibitem[{{Anders} \& {Grevesse}(1989)}]{1989GeCoA..53..197A}
{Anders} E., {Grevesse} N., 1989, \gca, 53, 197

\bibitem[{{Aracil} {et~al}\mbox{.}(2006){Aracil}, {Tripp}, {Bowen},
  {Prochaska}, {Chen}, \& {Frye}}]{2006MNRAS.367..139A}
{Aracil} B., {Tripp} T.~M., {Bowen} D.~V., {Prochaska} J.~X., {Chen} H.-W.,
  {Frye} B.~L., 2006, \mnras, 367, 139

\bibitem[{{Barai} {et~al}\mbox{.}(2013){Barai}, {Viel}, {Borgani}, {Tescari},
  {Tornatore}, {Dolag}, {Killedar}, {Monaco}, {D'Odorico}, \&
  {Cristiani}}]{2013MNRAS.430.3213B}
{Barai} P. {et~al.}, 2013, \mnras, 430, 3213

\bibitem[{{Becker} \& {Bolton}(2013)}]{2013arXiv1307.2259B}
{Becker} G.~D., {Bolton} J.~S., 2013, ArXiv e-prints

\bibitem[{{Bielby} {et~al}\mbox{.}(2013){Bielby}, {Hill}, {Shanks}, {Crighton},
  {Infante}, {Bornancini}, {Francke}, {H{\'e}raudeau}, {Lambas}, {Metcalfe},
  {Minniti}, {Padilla}, {Theuns}, {Tummuangpak}, \&
  {Weilbacher}}]{2013MNRAS.430..425B}
{Bielby} R. {et~al.}, 2013, \mnras, 430, 425

\bibitem[{{Bielby} {et~al}\mbox{.}(2011){Bielby}, {Shanks}, {Weilbacher},
  {Infante}, {Crighton}, {Bornancini}, {Bouch{\'e}}, {H{\'e}raudeau}, {Lambas},
  {Lowenthal}, {Minniti}, {Padilla}, {Petitjean}, \&
  {Theuns}}]{2011MNRAS.414....2B}
{Bielby} R.~M. {et~al.}, 2011, \mnras, 414, 2

\bibitem[{{Bolton} {et~al}\mbox{.}(2012){Bolton}, {Schlegel}, {Aubourg},
  {Bailey}, {Bhardwaj}, {Brownstein}, {Burles}, {Chen}, {Dawson}, {Eisenstein},
  {Gunn}, {Knapp}, {Loomis}, {Lupton}, {Maraston}, {Muna}, {Myers}, {Olmstead},
  {Padmanabhan}, {P{\^a}ris}, {Percival}, {Petitjean}, {Rockosi}, {Ross},
  {Schneider}, {Shu}, {Strauss}, {Thomas}, {Tremonti}, {Wake}, {Weaver}, \&
  {Wood-Vasey}}]{2012AJ....144..144B}
{Bolton} A.~S. {et~al.}, 2012, \aj, 144, 144

\bibitem[{{Bovy} {et~al}\mbox{.}(2011){Bovy}, {Hennawi}, {Hogg}, {Myers},
  {Kirkpatrick}, {Schlegel}, {Ross}, {Sheldon}, {McGreer}, {Schneider}, \&
  {Weaver}}]{2011ApJ...729..141B}
{Bovy} J. {et~al.}, 2011, \apj, 729, 141

\bibitem[{{Cooke} {et~al}\mbox{.}(2011){Cooke}, {Pettini}, {Steidel}, {Rudie},
  \& {Nissen}}]{2011MNRAS.417.1534C}
{Cooke} R., {Pettini} M., {Steidel} C.~C., {Rudie} G.~C., {Nissen} P.~E., 2011,
  \mnras, 417, 1534

\bibitem[{{Cowie} {et~al}\mbox{.}(1995){Cowie}, {Songaila}, {Kim}, \&
  {Hu}}]{1995AJ....109.1522C}
{Cowie} L.~L., {Songaila} A., {Kim} T.-S., {Hu} E.~M., 1995, \aj, 109, 1522

\bibitem[{{Crighton} {et~al}\mbox{.}(2011){Crighton}, {Bielby}, {Shanks},
  {Infante}, {Bornancini}, {Bouch{\'e}}, {Lambas}, {Lowenthal}, {Minniti},
  {Morris}, {Padilla}, {P{\'e}roux}, {Petitjean}, {Theuns}, {Tummuangpak},
  {Weilbacher}, {Wisotzki}, \& {Worseck}}]{2011MNRAS.414...28C}
{Crighton} N.~H.~M. {et~al.}, 2011, \mnras, 414, 28

\bibitem[{{Crighton}, {Hennawi} \& {Prochaska}(2013){Crighton}, {Hennawi}, \&
  {Prochaska}}]{2013ApJ...776L..18C}
{Crighton} N.~H.~M., {Hennawi} J.~F., {Prochaska} J.~X., 2013, \apjl, 776, L18

\bibitem[{{Dawson} {et~al}\mbox{.}(2013){Dawson}, {Schlegel}, {Ahn},
  {Anderson}, {Aubourg}, {Bailey}, {Barkhouser}, {Bautista}, {Beifiori},
  {Berlind}, {Bhardwaj}, {Bizyaev}, {Blake}, {Blanton}, {Blomqvist}, {Bolton},
  {Borde}, {Bovy}, {Brandt}, {Brewington}, {Brinkmann}, {Brown}, {Brownstein},
  {Bundy}, {Busca}, {Carithers}, {Carnero}, {Carr}, {Chen}, {Comparat},
  {Connolly}, {Cope}, {Croft}, {Cuesta}, {da Costa}, {Davenport}, {Delubac},
  {de Putter}, {Dhital}, {Ealet}, {Ebelke}, {Eisenstein}, {Escoffier}, {Fan},
  {Filiz Ak}, {Finley}, {Font-Ribera}, {G{\'e}nova-Santos}, {Gunn}, {Guo},
  {Haggard}, {Hall}, {Hamilton}, {Harris}, {Harris}, {Ho}, {Hogg}, {Holder},
  {Honscheid}, {Huehnerhoff}, {Jordan}, {Jordan}, {Kauffmann}, {Kazin},
  {Kirkby}, {Klaene}, {Kneib}, {Le Goff}, {Lee}, {Long}, {Loomis}, {Lundgren},
  {Lupton}, {Maia}, {Makler}, {Malanushenko}, {Malanushenko}, {Mandelbaum},
  {Manera}, {Maraston}, {Margala}, {Masters}, {McBride}, {McDonald}, {McGreer},
  {McMahon}, {Mena}, {Miralda-Escud{\'e}}, {Montero-Dorta}, {Montesano},
  {Muna}, {Myers}, {Naugle}, {Nichol}, {Noterdaeme}, {Nuza}, {Olmstead},
  {Oravetz}, {Oravetz}, {Owen}, {Padmanabhan}, {Palanque-Delabrouille}, {Pan},
  {Parejko}, {P{\^a}ris}, {Percival}, {P{\'e}rez-Fournon},
  {P{\'e}rez-R{\`a}fols}, {Petitjean}, {Pfaffenberger}, {Pforr}, {Pieri},
  {Prada}, {Price-Whelan}, {Raddick}, {Rebolo}, {Rich}, {Richards}, {Rockosi},
  {Roe}, {Ross}, {Ross}, {Rossi}, {Rubi{\~n}o-Martin}, {Samushia},
  {S{\'a}nchez}, {Sayres}, {Schmidt}, {Schneider}, {Sc{\'o}ccola}, {Seo},
  {Shelden}, {Sheldon}, {Shen}, {Shu}, {Slosar}, {Smee}, {Snedden}, {Stauffer},
  {Steele}, {Strauss}, {Streblyanska}, {Suzuki}, {Swanson}, {Tal}, {Tanaka},
  {Thomas}, {Tinker}, {Tojeiro}, {Tremonti}, {Vargas Maga{\~n}a}, {Verde},
  {Viel}, {Wake}, {Watson}, {Weaver}, {Weinberg}, {Weiner}, {West}, {White},
  {Wood-Vasey}, {Yeche}, {Zehavi}, {Zhao}, \& {Zheng}}]{2013AJ....145...10D}
{Dawson} K.~S. {et~al.}, 2013, \aj, 145, 10

\bibitem[{{Eisenstein} {et~al}\mbox{.}(2011){Eisenstein}, {Weinberg}, {Agol},
  {Aihara}, {Allende Prieto}, {Anderson}, {Arns}, {Aubourg}, {Bailey},
  {Balbinot}, \& et~al.}]{2011AJ....142...72E}
{Eisenstein} D.~J. {et~al.}, 2011, \aj, 142, 72

\bibitem[{{Faucher-Gigu{\`e}re} \& {Kere{\v s}}(2011)}]{2011MNRAS.412L.118F}
{Faucher-Gigu{\`e}re} C.-A., {Kere{\v s}} D., 2011, \mnras, 412, L118

\bibitem[{{Faucher-Gigu{\`e}re} {et~al}\mbox{.}(2008){Faucher-Gigu{\`e}re},
  {Prochaska}, {Lidz}, {Hernquist}, \& {Zaldarriaga}}]{2008ApJ...681..831F}
{Faucher-Gigu{\`e}re} C.-A., {Prochaska} J.~X., {Lidz} A., {Hernquist} L.,
  {Zaldarriaga} M., 2008, \apj, 681, 831

\bibitem[{{Ferland} {et~al}\mbox{.}(1998){Ferland}, {Korista}, {Verner},
  {Ferguson}, {Kingdon}, \& {Verner}}]{1998PASP..110..761F}
{Ferland} G.~J., {Korista} K.~T., {Verner} D.~A., {Ferguson} J.~W., {Kingdon}
  J.~B., {Verner} E.~M., 1998, \pasp, 110, 761

\bibitem[{{Frank} {et~al}\mbox{.}(2007){Frank}, {Bentz}, {Stanek}, {Mathur},
  {Dietrich}, {Peterson}, \& {Atlee}}]{2007Ap&SS.312..325F}
{Frank} S., {Bentz} M.~C., {Stanek} K.~Z., {Mathur} S., {Dietrich} M.,
  {Peterson} B.~M., {Atlee} D.~W., 2007, \apss, 312, 325

\bibitem[{{Fumagalli} {et~al}\mbox{.}(2013){Fumagalli}, {O'Meara}, {Prochaska},
  \& {Worseck}}]{2013ApJ...775...78F}
{Fumagalli} M., {O'Meara} J.~M., {Prochaska} J.~X., {Worseck} G., 2013, \apj,
  775, 78 F13

\bibitem[{{Fumagalli} {et~al}\mbox{.}(2011){Fumagalli}, {Prochaska}, {Kasen},
  {Dekel}, {Ceverino}, \& {Primack}}]{2011MNRAS.418.1796F}
{Fumagalli} M., {Prochaska} J.~X., {Kasen} D., {Dekel} A., {Ceverino} D.,
  {Primack} J.~R., 2011, \mnras, 418, 1796

\bibitem[{{Gunn} \& {Peterson}(1965)}]{1965ApJ...142.1633G}
{Gunn} J.~E., {Peterson} B.~A., 1965, \apj, 142, 1633

\bibitem[{{Gunn} {et~al}\mbox{.}(2006){Gunn}, {Siegmund}, {Mannery}, {Owen},
  {Hull}, {Leger}, {Carey}, {Knapp}, {York}, {Boroski}, {Kent}, {Lupton},
  {Rockosi}, {Evans}, {Waddell}, {Anderson}, {Annis}, {Barentine}, {Bartoszek},
  {Bastian}, {Bracker}, {Brewington}, {Briegel}, {Brinkmann}, {Brown}, {Carr},
  {Czarapata}, {Drennan}, {Dombeck}, {Federwitz}, {Gillespie}, {Gonzales},
  {Hansen}, {Harvanek}, {Hayes}, {Jordan}, {Kinney}, {Klaene}, {Kleinman},
  {Kron}, {Kresinski}, {Lee}, {Limmongkol}, {Lindenmeyer}, {Long}, {Loomis},
  {McGehee}, {Mantsch}, {Neilsen}, {Neswold}, {Newman}, {Nitta}, {Peoples},
  {Pier}, {Prieto}, {Prosapio}, {Rivetta}, {Schneider}, {Snedden}, \&
  {Wang}}]{2006AJ....131.2332G}
{Gunn} J.~E. {et~al.}, 2006, \aj, 131, 2332

\bibitem[{{Haardt} \& {Madau}(2001)}]{2001cghr.confE..64H}
{Haardt} F., {Madau} P., 2001, in Clusters of Galaxies and the High Redshift
  Universe Observed in X-rays, {Neumann} D.~M., {Tran} J.~T.~V., eds.

\bibitem[{{Hacker} {et~al}\mbox{.}(2013){Hacker}, {Brunner}, {Lundgren}, \&
  {York}}]{2013MNRAS.434..163H}
{Hacker} T.~L., {Brunner} R.~J., {Lundgren} B.~F., {York} D.~G., 2013, \mnras,
  434, 163

\bibitem[{{Hao} {et~al}\mbox{.}(2007){Hao}, {Stanek}, {Dobrzycki}, {Matheson},
  {Bentz}, {Kuraszkiewicz}, {Garnavich}, {Howk}, {Calkins}, {Worthey},
  {Modjaz}, \& {Serven}}]{2007ApJ...659L..99H}
{Hao} H. {et~al.}, 2007, \apjl, 659, L99

\bibitem[{{Hinshaw} {et~al}\mbox{.}(2009){Hinshaw}, {Weiland}, {Hill},
  {Odegard}, {Larson}, {Bennett}, {Dunkley}, {Gold}, {Greason}, {Jarosik},
  {Komatsu}, {Nolta}, {Page}, {Spergel}, {Wollack}, {Halpern}, {Kogut},
  {Limon}, {Meyer}, {Tucker}, \& {Wright}}]{2009ApJS..180..225H}
{Hinshaw} G. {et~al.}, 2009, \apjs, 180, 225

\bibitem[{{Hu} {et~al}\mbox{.}(1995){Hu}, {Kim}, {Cowie}, {Songaila}, \&
  {Rauch}}]{1995AJ....110.1526H}
{Hu} E.~M., {Kim} T.-S., {Cowie} L.~L., {Songaila} A., {Rauch} M., 1995, \aj,
  110, 1526

\bibitem[{{Hummels} {et~al}\mbox{.}(2013){Hummels}, {Bryan}, {Smith}, \&
  {Turk}}]{2013MNRAS.430.1548H}
{Hummels} C.~B., {Bryan} G.~L., {Smith} B.~D., {Turk} M.~J., 2013, \mnras, 430,
  1548

\bibitem[{{Kim} {et~al}\mbox{.}(2002){Kim}, {Carswell}, {Cristiani},
  {D'Odorico}, \& {Giallongo}}]{2002MNRAS.335..555K}
{Kim} T.-S., {Carswell} R.~F., {Cristiani} S., {D'Odorico} S., {Giallongo} E.,
  2002, \mnras, 335, 555

\bibitem[{{Kobayashi}, {Tominaga} \& {Nomoto}(2011){Kobayashi}, {Tominaga}, \&
  {Nomoto}}]{2011ApJ...730L..14K}
{Kobayashi} C., {Tominaga} N., {Nomoto} K., 2011, \apjl, 730, L14

\bibitem[{{Lee} {et~al}\mbox{.}(2013){Lee}, {Bailey}, {Bartsch}, {Carithers},
  {Dawson}, {Kirkby}, {Lundgren}, {Margala}, {Palanque-Delabrouille}, {Pieri},
  {Schlegel}, {Weinberg}, {Y{\`e}che}, {Aubourg}, {Bautista}, {Bizyaev},
  {Blomqvist}, {Bolton}, {Borde}, {Brewington}, {Busca}, {Croft}, {Delubac},
  {Ebelke}, {Eisenstein}, {Font-Ribera}, {Ge}, {Hamilton}, {Hennawi}, {Ho},
  {Honscheid}, {Le Goff}, {Malanushenko}, {Malanushenko}, {Miralda-Escud{\'e}},
  {Myers}, {Noterdaeme}, {Oravetz}, {Pan}, {P{\^a}ris}, {Petitjean}, {Rich},
  {Rollinde}, {Ross}, {Rossi}, {Schneider}, {Simmons}, {Snedden}, {Slosar},
  {Spergel}, {Suzuki}, {Viel}, \& {Weaver}}]{2013AJ....145...69L}
{Lee} K.-G. {et~al.}, 2013, \aj, 145, 69

\bibitem[{{Lee}, {Suzuki} \& {Spergel}(2012){Lee}, {Suzuki}, \&
  {Spergel}}]{2012AJ....143...51L}
{Lee} K.-G., {Suzuki} N., {Spergel} D.~N., 2012, \aj, 143, 51

\bibitem[{{Lehner} {et~al}\mbox{.}(2013){Lehner}, {Howk}, {Tripp}, {Tumlinson},
  {Prochaska}, {O'Meara}, {Thom}, {Werk}, {Fox}, \&
  {Ribaudo}}]{2013ApJ...770..138L}
{Lehner} N. {et~al.}, 2013, \apj, 770, 138

\bibitem[{{Lundgren} {et~al}\mbox{.}(2012){Lundgren}, {Brammer}, {van Dokkum},
  {Bezanson}, {Franx}, {Fumagalli}, {Momcheva}, {Nelson}, {Skelton}, {Wake},
  {Whitaker}, {da Cunha}, {Erb}, {Fan}, {Kriek}, {Labb{\'e}}, {Marchesini},
  {Patel}, {Rix}, {Schmidt}, \& {van der Wel}}]{2012ApJ...760...49L}
{Lundgren} B.~F. {et~al.}, 2012, \apj, 760, 49

\bibitem[{{Lynch} \& {Charlton}(2007)}]{2007ApJ...666...64L}
{Lynch} R.~S., {Charlton} J.~C., 2007, \apj, 666, 64

\bibitem[{{Lynds}(1971)}]{1971ApJ...164L..73L}
{Lynds} R., 1971, \apjl, 164, L73

\bibitem[{{Mac Low} \& {Ferrara}(1999)}]{1999ApJ...513..142M}
{Mac Low} M.-M., {Ferrara} A., 1999, \apj, 513, 142

\bibitem[{{Martin}(2005)}]{2005ApJ...621..227M}
{Martin} C.~L., 2005, \apj, 621, 227

\bibitem[{{McDonald} {et~al}\mbox{.}(2006){McDonald}, {Seljak}, {Burles},
  {Schlegel}, {Weinberg}, {Cen}, {Shih}, {Schaye}, {Schneider}, {Bahcall},
  {Briggs}, {Brinkmann}, {Brunner}, {Fukugita}, {Gunn}, {Ivezi{\'c}}, {Kent},
  {Lupton}, \& {Vanden Berk}}]{2006ApJS..163...80M}
{McDonald} P. {et~al.}, 2006, \apjs, 163, 80

\bibitem[{{Meiring} {et~al}\mbox{.}(2013){Meiring}, {Tripp}, {Werk}, {Howk},
  {Jenkins}, {Prochaska}, {Lehner}, \& {Sembach}}]{2013ApJ...767...49M}
{Meiring} J.~D., {Tripp} T.~M., {Werk} J.~K., {Howk} J.~C., {Jenkins} E.~B.,
  {Prochaska} J.~X., {Lehner} N., {Sembach} K.~R., 2013, \apj, 767, 49

\bibitem[{{Meyer} \& {York}(1987)}]{1987ApJ...315L...5M}
{Meyer} D.~M., {York} D.~G., 1987, \apjl, 315, L5

\bibitem[{{Nicastro} {et~al}\mbox{.}(2002){Nicastro}, {Zezas}, {Drake},
  {Elvis}, {Fiore}, {Fruscione}, {Marengo}, {Mathur}, \&
  {Bianchi}}]{2002ApJ...573..157N}
{Nicastro} F. {et~al.}, 2002, \apj, 573, 157

\bibitem[{{Nielsen} {et~al}\mbox{.}(2013){Nielsen}, {Churchill}, {Kacprzak}, \&
  {Murphy}}]{2013arXiv1304.6716N}
{Nielsen} N.~M., {Churchill} C.~W., {Kacprzak} G.~G., {Murphy} M.~T., 2013,
  ArXiv e-prints

\bibitem[{{Noterdaeme} {et~al}\mbox{.}(2012){Noterdaeme}, {Petitjean},
  {Carithers}, {P{\^a}ris}, {Font-Ribera}, {Bailey}, {Aubourg}, {Bizyaev},
  {Ebelke}, {Finley}, {Ge}, {Malanushenko}, {Malanushenko},
  {Miralda-Escud{\'e}}, {Myers}, {Oravetz}, {Pan}, {Pieri}, {Ross},
  {Schneider}, {Simmons}, \& {York}}]{2012A&A...547L...1N}
{Noterdaeme} P. {et~al.}, 2012, \aap, 547, L1

\bibitem[{{O'Meara} {et~al}\mbox{.}(2013){O'Meara}, {Prochaska}, {Worseck},
  {Chen}, \& {Madau}}]{2013ApJ...765..137O}
{O'Meara} J.~M., {Prochaska} J.~X., {Worseck} G., {Chen} H.-W., {Madau} P.,
  2013, \apj, 765, 137

\bibitem[{{Oppenheimer} \& {Dav{\'e}}(2006)}]{2006MNRAS.373.1265O}
{Oppenheimer} B.~D., {Dav{\'e}} R., 2006, \mnras, 373, 1265

\bibitem[{{Oppenheimer} \& {Dav{\'e}}(2008)}]{2008MNRAS.387..577O}
{Oppenheimer} B.~D., {Dav{\'e}} R., 2008, \mnras, 387, 577

\bibitem[{{Oppenheimer} \& {Schaye}(2013)}]{2013MNRAS.434.1043O}
{Oppenheimer} B.~D., {Schaye} J., 2013, \mnras, 434, 1043

\bibitem[{{Palanque-Delabrouille} {et~al}\mbox{.}(2013){Palanque-Delabrouille},
  {Y{\`e}che}, {Borde}, {Le Goff}, {Rossi}, {Viel}, {Aubourg}, {Bailey},
  {Bautista}, {Blomqvist}, {Bolton}, {Bolton}, {Busca}, {Carithers}, {Croft},
  {Dawson}, {Delubac}, {Font-Ribera}, {Ho}, {Kirkby}, {Lee}, {Margala},
  {Miralda-Escud{\'e}}, {Muna}, {Myers}, {Noterdaeme}, {P{\^a}ris},
  {Petitjean}, {Pieri}, {Rich}, {Rollinde}, {Ross}, {Schlegel}, {Schneider},
  {Slosar}, \& {Weinberg}}]{2013arXiv1306.5896P}
{Palanque-Delabrouille} N. {et~al.}, 2013, \aap\ accepted

\bibitem[{{P{\^a}ris} {et~al}\mbox{.}(2012){P{\^a}ris}, {Petitjean}, {Aubourg},
  {Bailey}, {Ross}, {Myers}, {Strauss}, {Anderson}, {Arnau}, {Bautista},
  {Bizyaev}, {Bolton}, {Bovy}, {Brandt}, {Brewington}, {Browstein}, {Busca},
  {Capellupo}, {Carithers}, {Croft}, {Dawson}, {Delubac}, {Ebelke},
  {Eisenstein}, {Engelke}, {Fan}, {Filiz Ak}, {Finley}, {Font-Ribera}, {Ge},
  {Gibson}, {Hall}, {Hamann}, {Hennawi}, {Ho}, {Hogg}, {Ivezi{\'c}}, {Jiang},
  {Kimball}, {Kirkby}, {Kirkpatrick}, {Lee}, {Le Goff}, {Lundgren}, {MacLeod},
  {Malanushenko}, {Malanushenko}, {Maraston}, {McGreer}, {McMahon},
  {Miralda-Escud{\'e}}, {Muna}, {Noterdaeme}, {Oravetz},
  {Palanque-Delabrouille}, {Pan}, {Perez-Fournon}, {Pieri}, {Richards},
  {Rollinde}, {Sheldon}, {Schlegel}, {Schneider}, {Slosar}, {Shelden}, {Shen},
  {Simmons}, {Snedden}, {Suzuki}, {Tinker}, {Viel}, {Weaver}, {Weinberg},
  {White}, {Wood-Vasey}, \& {Y{\`e}che}}]{2012A&A...548A..66P}
{P{\^a}ris} I. {et~al.}, 2012, \aap, 548, A66

\bibitem[{{Penprase} {et~al}\mbox{.}(2010){Penprase}, {Prochaska}, {Sargent},
  {Toro-Martinez}, \& {Beeler}}]{2010ApJ...721....1P}
{Penprase} B.~E., {Prochaska} J.~X., {Sargent} W.~L.~W., {Toro-Martinez} I.,
  {Beeler} D.~J., 2010, \apj, 721, 1

\bibitem[{{Pettini} {et~al}\mbox{.}(2008){Pettini}, {Zych}, {Steidel}, \&
  {Chaffee}}]{2008MNRAS.385.2011P}
{Pettini} M., {Zych} B.~J., {Steidel} C.~C., {Chaffee} F.~H., 2008, \mnras,
  385, 2011

\bibitem[{{Pieri} {et~al}\mbox{.}(2010{\natexlab{a}}){Pieri}, {Frank},
  {Mathur}, {Weinberg}, {York}, \& {Oppenheimer}}]{2010ApJ...716.1084P}
{Pieri} M.~M., {Frank} S., {Mathur} S., {Weinberg} D.~H., {York} D.~G.,
  {Oppenheimer} B.~D., 2010{\natexlab{a}}, \apj, 716, 1084

\bibitem[{{Pieri} {et~al}\mbox{.}(2010{\natexlab{b}}){Pieri}, {Frank},
  {Weinberg}, {Mathur}, \& {York}}]{2010ApJ...724L..69P}
{Pieri} M.~M., {Frank} S., {Weinberg} D.~H., {Mathur} S., {York} D.~G.,
  2010{\natexlab{b}}, \apjl, 724, L69 P10

\bibitem[{{Pieri} \& {Haehnelt}(2004)}]{2004MNRAS.347..985P}
{Pieri} M.~M., {Haehnelt} M.~G., 2004, \mnras, 347, 985

\bibitem[{{Pieri} \& {Martel}(2007)}]{2007ApJ...662L...7P}
{Pieri} M.~M., {Martel} H., 2007, \apjl, 662, L7

\bibitem[{{Pieri}, {Martel} \& {Grenon}(2007){Pieri}, {Martel}, \&
  {Grenon}}]{2007ApJ...658...36P}
{Pieri} M.~M., {Martel} H., {Grenon} C., 2007, \apj, 658, 36

\bibitem[{{Pieri}, {Schaye} \& {Aguirre}(2006){Pieri}, {Schaye}, \&
  {Aguirre}}]{2006ApJ...638...45P}
{Pieri} M.~M., {Schaye} J., {Aguirre} A., 2006, \apj, 638, 45

\bibitem[{{Pinsonneault}, {Martel} \& {Pieri}(2010){Pinsonneault}, {Martel}, \&
  {Pieri}}]{2010ApJ...725.2087P}
{Pinsonneault} S., {Martel} H., {Pieri} M.~M., 2010, \apj, 725, 2087

\bibitem[{{Prochaska}, {O'Meara} \& {Worseck}(2010){Prochaska}, {O'Meara}, \&
  {Worseck}}]{2010ApJ...718..392P}
{Prochaska} J.~X., {O'Meara} J.~M., {Worseck} G., 2010, \apj, 718, 392

\bibitem[{{Rahmati} \& {Schaye}(2014)}]{2014MNRAS.438..529R}
{Rahmati} A., {Schaye} J., 2014, \mnras, 438, 529

\bibitem[{{Rakic} {et~al}\mbox{.}(2012){Rakic}, {Schaye}, {Steidel}, \&
  {Rudie}}]{2012ApJ...751...94R}
{Rakic} O., {Schaye} J., {Steidel} C.~C., {Rudie} G.~C., 2012, \apj, 751, 94

\bibitem[{{Reddy} {et~al}\mbox{.}(2008){Reddy}, {Steidel}, {Pettini},
  {Adelberger}, {Shapley}, {Erb}, \& {Dickinson}}]{2008ApJS..175...48R}
{Reddy} N.~A., {Steidel} C.~C., {Pettini} M., {Adelberger} K.~L., {Shapley}
  A.~E., {Erb} D.~K., {Dickinson} M., 2008, \apjs, 175, 48

\bibitem[{{Rigby}, {Charlton} \& {Churchill}(2002){Rigby}, {Charlton}, \&
  {Churchill}}]{2002ApJ...565..743R}
{Rigby} J.~R., {Charlton} J.~C., {Churchill} C.~W., 2002, \apj, 565, 743

\bibitem[{{Ross} {et~al}\mbox{.}(2012){Ross}, {Myers}, {Sheldon}, {Y{\`e}che},
  {Strauss}, {Bovy}, {Kirkpatrick}, {Richards}, {Aubourg}, {Blanton}, {Brandt},
  {Carithers}, {Croft}, {da Silva}, {Dawson}, {Eisenstein}, {Hennawi}, {Ho},
  {Hogg}, {Lee}, {Lundgren}, {McMahon}, {Miralda-Escud{\'e}},
  {Palanque-Delabrouille}, {P{\^a}ris}, {Petitjean}, {Pieri}, {Rich}, {Roe},
  {Schiminovich}, {Schlegel}, {Schneider}, {Slosar}, {Suzuki}, {Tinker},
  {Weinberg}, {Weyant}, {White}, \& {Wood-Vasey}}]{2012ApJS..199....3R}
{Ross} N.~P. {et~al.}, 2012, \apjs, 199, 3

\bibitem[{{Rudie} {et~al}\mbox{.}(2012){Rudie}, {Steidel}, {Trainor}, {Rakic},
  {Bogosavljevi{\'c}}, {Pettini}, {Reddy}, {Shapley}, {Erb}, \&
  {Law}}]{2012ApJ...750...67R}
{Rudie} G.~C. {et~al.}, 2012, \apj, 750, 67

\bibitem[{{Scannapieco}, {Ferrara} \& {Madau}(2002){Scannapieco}, {Ferrara}, \&
  {Madau}}]{2002ApJ...574..590S}
{Scannapieco} E., {Ferrara} A., {Madau} P., 2002, \apj, 574, 590

\bibitem[{{Schaye} {et~al}\mbox{.}(2003){Schaye}, {Aguirre}, {Kim}, {Theuns},
  {Rauch}, \& {Sargent}}]{2003ApJ...596..768S}
{Schaye} J., {Aguirre} A., {Kim} T.-S., {Theuns} T., {Rauch} M., {Sargent}
  W.~L.~W., 2003, \apj, 596, 768

\bibitem[{{Schaye}, {Carswell} \& {Kim}(2007){Schaye}, {Carswell}, \&
  {Kim}}]{2007MNRAS.379.1169S}
{Schaye} J., {Carswell} R.~F., {Kim} T.-S., 2007, \mnras, 379, 1169

\bibitem[{{Schaye} {et~al}\mbox{.}(2010){Schaye}, {Dalla Vecchia}, {Booth},
  {Wiersma}, {Theuns}, {Haas}, {Bertone}, {Duffy}, {McCarthy}, \& {van de
  Voort}}]{2010MNRAS.402.1536S}
{Schaye} J. {et~al.}, 2010, \mnras, 402, 1536

\bibitem[{{Schaye} {et~al}\mbox{.}(2000){Schaye}, {Rauch}, {Sargent}, \&
  {Kim}}]{2000ApJ...541L...1S}
{Schaye} J., {Rauch} M., {Sargent} W.~L.~W., {Kim} T.-S., 2000, \apjl, 541, L1

\bibitem[{{Shen} {et~al}\mbox{.}(2013){Shen}, {Madau}, {Guedes}, {Mayer},
  {Prochaska}, \& {Wadsley}}]{2013ApJ...765...89S}
{Shen} S., {Madau} P., {Guedes} J., {Mayer} L., {Prochaska} J.~X., {Wadsley}
  J., 2013, \apj, 765, 89

\bibitem[{{Simcoe}, {Sargent} \& {Rauch}(2004){Simcoe}, {Sargent}, \&
  {Rauch}}]{2004ApJ...606...92S}
{Simcoe} R.~A., {Sargent} W.~L.~W., {Rauch} M., 2004, \apj, 606, 92

\bibitem[{{Simcoe} {et~al}\mbox{.}(2006){Simcoe}, {Sargent}, {Rauch}, \&
  {Becker}}]{2006ApJ...637..648S}
{Simcoe} R.~A., {Sargent} W.~L.~W., {Rauch} M., {Becker} G., 2006, \apj, 637,
  648

\bibitem[{{Smee} {et~al}\mbox{.}(2013){Smee}, {Gunn}, {Uomoto}, {Roe},
  {Schlegel}, {Rockosi}, {Carr}, {Leger}, {Dawson}, {Olmstead}, {Brinkmann},
  {Owen}, {Barkhouser}, {Honscheid}, {Harding}, {Long}, {Lupton}, {Loomis},
  {Anderson}, {Annis}, {Bernardi}, {Bhardwaj}, {Bizyaev}, {Bolton},
  {Brewington}, {Briggs}, {Burles}, {Burns}, {Castander}, {Connolly},
  {Davenport}, {Ebelke}, {Epps}, {Feldman}, {Friedman}, {Frieman}, {Heckman},
  {Hull}, {Knapp}, {Lawrence}, {Loveday}, {Mannery}, {Malanushenko},
  {Malanushenko}, {Merrelli}, {Muna}, {Newman}, {Nichol}, {Oravetz}, {Pan},
  {Pope}, {Ricketts}, {Shelden}, {Sandford}, {Siegmund}, {Simmons}, {Smith},
  {Snedden}, {Schneider}, {SubbaRao}, {Tremonti}, {Waddell}, \&
  {York}}]{2013AJ....146...32S}
{Smee} S.~A. {et~al.}, 2013, \aj, 146, 32

\bibitem[{{Springel}(2005)}]{2005MNRAS.364.1105S}
{Springel} V., 2005, \mnras, 364, 1105

\bibitem[{{Steidel} {et~al}\mbox{.}(2010){Steidel}, {Erb}, {Shapley},
  {Pettini}, {Reddy}, {Bogosavljevi{\'c}}, {Rudie}, \&
  {Rakic}}]{2010ApJ...717..289S}
{Steidel} C.~C., {Erb} D.~K., {Shapley} A.~E., {Pettini} M., {Reddy} N.,
  {Bogosavljevi{\'c}} M., {Rudie} G.~C., {Rakic} O., 2010, \apj, 717, 289

\bibitem[{{Stocke} {et~al}\mbox{.}(2013){Stocke}, {Keeney}, {Danforth},
  {Shull}, {Froning}, {Green}, {Penton}, \& {Savage}}]{2013ApJ...763..148S}
{Stocke} J.~T., {Keeney} B.~A., {Danforth} C.~W., {Shull} J.~M., {Froning}
  C.~S., {Green} J.~C., {Penton} S.~V., {Savage} B.~D., 2013, \apj, 763, 148

\bibitem[{{Tripp} {et~al}\mbox{.}(2002){Tripp}, {Jenkins}, {Williger}, {Heap},
  {Bowers}, {Danks}, {Dav{\'e}}, {Green}, {Gull}, {Joseph}, {Kaiser},
  {Lindler}, {Weymann}, \& {Woodgate}}]{2002ApJ...575..697T}
{Tripp} T.~M. {et~al.}, 2002, \apj, 575, 697

\bibitem[{{Tripp} {et~al}\mbox{.}(2011){Tripp}, {Meiring}, {Prochaska},
  {Willmer}, {Howk}, {Werk}, {Jenkins}, {Bowen}, {Lehner}, {Sembach}, {Thom},
  \& {Tumlinson}}]{2011Sci...334..952T}
{Tripp} T.~M. {et~al.}, 2011, Science, 334, 952

\bibitem[{{Tumlinson} {et~al}\mbox{.}(2011){Tumlinson}, {Thom}, {Werk},
  {Prochaska}, {Tripp}, {Weinberg}, {Peeples}, {O'Meara}, {Oppenheimer},
  {Meiring}, {Katz}, {Dav{\'e}}, {Ford}, \& {Sembach}}]{2011Sci...334..948T}
{Tumlinson} J. {et~al.}, 2011, Science, 334, 948

\bibitem[{{York} {et~al}\mbox{.}(2000){York}, {Adelman}, {Anderson},
  {Anderson}, {Annis}, {Bahcall}, {Bakken}, {Barkhouser}, {Bastian}, {Berman},
  {Boroski}, {Bracker}, {Briegel}, {Briggs}, {Brinkmann}, {Brunner}, {Burles},
  {Carey}, {Carr}, {Castander}, {Chen}, {Colestock}, {Connolly}, {Crocker},
  {Csabai}, {Czarapata}, {Davis}, {Doi}, {Dombeck}, {Eisenstein}, {Ellman},
  {Elms}, {Evans}, {Fan}, {Federwitz}, {Fiscelli}, {Friedman}, {Frieman},
  {Fukugita}, {Gillespie}, {Gunn}, {Gurbani}, {de Haas}, {Haldeman}, {Harris},
  {Hayes}, {Heckman}, {Hennessy}, {Hindsley}, {Holm}, {Holmgren}, {Huang},
  {Hull}, {Husby}, {Ichikawa}, {Ichikawa}, {Ivezi{\'c}}, {Kent}, {Kim},
  {Kinney}, {Klaene}, {Kleinman}, {Kleinman}, {Knapp}, {Korienek}, {Kron},
  {Kunszt}, {Lamb}, {Lee}, {Leger}, {Limmongkol}, {Lindenmeyer}, {Long},
  {Loomis}, {Loveday}, {Lucinio}, {Lupton}, {MacKinnon}, {Mannery}, {Mantsch},
  {Margon}, {McGehee}, {McKay}, {Meiksin}, {Merelli}, {Monet}, {Munn},
  {Narayanan}, {Nash}, {Neilsen}, {Neswold}, {Newberg}, {Nichol}, {Nicinski},
  {Nonino}, {Okada}, {Okamura}, {Ostriker}, {Owen}, {Pauls}, {Peoples},
  {Peterson}, {Petravick}, {Pier}, {Pope}, {Pordes}, {Prosapio},
  {Rechenmacher}, {Quinn}, {Richards}, {Richmond}, {Rivetta}, {Rockosi},
  {Ruthmansdorfer}, {Sandford}, {Schlegel}, {Schneider}, {Sekiguchi}, {Sergey},
  {Shimasaku}, {Siegmund}, {Smee}, {Smith}, {Snedden}, {Stone}, {Stoughton},
  {Strauss}, {Stubbs}, {SubbaRao}, {Szalay}, {Szapudi}, {Szokoly}, {Thakar},
  {Tremonti}, {Tucker}, {Uomoto}, {Vanden Berk}, {Vogeley}, {Waddell}, {Wang},
  {Watanabe}, {Weinberg}, {Yanny}, {Yasuda}, \& {SDSS
  Collaboration}}]{2000AJ....120.1579Y}
{York} D.~G. {et~al.}, 2000, \aj, 120, 1579

\end{thebibliography}
\end{document}